\documentclass[nofootinbib,
eqsecnum,tightenlines,11pt]{revtex4}

\usepackage{hyperref}
\usepackage{graphicx}
\usepackage{amsmath,amssymb,amsfonts,amsthm,stmaryrd,mathtools}
\usepackage{mathrsfs}
\usepackage{color}
\usepackage{multirow,bigdelim}
\usepackage{dsfont}
\allowdisplaybreaks[0]

\makeatletter
\newcounter {subsubsubsection}[subsubsection]
\renewcommand\thesubsubsubsection{\thesubsubsection .\@alph\c@subsubsubsection}
\newcommand\subsubsubsection{\@startsection{subsubsubsection}{4}{\z@}%
                                     {-3.25ex\@plus -1ex \@minus -.2ex}%
                                     {1.5ex \@plus .2ex}%
                                     {\centering\normalfont\small\textit}}
\newcommand*\l@subsubsubsection{\@dottedtocline{3}{10.0em}{4.1em}}
\newcommand*{\subsubsubsectionmark}[1]{}
\makeatother

\newtheorem{Definition}{Definition}[section]

\newcommand{\pic}[1]{
\vcenter{\hbox{\includegraphics[scale=0.6]{#1-eps-converted-to.pdf}}}
}

\def\be#1\ee{\begin{align}#1\end{align}}

\def\ba{\begin{eqnarray}}
\def\ea{\end{eqnarray}}
\def\la{\langle}
\def\ra{\rangle}

\def\lp{\ell_\text{Pl}}

\def\f{\frac}

\def\openone{\mathds{1}}

\def\E{{\mathrm{\tiny{E}}}}
\def\lc{\ell_\text{c}}
\def\nn{\nonumber}
\def\q{\qquad}
\def\hs{\hspace{-0.3cm}}
\def\pound{\raisebox{0.03cm}{\text{\,\small{$\#$}\,}}}

\def\tr{\mathrm{tr}}
\def\i{\mathrm{i}}
\def\k{\mathrm{k}}

\def\de{\mathrm{d}}
\def\SU{\mathrm{SU}}

\def\su{\mathfrak{su}}

\def\bd{\mathbf{d}}

\def\bO{\mathbf{O}}
\def\Q{\mathcal{Q}}
\def\D{\mathcal{D}}
\def\C{\mathcal{C}}
\def\O{\mathcal{O}}
\def\V{\mathcal{V}}
\def\H{\mathcal{H}}
\def\E{\mathcal{E}}
\def\R{\mathcal{R}}
\def\G{\mathcal{G}}
\def\Z{\mathcal{Z}}
\def\K{\mathcal{K}}
\begin{document}

\title{Quantum gravity kinematics from extended TQFTs}

\author{Bianca Dittrich}
\author{Marc Geiller}
\affiliation{Perimeter Institute for Theoretical Physics,\\ 31 Caroline Street North, Waterloo, Ontario, Canada N2L 2Y5}

\begin{abstract}
In this paper, we show how extended topological quantum field theories (TQFTs) can be used to obtain a kinematical setup for quantum gravity, i.e. a kinematical Hilbert space together with a representation of the observable algebra including operators of quantum geometry. In particular, we consider the holonomy-flux algebra of $(2+1)$-dimensional Euclidean loop quantum gravity, and construct a new representation of this algebra that incorporates a positive cosmological constant. 

The vacuum state underlying our representation is defined by the Turaev--Viro TQFT. This vacuum state can be thought of as being peaked on connections with homogeneous curvature. We therefore construct here a generalization, or more precisely a quantum deformation at root of unity, of the previously-introduced SU(2) BF representation.

The extended Turaev--Viro TQFT provides a description of the excitations on top of the vacuum. These curvature and torsion excitations are classified by the Drinfeld centre category of the quantum deformation of SU(2), and are essential in order to allow for a representation of the holonomies and fluxes. The holonomies and fluxes are generalized to ribbon operators which create and interact with the excitations. These excitations agree with the ones induced by massive and spinning particles, and therefore the framework presented here allows automatically for a description of the coupling of such matter to $(2+1)$-dimensional gravity with a cosmological constant.

The new representation constructed here presents a number of advantages over the representations which exist so far. In particular, it possesses a very useful finiteness property which guarantees the discreteness of spectra for a wide class of quantum (intrinsic and extrinsic) geometrical operators. Also, the notion of basic excitations leads to a so-called fusion basis which offers exciting possibilities for the construction of states with interesting global properties, as well as states with certain stability properties under coarse graining.

In addition, the work presented here showcases how the framework of extended TQFTs, as well as techniques from condensed matter, can help design new representations, and construct and understand the associated notion of basic excitations. This is essential in order to find the best starting point for the construction of the dynamics of quantum gravity, and will enable the study of possible phases of spin foam models and group field theories from a new perspective.
\end{abstract}

\maketitle

\newpage

\tableofcontents

\newpage

\section{Introduction}

\noindent One of the key conceptual lessons of Einstein's general relativity is that gravity is encoded in the very geometry of spacetime. This suggests in turn that quantum gravity might be realized as a theory of quantum geometry. This concept is a cornerstone of loop quantum gravity (LQG hereafter). LQG, which comes in complementary canonical and covariant formulations \cite{lqg1,lqg2,lqg3,baez,oriti-thesis,perez-review,AGN}, provides a quantization of geometrical observables (more precisely the spatial metric and the extrinsic curvature of the spatial hypersurfaces), and in this sense defines a realization of quantum geometry. The geometrical observables are encoded in holonomies of an $\SU(2)$ connection (called the Ashtekar--Barbero connection \cite{ashtekar-variables,ashtekar-variables2,barbero,immirzi}) and canonically-conjugated flux variables. Therefore, realizations of quantum geometry can be understood as being given by representations of the holonomy-flux algebra formed by these variables.

Several such realizations of quantum geometry are now known. Each realization is based on and can be characterized by a different kinematical vacuum state. So far, all representations have in common that this vacuum is totally squeezed, i.e. sharply peaked either on the flux operators or on the holonomy operators. The operators of quantum geometry act on this vacuum state, and generate thereby all the excitations of the Hilbert space underlying the representation of the holonomy-flux algebra. The vacuum is therefore cyclic with respect to the set of  holonomy and flux operators.

The Ashtekar--Lewandowski (AL) representation \cite{ali1,ali2,ali3,ali4} was the first one to be constructed and, importantly, it allows to take into account invariance under spatial diffeomorphisms: The vacuum underlying this representation is diffeomorphism-invariant, which allows the construction of a (spatially) diffeomorphism-invariant Hilbert space. In the AL representation the vacuum is a totally squeezed state sharply peaked on vanishing expectation values of the fluxes, whereas the holonomy variables have maximal uncertainty. The AL vacuum state therefore describes a totally degenerate spatial geometry. This renders the construction of states describing large scale geometries very cumbersome. This difficulty motivated Koslowski and Sahlmann to introduce a variant of the AL representation, which essentially amounts to adding to the action of the fluxes a term describing a background flux (field) $E$ \cite{KS1,KS2,KS3}. By doing so, the vacuum state is then sharply peaked on this background flux field, while the holonomies are still maximally uncertain. Although this vacuum is not (spatially) diffeomorphism-invariant anymore, an action of diffeomorphisms can still be defined \cite{varadarajan1,varadarajan2,varadarajan3}.
  
An alternative path has recently been pointed out in \cite{bodendorferKodama}. It consists in using a (classical) canonical transformation affecting only the fluxes by adding to them a curvature-dependent term multiplied by an arbitrary constant $\theta$ \cite{PerezTheta}. By doing so, the new flux variables and the Ashtekar--Barbero connection still form a canonically-conjugated pair of variables, and one can therefore consider e.g. the usual AL representation based on these new variables. The corresponding AL vacuum state is then given by vanishing expectation values in the new fluxes, which, due to the $\theta$ shift, describes a geometry with certain symmetry properties\footnote{The geometry is not maximally symmetric since this would involve a condition on the extrinsic curvature, which includes the connection. The connection variables are however put into a maximally uncertain state.}. A disadvantage of this proposal is however that the basic (spatial) geometrical operators, like the area or the volume, cannot anymore be expressed easily in terms of the new fluxes.

A third possibility, introduced by the authors in \cite{DG1}, constructs a new representation of quantum geometry by dualizing many ingredients of the AL representation. The construction has been completed in \cite{DG2}, and in \cite{BDG} with Bahr. This representation is based on a kinematical vacuum state which is sharply peaked on flat connections, whereas the fluxes have maximal uncertainty\footnote{In order to have a well-defined notion of maximal uncertainty for the fluxes, one only allows for the action of exponentiated fluxes.}. Since the space of flat connections has a much richer structure than the space of vanishing fluxes, certain details of the construction (in particular the definition of cylindrically consistent gauge covariant flux observables) are much more involved\footnote{The AL Hilbert space can be expressed as an $L^2$ space over an extended configuration space of connections \cite{lqg3}. For the representation \cite{DG1}, one would expect some measure space over an extended configuration space of fluxes. However, the fluxes are inherently non-commutative, which would therefore require a configuration space with a non-commutative multiplication. This has however only been worked out for the AL case so far \cite{NCFlux1,NCFlux2}.} than in the AL case (where one usually works with fluxes which are not gauge covariant). The vacuum state is also invariant under spatial diffeomorphisms, and one can therefore achieve a diffeomorphism-invariant representation. This representation is better suited for semi-classical constructions, and in particular for connecting canonical loop quantum gravity to spin foams. The reason is that spin foams are based on BF theory \cite{BF}, a topological field theory whose physical states are also states with (locally) flat connections. For this reason, we refer to the representation worked out in \cite{DG1,DG2,BDG} as the ``BF representation''.


The fact that there are now several available realization of quantum geometry raises the question of whether there exist further possible generalizations. It was actually proposed in \cite{DittStein} to build representations by choosing a suitable topological quantum field theory (TQFT). The physical state (which is unique for a hypersurface with trivial topology) of this TQFT serves as a vacuum state, and the TQFT also determines the possible excitations which the representation can have. These excitations appear in the form of defects. These can be studied in the framework of extended TQFTs, which can broadly be understood as constructing TQFTs with boundaries. A particular class of boundaries, say for a two-dimensional hypersurface, are punctures obtained by removing disks from the hypersurface. These punctures will carry the defect excitations. A state with defect excitations is also a physical state but now, due to the punctures, on a topologically non-trivial hypersurface.

So far, we have not mentioned the (holonomy-flux) observable algebra. This is the next question which one has to address: Can one define operators, including creation and annihilation operators for the defect excitations, that realize a representation of the observable algebra? To get a representation of the full holonomy-flux observable algebra (which includes holonomies and fluxes that are allowed to act anywhere), one needs to enlarge the framework of extended TQFTs by a so-called inductive limit. This procedure constructs, out of the Hilbert spaces with a fixed number of defects at fixed positions, a continuum Hilbert space where defects (and thus the associated excitations) can appear in any number and at any position.

Both the AL and the BF representations can be understood in these terms. Indeed, in the case of the AL representation, the underlying TQFT is a trivial one imposing vanishing fluxes, and the ``defects'' describe excitations of spatial geometry. In the case of the BF representation, the underlying TQFT is BF theory, and the defects are given by curvature excitations\footnote{Torsion excitations are also possible, but in the case of $\SU(2)$ they lead to non-normalizable states.}.

Having at hand different representations enlarges hugely our toolbox for describing different regimes of quantum gravity. A key open problem is a construction of the dynamics and the continuum limit of the theory, two issues which are deeply intertwined with each other \cite{BD14}. To this end, it would be helpful to have a representation which is based on a vacuum state which is as close as possible to a \textit{physical} state, i.e. a state satisfying the constraints of the theory. The BF representation delivers such a state in $(2+1)$ dimensions with a vanishing cosmological constant. This also holds in $(3+1)$ dimensions for a certain operator ordering of the Hamiltonian constraint\footnote{Actually, only the so-called Euclidean part of the Hamiltonian constraint.}. However, certain issues remain, as the geometry encoded by this vacuum is rather a generalized (so-called) twisted one \cite{DittSpez,DittRyan1,Twist,DittRyan2}. We are thus naturally led to the question of whether there are TQFTs leading to a better-suited vacuum state.

From this point of view, it is therefore helpful to understand all possible $(2+1)$- and $(3+1)$-dimensional TQFTs which admit a geometrical interpretation. A related question is to understand all possible phases in condensed matter, where there is lots of recent progress in $(2+1)$ dimensions (see for example \cite{CondMatt1,CondMatt2}). One possible way to find TQFTs, in particular with a geometrical interpretation, is to study the coarse graining of spin foams, as the end points of the coarse graining flows give typically topological models \cite{DittrichKaminski,QGSpinnets,DecTNW,CDtoappear}.

Apart from finding new TQFTs, we can also consider TQFTs which are known, as for example the Turaev--Viro (TV hereafter) model in $(2+1)$ dimensions \cite{TV,BW} and the Crane--Yetter model in $(3+1)$ dimensions \cite{CY,CKY}. These models can both be understood as so-called quantum deformations of BF theory, where the classical gauge group $\SU(2)$ is replaced by its quantum deformation $\mathcal{U}_q\big(\su(2)\big)$ with deformation parameter $q$ a root of unity.

The aim of the present paper is to develop a representation of the holonomy-flux algebra based on the TV vacuum. This will in particular show that the general strategy proposed in \cite{DittStein} does work in the example of the TV TQFT, and that this strategy and the techniques outlined in this paper may also be applied to other TQFTs.

The TV model describes Euclidean quantum gravity with a positive cosmological constant. There exists a tight relationship between the quantum deformation parameter and the cosmological constant, and a similar role of the quantum deformation is believed to hold also in $(3+1)$ dimensions \cite{Carlo,SmolinTop,SmolinQ,Girelli1,BDGirelli,AldoLambda,Aldo2,Aldo3}. Another important feature of the quantum deformation at root of unity is that one can expect the Hilbert spaces associated to a fixed graph (which here will be replaced by Hilbert spaces on manifolds with fixed punctures) to be finite-dimensional. This will avoid certain technical inconveniences of the (undeformed) BF representation, in particular the need to resort to a Bohr compactification of the dual of the group $\SU(2)$. Furthermore, the TV model in its extended form has been quite recently developed  mathematically \cite{BalKir,Balsam1,Balsam2,BalsamThesis} and has also found widespread applications in condensed matter and quantum computation \cite{LevinWen,KKR,Lan,Wu}. In particular, the structure of the excitations for this model is very well understood. For this reason, we will concentrate here on the $(2+1)$-dimensional case, and leave the $(3+1)$-dimensional one for future development.

As mentioned above, the $(2+1)$-dimensional TV model describes quantum gravity with a cosmological constant. Therefore, the physical states of the theory are peaked on homogeneous curvature. Hence, one could think that a corresponding representation can be achieved from the BF representation, in which the vacuum is peaked on vanishing curvature, by using a similar construction as when shifting from the AL representation to the KS one. There, one shifts the momenta (given by the fluxes in the AL representation) by a constant term which is encoding the background flux field. However, homogeneous curvature means that the curvature is a functional of the fluxes themselves\footnote{In $(2+1)$ dimensions one has $F=\Lambda(e\wedge e)$, where $F$ is the curvature two-form and $e$ the triad which can built from the fluxes.}, and thus we face a much more complicated situation than in the KS construction (where the background flux field cannot depend on the connection). Additionally, we will still have a diffeomorphism-invariant vacuum state, in contrast to the KS representation.

Lots of previous work has aimed at constructing the physical Hilbert space, or in other words the vacuum state peaked on homogeneous curvature, starting from the AL representation \cite{Perez:2010pm,Noui:2011im,PranzettiLambda}. This has not yet been attempted by starting from the BF representation, since this formulation is a very recent one. However, at the classical level, it has been shown that Regge calculus with homogeneously curved building blocks arises from coarse graining Regge calculus with flat building blocks and a cosmological constant term \cite{Improved,NewRegge}.
  
While the derivation of the quantum deformed structure from the canonical quantization (in either the AL or BF representation) of gravity with a cosmological constant is still open, here we will simply assume the quantum group structure from the onset. Since we aim at representing the full holonomy-flux algebra, we will not only consider the vacuum state, but also excitations on top of this vacuum. It turns out that these excitations do agree with the ones that would be induced by coupling particles to gravity\footnote{Here we interpolate from the flat case, i.e. the coupling of particles to the Ponzano--Regge model \cite{PR1,NouiPerez,Noui} to the homogeneously curved case. To our knowledge, an explicit coupling of particles to the TV model has not yet been discussed.}. Notice that this does \textit{not} show that we have coupled matter to gravity. Rather, it turns out that particles lead to the most elementary excitations of quantum geometry, which coincide with the defect excitations in the BF (for the flat case) or TV model.

In constructing a realization of quantum geometry based on the TV TQFT, we will rely on a broad range of previous works from condensed matter and mathematical physics. Indeed, the Hilbert space (for a fixed number of punctures) is closely related to so-called string net models \cite{LevinWen,KKR}, a mathematical exposition of which can be found in \cite{Kir}. The (defect) excitations allow for anyonic quantum computation as is explained in \cite{KKR}. The classification of these excitations goes back to an argument by Ocneanu \cite{Ocneanu1,Ocneanu2}, and is discussed in \cite{Lan} and \cite{Kir}. Mathematically, this classification relies on the definition of the Drinfeld centre of a fusion category, which has been explored in \cite{Mueger1,Mueger2}. The relation between the inner product of the Hilbert space and the (extended) TV TQFT is discussed in \cite{BalsamThesis,Kir}. We are going to see that the holonomy and flux operators appear as ribbon operators. Closed ribbon operators are discussed in \cite{LevinWen}. Open ribbon operators are mentioned in \cite{Lan} and are defined in the context of a fixed (dual) lattice in \cite{Wu}. However the detailed definition of open ribbon operators via fusion of punctures is only (as far as we are aware of) developed in the present work. Likewise, the entire setup of the inductive limit which is allowing for an arbitrary number of punctures and ensuring cylindrical consistency, is new.

The aim of the present paper is to be as self-contained as possible, and we will therefore provide a review of some of the material mentioned in the previous paragraph. We will also provide a quantum geometry interpretation (as far as it is available) of the structures and operators which will appear in this work. We hope that this paper will introduce a number of techniques from condensed matter and extended TQFTs into the (loop) quantum gravity community. In particular, we believe that these techniques can be generalized and might lead to an entire class of new realizations of quantum geometry.

Apart from providing an important case study of how to construct a realization of quantum geometry starting from a TQFT, the TV representation possesses a number of advantages, both over the AL and the BF representation. We here list a few of these.
\begin{itemize}
\item The Hilbert spaces based on a fixed number of punctures (which are analogous to the Hilbert spaces based on fixed triangulations in the BF representation, or the Hilbert spaces based on fixed graphs in the AL representation) are finite-dimensional. Consequently, the spectra of operators which do not introduce new punctures are discrete\footnote{This expectation (in the context of an AL-based representation of quantum group deformed holonomies) has also been pointed out in \cite{RovVid}. This paper proposes an action for the holonomy operators but not for the fluxes. Also, the continuum description of an AL-based representation of a quantum group deformed holonomy-flux algebra is so far missing, but will be forthcoming in \cite{GirelliAND}.}. This is a major advantage compared to the BF representation where spectra of geometric operators are continuous due to a compactification of the dual of the Lie group which leads to an aperiodic winding of the spectra. This turns a ``discrete looking'' spectrum into a continuous one (see \cite{BDG,chaos1,chaos2}). The natural cutoff on the spins provided by the quantum group deformation (at root of unity) also avoids possible (infrared) divergencies.

\item In the context of constrained quantization, a finite-dimensional Hilbert space offers also many advantages\footnote{This would be relevant for the four-dimensional case, assuming that we can generalize the current construction to four dimensions. For the three-dimensional case one can consider a setup where one starts with a kinematical Hilbert space based on some positive cosmological constant, but seeks for a dynamics with a different cosmological constant. Also, one would have to go through the usual constrained quantization if one couples matter, and in this case one has only finite-dimensional Hilbert spaces if the matter content does not destroy this property.}, as the construction of a so-called physical inner product is much simpler in a finite dimensional setup (see e.g. \cite{master1} for a discussion). Here, we have however to assume that the constraints can be quantized in such a way that they do not change the underlying discrete structure (here the number of punctures) of the states.

This finiteness with quantum group structure is also a huge advantage if one want to consider the coarse graining of the models. First, it allows for a numerical implementation of so-called tensor network algorithms for coarse graining \cite{Levin,S3,DecTNW,CDtoappear}. For this reason, quantum group models are used in \cite{QGSpinnets,SteinhausM,DittSteinToappear}. Second, spin foam models based on the undeformed $\SU(2)$ group feature divergencies due to the unbounded summation over spins (see \cite{Linquing} and references therein). These divergencies are avoided if one works with a quantum deformation at root of unity.

\item In the case of the BF representation, the space of fluxes (corresponding to the dual of the group $\SU(2)$) is compactified. This requires the exponentiation of the fluxes, and with this the introduction of an additional exponentiation parameter (usually called $\mu$). In the four-dimensional case this parameter can be absorbed into the Barbero--Immirzi parameter (see \cite{BDG}), and thus one still has the same number of additional parameters. In three spacetime dimensions however, one can define the theory without a Barbero--Immirzi parameter, and one has thus a priori (i.e. on the kinematical Hilbert space) an additional parameter, which needs to be fixed via some additional physical principle. In the quantum group case the compactification is provided naturally, and we do not need an additional parameter. In some sense this parameter is rather provided by the cosmological constant, whose inverse represents a cutoff on the admissible spins.

\item We provide here the first continuum construction of a holonomy-flux representation that incorporates a (positive) cosmological constant. This is based on a vacuum state describing homogeneously curved geometries. A full continuum description has so far not been achieved even with an AL-like vacuum, but will appear soon in \cite{GirelliAND}.

\item The framework presented here allows for a very natural identification of ``basic excitations''. These basic excitations turn out to be described with the same quantum numbers as particles coupled to gravity. Therefore, we have a natural starting point for describing the coupling of point particles (and possibly other types of matter) to gravity with a cosmological constant, which to our knowledge has so far not been discussed in the literature. We will also see that structures such as the Drinfeld double will appear quite naturally here, while they have been derived in a more complicated fashion for the discussion of point particles coupled to gravity without cosmological constant \cite{Freidel2,Noui,MeusburgerNoui}.

\item Finally, the framework presented here introduces a new kind of basis, based on the ``basic excitations'' and a coarse graining of these basic excitations. This is the so-called fusion basis, which offers a range of new possibilities to design states with global properties much more effortlessly than in the spin network basis.
\end{itemize}

This paper is organized as follows: In the next section, we briefly review the most important features of the BF representation and expose, by comparison, the main results of the TV representation developed in the rest of the paper. In section \ref{sec:graphical} we present the tools of graphical calculus which will enable us to depict and manipulate the various mathematical structures (such as states, excitations, \dots) playing a role in our construction. Section \ref{sec:graph hilbert space} presents the construction of the so-called graph Hilbert space based on punctured manifolds, together with its basis and inner product, and subsequently the characterization of the vacuum state in the case of spherical topology. In section \ref{section:2p}, we focus on the two-punctured sphere in order to explain the structure of the basic quasi-particle excitations on top of the vacuum, and introduce explicitly various mathematical notions (such as the half-braiding of the Drinfeld center) which enable the manipulation of these excitations. Section \ref{tpsphere} is devoted to the study of the three-punctured sphere, which is necessary in order to define the fusion of quasi-particle excitations and in turn the action via fusion of the ribbon creation operators. The detailed properties and the action of open and closed ribbon operators are discussed in section \ref{sec:creation-operators}, where it is also explained how this forms a representation of the holonomy-flux algebra. Finally, we present the conclusion and some perspectives in section \ref{sec:conclusion}.

\section{Outline of BF and TV representations}

\noindent In the first part of this section we shortly review the construction of the BF representation \cite{DG1,DG2,BDG}. This will make easier the understanding of the main similarities and differences with the new TV representation, which we outline in the second part of this section.

\subsection{BF representation and flat curvature vacuum}
\label{BFrep}

\noindent The continuum Hilbert spaces of both the AL and the BF representation are built as an inductive limit of Hilbert spaces. This inductive limit construction will also hold for the TV representation. To construct an inductive limit Hilbert space, one considers a family of Hilbert spaces $\H_\Delta$ labeled (in the case of the BF representation) by embedded triangulations. These triangulations are also equipped with a flagged root (see \cite{DG2} for a precise definition) and carry a partial order $\prec$. Furthermore, this partial order is also directed in the sense that a triangulation $\Delta'$ is finer than $\Delta$, which we denote by $\Delta\prec\Delta'$, iff $\Delta'$ can be obtained from $\Delta$ by a sequence of refinement operations (specified in \cite{DG2,BDG}). Additionally, given two triangulations $\Delta$ and $\Delta'$, one can always find a triangulation $\Delta''$ which is a refinement of both $\Delta$ and $\Delta'$. 

Together with this family of Hilbert spaces, we need embedding maps $\E_{\Delta\Delta'}:\H_\Delta\rightarrow\H_{\Delta'}$ which embed isometrically the states in the Hilbert space based on a coarser triangulation into the Hilbert space based on a finer triangulation. These embedding maps need to satisfy a number of consistency relations, the most important one being
\be
\E_{\Delta',\Delta''}\circ\E_{\Delta,\Delta'}=\E_{\Delta, \Delta''}
\ee
for any triple $\Delta\prec\Delta'\prec\Delta''$. This condition implements the fact that the embedding map of a Hilbert space $\H_\Delta$ into a finer $\H_{\Delta''}$ does not depend on the number of intermediate steps which one might take in order to arrive at this map.

The inductive limit Hilbert space is formally defined as the (closure of the) disjoint union of all Hilbert spaces modulo the following equivalence relation: Two states $\psi_\Delta$ and $\psi'_{\Delta'}$ are equivalent if there exists an embedding map $\E_{\Delta,\Delta'}$ such that
\be
\E_{\Delta,\Delta'}(\psi_\Delta)=\psi'_{\Delta'}.
\ee
This means that states are equivalent if they can be made equal under some refinement operation.

The embedding maps can be interpreted as (locally) specifying a vacuum state \cite{BD12b}. Indeed, excitations are restricted to occur at the vertices of the triangulation in $(2+1)$ spacetime dimensions or at the edges of the triangulation in $(3+1)$ dimensions. Thus, regions without vertices or edges do not carry any excitations and are (implicitly) in a vacuum state. If a refinement of the triangulation adds vertices or edges to this region, the corresponding refinement map has to result in an equivalent state, i.e. a state which still assigns a (quasi-local) vacuum to this region. Since a refinement adds additional variables, the refinement map has to assign to the additional variables  the vacuum configuration, thereby making the vacuum state as a function of these additional variables explicit. 

This interpretation does not only hold for the BF representation, but also for the AL one where the vacuum corresponds to vanishing spatial geometry. It will also hold for the TV representation which we construct in this work. In fact, this notion can be generalized to a physical vacuum and dynamical embedding maps, which might not necessarily be described by a topological field theory \cite{BD12b,BD14} (this would be relevant for $(3+1)$ or $(2+1)$ dimensions when coupling to a matter field).

In addition to the inductive limit Hilbert space construction (which as we explained exists for the other representations as well), the BF representation is based on the following specific ingredients:
\begin{itemize} 
\item \textbf{Hilbert space.} The Hilbert space $\H_\Delta$ associated to a fixed (rooted) triangulation consists of functions of holonomies on the graph $\Gamma$ (dual to $\Delta$) that are gauge-invariant everywhere except at the root. Such functions can be expressed as functions of holonomies assigned to a basis of independent cycles $\ell$ of $\Gamma$ (this can be most conveniently achieved by using a gauge-fixing along a spanning tree in the dual graph, and then the leaves with respect to the tree define a basis of independent cycles). We denote the states by $\psi\{g_\ell\}$. To make the above-mentioned embedding maps well-defined and isometric, we have to equip the configuration space of holonomies with a discrete topology. This means that the inner product on $\H_\Delta$ is given by
\be
\la\psi_1|\psi_2\ra=\int\overline{\psi_1\{g_\ell\}}\psi_2\{g_\ell\}\prod_\ell\bd\mu_\text{d}(g_\ell),
\ee
where $\mu_\text{d}$ is the discrete measure.

\item \textbf{Global vacuum.} The global vacuum $\psi_\emptyset$ for a fixed $\H_\Delta$ is given by
\be
\psi_\emptyset=\prod_\ell\delta(g_\ell,\openone),
\ee
where $\delta(g_\ell,\openone)=1$ iff $g_\ell=\openone$, i.e. $\delta(\cdot,\cdot)$ is the Kronecker delta. Note that this state is well-defined and normalized in the discrete topology inner product which we have defined for $\H_\Delta$.

\item \textbf{Embedding maps.}  A refinement $\Delta\rightarrow\Delta'$ of a triangulation leads to \textit{additional} independent cycles $\{\ell'\}$ in the refined dual graph $\Gamma'$ as compared to $\Gamma$ (with a basis of independent cycles $\{\ell\}$). The embedding maps are then given by
\be
\E_{\Delta,\Delta'}(\psi_\Delta)\{g_\ell;g_{\ell'}\}=\psi_\Delta\{g_\ell\}\prod_{\ell'}\delta(g_{\ell'},\openone).
\ee
Therefore, the additional holonomy observables $\{g_{\ell'}\}_{\ell'}$ are put into the vacuum state, which is given by the (Kronecker) delta function peaked on vanishing curvature.

\item \textbf{Gauge invariance.} In the case of the BF representation, one works with states that are gauge-invariant at every node of the dual graph except at the root. The reason for working with such (almost) gauge-invariant states is that due to the discrete topology on the gauge group, gauge averaging leads to non-normalizable states. For the root one can define a group-averaging (rigging map) procedure that is compatible with the embedding maps defined above (see \cite{BDG} for details).

\item \textbf{Representation of holonomy operators.} The holonomies act (in the holonomy representation) as multiplication operators. They leave the vacuum invariant and therefore do not lead to (curvature) excitations. In order to preserve gauge invariance, (non-trace) holonomies are only allowed to start and end at the root.

\item \textbf{Representation of exponentiated fluxes.} In the AL representation, fluxes act as (Lie) derivatives on the holonomies. This action is not well defined with the discrete topology of the BF representation, and there we rather have to work with the exponentiated action of the flux operators. This is in turn the generator of translations, whose action at the level of the group is given by multiplication from the left or right on the (group) argument of the wave function.

In order to preserve gauge invariance, we have to parallel transport the (exponentiated) fluxes to the root. For a multiplication from the right we therefore have e.g.
\be
R^\alpha_i\psi\{g_\ell\}=\psi(g_1,\dots,g_ih_{rs(i)}\alpha h^{-1}_{rs(i)},\dots,g_{|\ell|}),
\ee
where $h_{rs(i)}$ is the holonomy from the root to the source node of the link $i$. Furthermore, one can compose fluxes along edges of the triangulation (in $(2+1)$ dimensions) or triangles of the triangulation (in $(3+1)$ dimensions), as described in detail in \cite{DG2}.

Since an exponentiated flux acts by multiplication, it shifts the argument of the vacuum state which is given by the Kronecker delta. This therefore leads to a curvature excitation for the faces that border the link associated to the shifted group argument. In $(2+1)$ dimensions, this means that we get (opposite) curvatures associated to the two vertices of the edge dual to the link.

A composed flux along a number of edges (giving a so-called co-path) will lead to excitations only at the boundaries of this co-path. In $(2+1)$ dimensions, these operators constitute a special case of the so-called ribbon operators introduced by Kitaev in the case of finite groups \cite{Kitaev1}. In the case of \cite{Kitaev1} the ribbons also include an action of the holonomy operator. With a finite group, it is less problematic to allow for torsion degrees of freedom (i.e. violations of gauge invariance), and thus to have holonomies starting and ending at arbitrary nodes. Those ribbon operators can lead to both curvature excitations and torsion excitations at the ``ends'' of the ribbon (where the ``ends'' of a ribbon now include a vertex of the triangulation, at which one can have curvature excitation, and a node of the dual graph, at which one can have a torsion excitation).

\item \textbf{Cylindrical consistency of operators.} To a given (holonomy or exponentiated flux) operator $\O_\Delta$ associated to a fixed triangulation $\Delta$, one can assign a refined operator $\O_{\Delta'}$ such that the action of the operator intertwines with the refinement map in the sense that
\be
\E_{\Delta, \Delta'}\circ\O_\Delta=\O_{\Delta'}\circ\E_{\Delta,\Delta'}.
\ee
This makes the operators well-defined on the inductive limit (and therefore continuum) Hilbert space $\H_\infty$. For the refinement of the flux operators, the notion of composition of fluxes (leading in $(2+1)$ dimensions to the notion of ribbons) is essential.
\end{itemize}

Now that we have recalled the essential features of the BF representation, we explain the main results about the TV representation which we construct in the rest of this paper.

\subsection{Outline of the TV representation}

\noindent Our goal is to construct a representation of a (possibly deformed) holonomy-flux algebra, with a vacuum state given by the physical state of the TV model \cite{TV}. The TV model is based on the fusion category $\SU(2)_\k$, the objects of which are representations of the quantum group $\mathcal{U}_q\big(\su(2)\big)$ (with $q=e^{2\pi\i/(\k+2)}$ a root of unity), which can in turn be tensored (or ``fused'') with each other. We will give more details about the properties of this fusion category in section \ref{sec:graphical}. Since the quantum group $\mathcal{U}_q\big(\su(2)\big)$ at root of unity does not posses a group representation, we are forced to work in the spin representation. This latter is however much easier to define and manipulate than in the BF case (see \cite{BDG}), as in contrast to $\SU(2)$ there are just finitely many (admissible) representations in $\SU(2)_\k$.

We are going to define Hilbert spaces spanned by a spin network basis, which is represented by a certain class of graphs whose edges are labelled by representations of $\SU(2)_\k$. Furthermore, we will define operators and equivalence relations on this basis by using a certain graphical calculus introduced in section \ref{sec:graphical}.

As mentioned above, we will also use for the TV representation an inductive limit construction. For the BF representation, the inductive limit was constructed with the help of a partially ordered set of triangulations. We will proceed differently for the TV representations. Here, the partially ordered set will be given by (spatial) manifolds $\Sigma_p$ where $p$ denotes the number of punctures (we will  restrict to spheres in the main text, and comment on the higher genus case in appendix \ref{appendix:higher-genus}). The reason is that the excitations are restricted to occur at the punctures. Thus, if we want to represent a more complicated state, defined by having more distributed excitations, we just need to add more punctures.

In some sense, the Hilbert space $\H_{\Sigma_p}$ associated to a fixed position and number of punctures will therefore already be a continuum Hilbert space (it is in fact the continuum Hilbert space associated to the corresponding TQFT on the hypersurface $\Sigma_p$). It will however only allow for excitations located at the punctures.  Nevertheless, the states will be represented by a combinatorial (discrete) structure, namely the labelled graphs. These graphs will be dual to a triangulation of the manifold with punctures.

To reconcile this with the notion of a continuum Hilbert space, we will allow for all possible triangulations and associated dual graphs, as long as these have the same punctures (which are embedded in $\Sigma$). Correspondingly, we will define equivalence relations on the states which are nothing else than analogs of refining moves and now also coarsening operations on the triangulation. For this to work, it is also essential that regions of $\Sigma_p$ not being crossed by strands of the graph can be interpreted as being in a (local) vacuum state. Therefore, the inductive limit over triangulations in the BF case will here, in the TV case, split into two parts. The first part is encoded into the equivalences between graph states with the same punctures, and the second part will be encoded in embedding maps that do add further punctures to a manifold $\Sigma_p$.

This brings us to the discussion of the vacuum state and the embedding maps. The vacuum state will again be a state without curvature\footnote{Note that there is here a subtlety related to the notion of ``absence of curvature". Since the TV model is a quantization of Euclidean gravity with a positive cosmological constant, its physical state, which serves as our vacuum, is a state with homogeneous curvature (i.e. locally representing a three-sphere). As encoded in the Chern--Simons formulation, this homogeneous curvature can always be reabsorbed in order to define a new (effective) connection which is flat. It is this notion of flatness which we are going to encode later on in terms of $F$-symbols. In this sense, when speaking about curvature, we therefore mean curvature in excess of the homogeneous curvature encoded by the vacuum state.} Later on, this will be translated explicitly into two properties: $(i)$ Wilson loops (in an $\SU(2)_\k$ representation) will act trivially, and $(ii)$ a strand of the graph can be deformed by moving this strand over a puncture with no excitation. This latter property can be understood as being able to deform a Wilson line, which is indeed only possible if there is no (excess) curvature. The vacuum state will also be without torsion\footnote{This torsion is defined as a violation of the Gauss constraints given by the TV model. Note that when translating these Gauss constraints in terms of fluxes they appear to be deformed as compared to the Gauss constraints in the flat case \cite{Aldo2,BDGL,CharlesLivineG}.}. Note that, in contrast to the BF representation, we will allow for the possibility of having excitations with torsion at the punctures (again this is made possible by the ``finiteness'' of $\SU(2)_\k$ as compared to $\SU(2)$). The embedding maps will be defined in such a way that any new puncture does carry vanishing curvature and torsion. The notion of  ``vanishing curvature and torsion'' will also be specified by constraints which do agree with the constraints imposed by the TV model.

Having discussed the vacuum state, we now come to the excitations and to the operators which may generate these excitations. There is a beautiful argument, going back to Ocneanu \cite{Ocneanu1,Ocneanu2}, which allows to define the notion of ``elementary excitation'' associated to a puncture. These excitations are indeed given by the violation of curvature and torsion constraints and turn out to be labelled by objects of the so-called Drinfeld centre of the fusion category, which itself is a fusion category \cite{Ocneanu1,Ocneanu2,Mueger1,Mueger2,Lan}. In our case, the labels can be interpreted in terms of mass and spin, denoting the amount of curvature and torsion constraint violation respectively.

Next, we define operators which do generate these excitations. As in the case of the $(2+1)$-dimensional BF representation these are given by ribbon operators, but are now however labelled by an object of the Drinfeld centre. We will argue that these ribbon operators provide indeed generalizations of the holonomy and flux operators. The (cylindrical) consistency of operators splits again into two parts. First, the operators need to be well-defined on a fixed Hilbert space $\H_{\Sigma_p}$, i.e. compatible with the equivalences of graph states which implement changes of the triangulation. This will be imposed by a so-called naturality condition of the half-braiding, a structure which comes with the Drinfeld centre. Second, the ribbon operators need to be compatible with the addition of punctures. This will follow from the ability of gluing ribbons together along punctures.

In summary, these ingredients will define a representation of a ($q$-deformed) holonomy-flux algebra based on the vacuum defined by the TV state sum model \cite{TV}. The excitations agree with the ones defined by the extended TQFT based on this model \cite{Kir}. The ribbon operators, which represent (combinations of) holonomies and fluxes allow us the generate, manipulate, and measure these excitations.

Having presented an overview of our results, we now turn to the details of the construction.

\section{Graphical calculus}
\label{sec:graphical}

\noindent In this section, we introduce the basic algebraic data which we are going to use for our construction, along with graphical rules which will enable us to efficiently depict and manipulate various structures (such as states, operators, morphisms, and inner products). As explained in the introduction, we are interested in data coming from the quantum group $\SU(2)_\k$ at root of unity, and more precisely in the fact that its representation theory gives rise to a modular fusion category. In order to avoid wandering through lengthy mathematical definitions, we are (safely) going to overlook the unnecessary details and instead focus on the physical meaning of the various axioms defining such a category. However, as a result of this simplification, one should remember that most of the properties introduced below will only hold for $\SU(2)_\k$ and not for arbitrary fusion categories.

We refer the reader to appendix \ref{appendix1} for details about the structure of $\SU(2)_\k$, and to \cite{EGNO} for generalities about fusion categories and their relation to quantum groups. A review of a different version of graphical calculus for $\SU(2)_\k$ can be found in \cite{Reshet,KaufLins}. Here we will mostly follow the notations of \cite{KKR}, although with certain slight adaptations.

\subsection{SU(2)${}_\text{k}$ as a modular fusion category}

\noindent Loosely speaking, a ribbon fusion category consists of a finite set of labels describing the objects of the category, so-called fusion coefficients and $F$-symbols describing the fusion structure for these labels, and an $R$-matrix describing their braiding\footnote{Technically, this only defines a braided fusion category. A ribbon fusion category requires an extra compatibility condition between the pivotal structure of the category and the braiding. Since this is satisfied for $\SU(2)_\k$ (which in fact gives rise to a spherical category) we ignore this detail and simply use the term ``ribbon fusion category''.}. A ribbon fusion category is sometimes also called a premodular fusion category, and when its topological $S$-matrix is non-degenerate it becomes a modular fusion category\footnote{Since the $S$-matrix is defined via the $R$-matrix, and therefore requires a notion of braiding, we will from now on omit the term ``ribbon'' when talking about a modular fusion category.}. We are going to explain why $\SU(2)_\k$ does indeed posses all this structure.

When the deformation parameter is a root of unity, the quantum group $\SU(2)_\k$ only has a finite number of irreducible representations with non-vanishing and non-cyclic quantum dimensions. These representations are labelled by half-integer spins $i,j,k,\dots$ which take values in the set $\mathcal{J}=\{0,\dots,\k/2\}$, and their quantum dimension is given by the quantum evaluation $d_j=[2j+1]$ defined in \eqref{qNumber}. This in itself already defines the structure of a category whose objects\footnote{In the condensed matter literature, these objects are often referred to as particles, anyons or charges. Here we will refrain from using such a nomenclature, and reserve the term ``anyon'' for the quasi-particle excitations on top of the vacuum which we will encounter later on.} are the representations and whose maps are morphisms between them. 

The fusion structure on this category comes from the existence of a well-defined tensor product between representations. More precisely, the recoupling of two representations is given by the fusion rule
\be\label{FusionRule}
i\times j=\sum_k\delta_{ijk}k,
\ee
where the $\delta_{ijk}$ are the so-called fusion coefficients, and where the sums should from now on be understood as running over the set $\mathcal{J}$ of spin labels. The fusion coefficients satisfy the conditions
\be\label{associativity fusion}
\delta_{i0k}=\delta_{0ik}=\delta_{ik},
\q\q
\sum_m\delta_{ijm}\delta_{mlk}=\sum_m\delta_{ikm}\delta_{mjl}.
\ee
The first condition simply means that the spin $j=0$ is the unit element of the fusion algebra, while the second one reflects the associativity of the fusion. Furthermore, the fusion coefficients satisfy the symmetry properties $\delta_{ijk}=\delta_{jki}=\delta_{kji}$. Explicitly, a triple of spins is said to be admissible, i.e. one has $\delta_{ijk}=1$, if and only if the following four conditions are satisfied:
\be\label{Admissibility}
i\leq j+k,
\q
j\leq i+k,
\q
k\leq i+j,
\q
i+j+k\leq\k,
\q
i+j+k\in\mathbb{N}.
\ee
If a triple is non-admissible then $\delta_{ijk}=0$. Since $\delta_{ijk}\in\{0,1\}$, we say that $\SU(2)_\k$ has no fusion multiplicities.

As a remark, let us point out that one can introduce the fusion matrices $\mathbf{N}_i$ whose coefficients are $(\mathbf{N}_i)_{jk}\coloneqq\delta_{ijk}$. Using \eqref{associativity fusion}, one can then compute that the product of two such matrices is given by $\mathbf{N}_i\mathbf{N}_j=\sum_k\delta_{ijk}\mathbf{N}_k$, which shows that the fusion matrices form a representation of the fusion algebra. We are later on going to encounter an action of these matrices on one-dimensional representations of the fusion algebra.

We are now ready to introduce the first elements of graphical calculus. For this, we simply assign to each representation of spin $j$ an unoriented and smoothly deformable strand
\be\label{StrandDeformation}
\pic{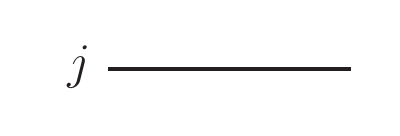}=\pic{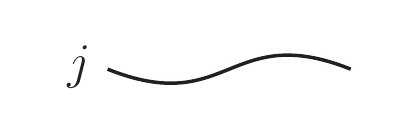}.
\ee
The absence of orientation comes from the fact that the representations are self-dual. This has an important subtle consequence, since one has to keep track of sign factors $\alpha_j\coloneqq(-1)^{2j}$ called the Frobenius--Schur indicators. In particular, the quantum trace of a single strand is given by
\be\label{Dimension}
\pic{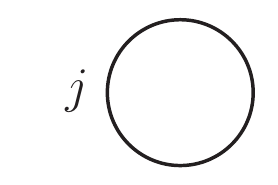}=(-1)^{2j}d_j\coloneqq v_j^2,
\ee
and is therefore not necessarily positive (although always real). We will make the choice of square root $v_j=(-1)^j \sqrt{d_j}$ (with $(-1)^{1/2}=\i$) for all $v_j^2$. As we will see in calculations later on, the $v_j$'s will typically appear in combinations of the form e.g. $\delta_{ijk}v_iv_jv_k$, which are real.

As usual, the fusion or splitting of representations in this graphical calculus is represented by trivalent nodes of the type
\be
\pic{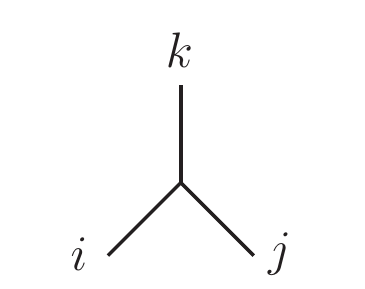},
\ee
and a trivial representation can always be grafted to a strand in the sense that
\be
\pic{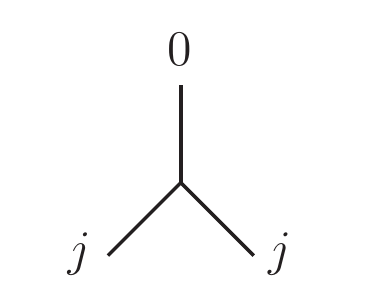}=\pic{HLine-j}.
\ee
This simply means that the strand with spin label $j=0$ is invisible, and therefore can always freely be inserted in order to fuse two parallel strands using $F$-symbols.

The above-mentioned $F$-symbols appear when considering the recoupling of four representations, where they describe the change of basis between $\text{Hom}\big(i,j\times(k\times l)\big)$ and $\text{Hom}\big(i,(j\times k)\times l\big)$. They are defined graphically by the relation
\be\label{F-move}
\pic{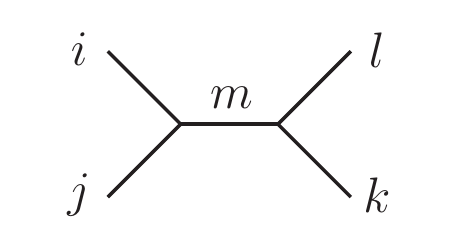}=\sum_nF^{ijm}_{kln}\pic{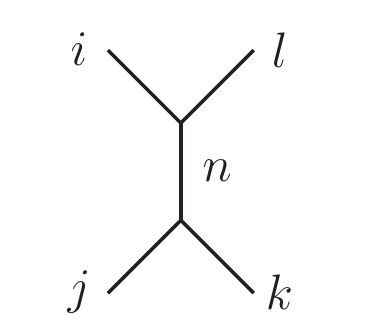},
\ee
which we will refer to as the $F$-move. In appendix \ref{appendix1} we give the explicit definition of these $F$-symbols in terms of the quantum $6j$-symbols of recoupling theory. Geometrically, if we think of the trivalent nodes as being dual to triangles, and of the strands as being dual to edges, it is well-known that equation \eqref{F-move} is the algebraic counterpart of the 2--2 Pachner move. What is important for our purposes is that the $F$-symbols satisfy the following relations:
\begin{subequations}
\be
\text{Physicality:}\q&F^{ijm}_{kln}=F^{ijm}_{kln}\delta_{ijm}\delta_{iln}\delta_{kjn}\delta_{klm},\\
\text{Tetrahedral symmetry:}\q&F^{ijm}_{kln}=F^{jim}_{lkn}=F^{lkm}_{jin}=F^{imj}_{knl}\f{v_mv_n}{v_jv_l},\\
\text{Orthogonality:}\q&\sum_nF^{ijm}_{kln}F^{ijp}_{kln}=\delta_{mp}\delta_{ijm}\delta_{klm},\label{F-orthogonality}\\
\text{Reality:}\q&\big(F^{ijm}_{kln}\big)^*=F^{ijm}_{kln},\\
\text{Normalization:}\q&F^{ii0}_{jjk}=\f{v_k}{v_iv_j}\delta_{ijk},\label{F-normalization}\\
\text{Pentagon identity:}\q&\sum_nF^{ijm}_{kln}F^{pql}_{nir}F^{rqn}_{kjs}=F^{ijm}_{spr}F^{pql}_{kms}.
\ee
\end{subequations}
The physicality condition is simply a condition of admissibility on the spin labels entering the $F$-symbols, and can be read off of the defining equation \eqref{F-move}. If one considers \eqref{F-move} in the triangulation picture, the physicality conditions implement the triangle inequalities (with the spin attached to a strand giving the length of the dual edge). The reality of the $F$-symbols, which simply comes from their definition \eqref{FDefinition}, implies that the fusion category is in fact unitary. In the geometrical setting where $F$-symbols are attached to tetrahedra, the pentagon identity is often referred to as the Biedenharn--Elliot identity. With all this structure at hand, we have so far defined a (unitary) fusion category.

The ribbon structure of the category is simply inherited from the non-trivial braiding in $\SU(2)_\k$. More precisely, the planar graphs can have non-trivial crossings, and these can be undone by using the $R$-matrix as follows:
\be\label{braiding}
\pic{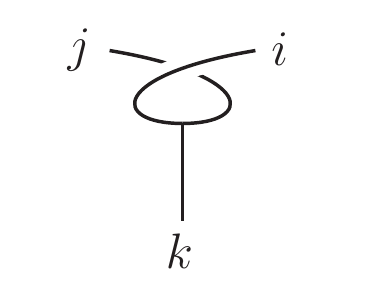}=R^{ij}_k\pic{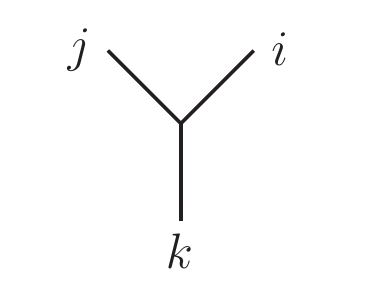},
\q\q
\pic{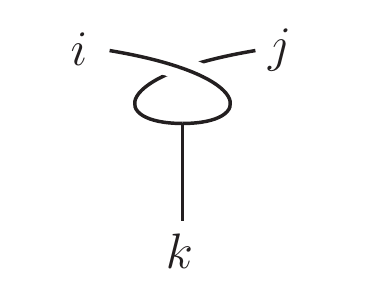}=\big(R^{ij}_k\big)^*\pic{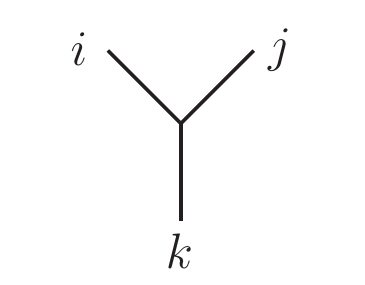}.
\ee
We give the explicit expression for the $R$-matrix of $\SU(2)_\k$ in appendix \ref{appendix1}. The $R$-matrix also has to satisfy certain consistency relations known as the hexagon identities, which ensure that the braiding is consistent with the $F$-moves. In the case of $\SU(2)_\k$, since $R^{ij}_k=R^{ji}_k$, there is only one such relation instead of two, and it is given by
\be\label{hexagon identity}
\text{Hexagon identity:}\q&R^{ki}_mF^{kim}_{ljp}R^{kj}_p=\sum_nF^{ikm}_{ljn}R^{kn}_lF^{jin}_{lkp}.
\ee
The existence of this non-trivial braiding structure is the reason for which we refer to the elements of the graphical representation as strands, and not simply as links. Furthermore, because of the semi-simplicity of $\SU(2)_\k$ (or of the category), it is sufficient to restrict ourselves to trivalent graphs. Now that we have the structure of a ribbon fusion (or premodular) category, we come to the last property of interest in this work, which is that of modularity.

Modularity is in fact a condition on the so-called topological $S$-matrix. A ribbon fusion category is said to be modular if $\det(S)\neq0$, which is the case for $\SU(2)_\k$. The $S$-matrix has entries defined via the braiding and the $R$-matrix by
\be\label{S matrix}
\D S_{ij}=s_{ij}=\pic{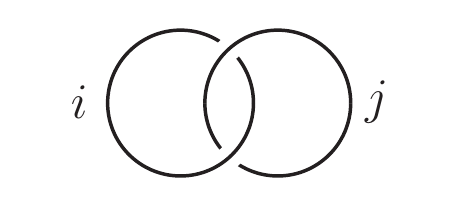}\hs,
\ee
where the elements $s_{ij}=(-1)^{2(i+j)}[(2i+1)(2j+1)]$ are the so-called non-normalized Verlinde coefficients, and where we have introduced the total quantum dimension
\be\label{total dimension}
\D\coloneqq\sqrt{\sum_jv_j^4}.
\ee
In particular, one has that $\D S_{0j}=v^2_j$. An important identity is that
\be\label{S through}
\pic{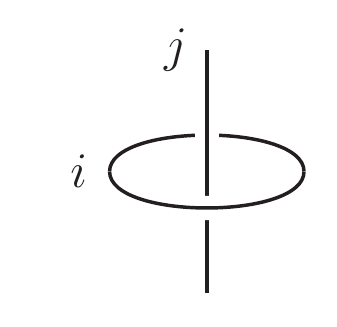}=\f{S_{ij}}{S_{0j}}\pic{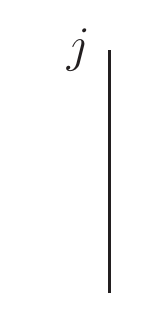},
\ee
which can be shown by realizing that both graphs must be proportional as elements of the same one-dimensional vector space, and by using the definitions \eqref{S matrix} and \eqref{Dimension} once the strand $j$ has been closed. More details about the definition and the properties of the $S$-matrix are given in appendix \ref{appendix1}.

We now have all the necessary ingredients to proceed with our construction. In the next subsection, we are going to give a few useful graphical evaluations which follow from the definitions given above, and introduce the concept of vacuum strands.

\subsection{Basic graphical identities}

\noindent The first useful identity to consider is that corresponding to the removal of a bubble. It is given by the so-called bubble-move
\be\label{bubble move}
\pic{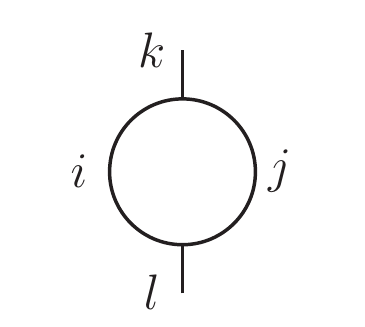}=\f{v_iv_j}{v_k}\delta_{kl}\delta_{ijk}\pic{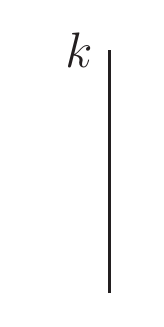}.
\ee
To prove this identity, notice first that its left-hand side, if read upwards, can be interpreted as an intertwining map from the representation space $l$ to the representation space $k$. However, due to Schur's lemma, a non-vanishing intertwining map between irreducible representations requires $k=l$. Therefore, we necessarily have that
\be\label{auxb01}
\pic{Bubble-ijkl}=\beta_{ijk}\delta_{kl}\pic{VLine-k}
\ee
for some coefficient $\beta_{ijk}$. To find this coefficient, we can now close the strand $k$ of the bubble configuration, and apply an $F$-move on a horizontal strand of spin 0 inside of the bubble. This gives
\be\label{auxb02}
\sum_m\f{v_m}{v_iv_j}\pic{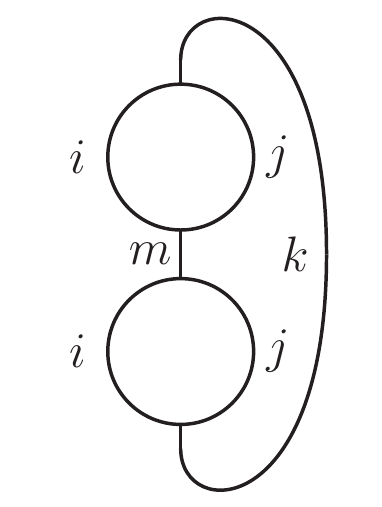}=\beta_{ijk}\delta_{kl}\pic{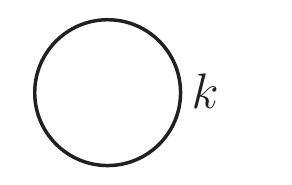},
\ee 
where we have used the normalization condition \eqref{F-normalization} for the $F$-symbol. Using \eqref{auxb01} repeatedly, as well as the quantum trace evaluation \eqref{Dimension}, equation \eqref{auxb02} becomes
\be
\f{v_k}{v_iv_j}\beta_{ijk}^2v_k^2=\beta_{ijk}v_k^2,
\ee
from which we get that $\beta_{ijk}=v_iv_j/v_k$.

As a consequence of the bubble-move, we therefore also find that
\be
\pic{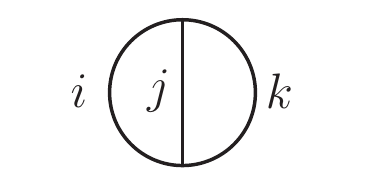}\hs=v_iv_jv_k\delta_{ijk}.
\ee
Now, this relation can be used to derive the following important result:
\be\label{vFusion}
v_i^2v_j^2=\pic{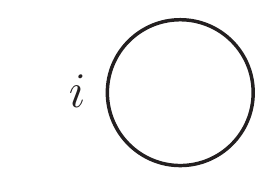}\pic{Loop-j}=\sum_k\f{v_k}{v_iv_j}\pic{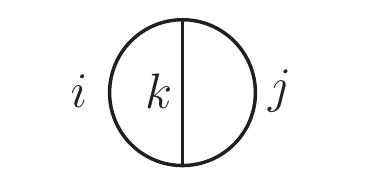}\hs=\sum_k\delta_{ijk}v_k^2,
\ee
where in the first step we have used an $F$-move with a strand of spin 0 between the two loops. This implies that the quantum trace evaluations \eqref{Dimension} obey the fusion rule, or, in other words, that the $v_j$'s provide a one-dimensional representation of the fusion algebra. Because of the fourth condition in \eqref{Admissibility}, one has $\alpha_i\alpha_j\alpha_k=1$ if $\delta_{ijk}=1$, which implies that the quantum dimensions themselves also satisfy the fusion rule, i.e.
\be\label{dFusion}
d_id_j=\sum_k\delta_{ijk}d_k.
\ee
Now, if we introduce the vector $\mathbf{d}$ whose components are given by the quantum dimensions, \eqref{dFusion} can be written as $d_id_j=\sum_k(\mathbf{N}_i)_{jk}d_k=(\mathbf{N}_i\mathbf{d})_j$. This means that $\mathbf{N}_i\mathbf{d}=d_i\mathbf{d}$, i.e. that $\mathbf{d}$ is an eigenvector for the fusion matrices $\mathbf{N}_i$ with eigenvalue $d_i$. We can also see that the total quantum dimension \eqref{total dimension} is nothing but the norm of this vector.

Next, the bubble-move can be combined with the $F$-move to obtain the relation
\be
\pic{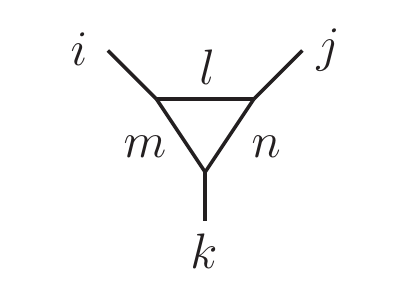}=\f{v_mv_n}{v_k}F^{ijk}_{nml}\pic{Ru2},
\ee
which is nothing but the algebraic expression of the 3--1 Pachner move. Moreover, as a special case of the $F$-move one has that the decomposition of the identity
\be\label{F split}
\sum_k\f{v_k}{v_iv_j}\pic{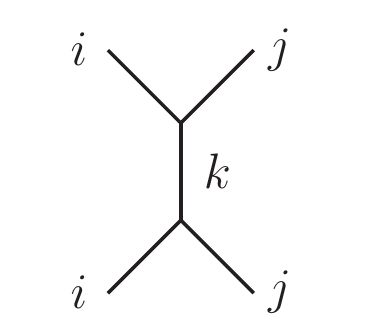}=\pic{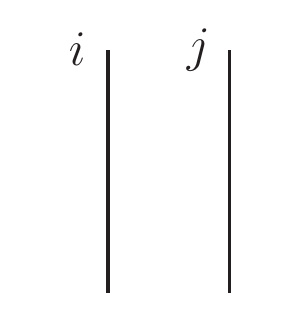},
\ee
which can further be combined with the braiding relation \eqref{braiding} to obtain the following resolutions of crossings:
\be\label{resolution of crossing}
\pic{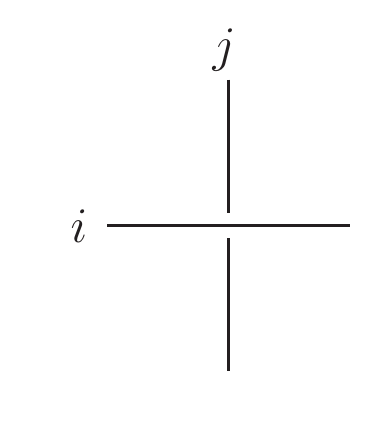}=\sum_k\f{v_k}{v_iv_j}R^{ij}_k\pic{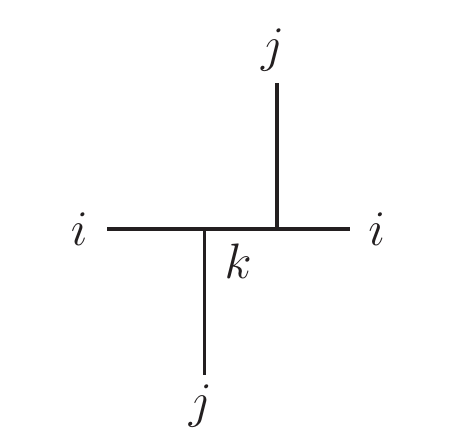},
\q
\pic{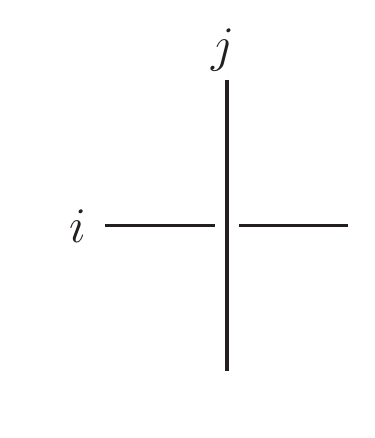}=\sum_k\f{v_k}{v_iv_j}\big(R^{ij}_k\big)^*\pic{Uncrossing}.
\ee
By compatibility between these two expressions (i.e. the possibility of resolving a crossing either vertically or horizontally) one then gets the conditions
\be\label{from R* to R}
\big(R^{ij}_k\big)^*=\sum_m\f{v_m}{v_k}F^{ijm}_{ijk}R^{ij}_m,
\q
R^{ij}_k=\sum_m\f{v_m}{v_k}F^{ijm}_{ijk}\big(R^{ij}_m\big)^*.
\ee
These conditions are indeed satisfied for $\SU(2)_\k$ \cite{CarterFlatobook}. It allows us to omit the orientation arrows of the strands even in the case where braiding is involved.

Throughout this work, we will keep in the main text only the proofs which require little space, and collect in appendix \ref{appendix2} the more lengthy results.

\subsection{Vacuum strands and loops}

\noindent Now, we introduce the important concept of vacuum strands. These are defined as the weighted sum of all possible standard strands divided by the total quantum dimension, i.e.
\be
\pic{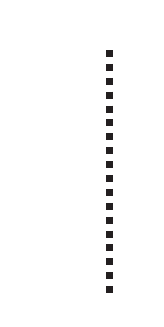}\coloneqq\f{1}{\D}\sum_jv_j^2\pic{VLine-j}.
\ee
If we think of the standard strands as representing graphically elements of the fusion algebra, we can naturally ask what is the graphical representation of the above-mentioned eigenvector $\mathbf{d}$. This is given precisely by these new dotted vacuum strands. By closing a vacuum strand so as to form a loop, one can see that
\be
\pic{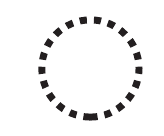}=\f{1}{\D}\sum_jv_j^2\pic{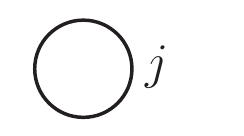}\hs=\D,
\ee
where we have used \eqref{total dimension} and \eqref{Dimension}. This means that, in the graphical representation, we can always introduce the identity in the form $1=(\text{vacuum loop})\times\D^{-1}$. We will make extensive use of this property later on.

Once they are closed so as to form loops, these vacuum strands have the remarkable property of being invisible to standard strands, which can slide over them as follows:
\be\label{vacuum sliding}
\pic{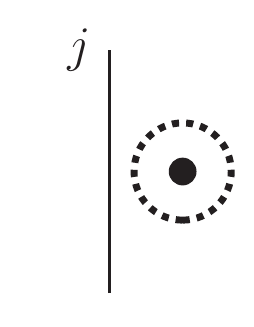}=\f{1}{\D}\sum_kv_k^2\pic{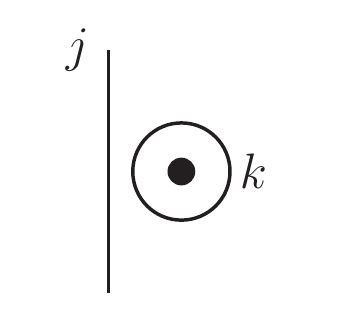}\hs=\f{1}{\D}\sum_{kl}\f{v_kv_l}{v_j}\pic{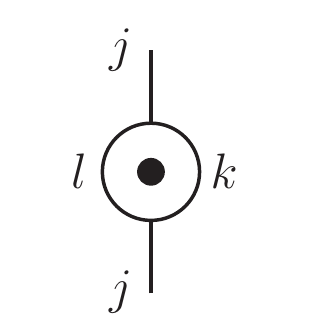}\hs=\f{1}{\D}\sum_lv_l^2\pic{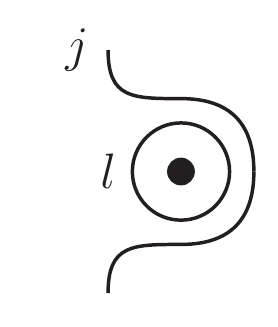}=\pic{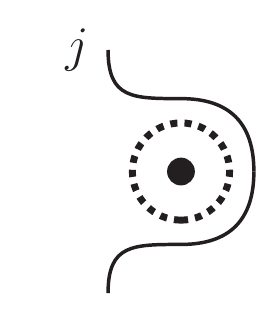}.
\ee
For this proof, we have used in the second equality an $F$-move with a strand of spin $0$ between the strands $j$ and $k$. It is important to notice that this sliding relation holds regardless of what is contained in the shaded region, even if it is a puncture in the manifold. In what follows we will always use a thick dot $\pic{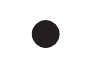}$ to represent punctures. Making use of this sliding property, we can now show that the stacking of two vacuum loops gives
\be\label{stacking vacuum loops}
\pic{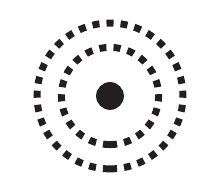}=\f{1}{\D}\sum_jv_j^2\pic{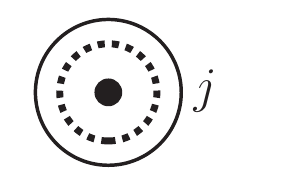}\hs=\f{1}{\D}\sum_jv_j^2\pic{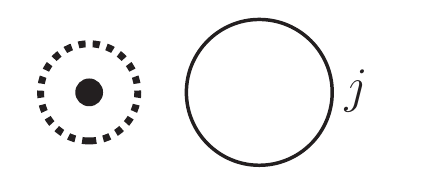}\hs=\f{1}{\D}\sum_jv_j^4\pic{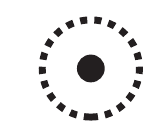}=\D\pic{VacuumStacking4}.
\ee
When a strand passes through a vacuum loop, one finds the following relation:
\be
\pic{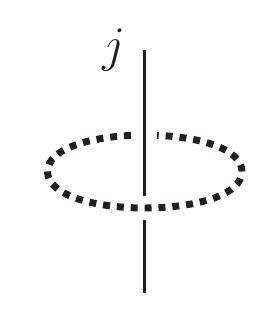}=\f{1}{\D}\sum_iv_i^2\pic{Sthrough}=\f{1}{v_j^2}\sum_iv_i^2S_{ij}\pic{VLine-j}=\f{\D}{v_j^2}\sum_iS_{0i}S_{ij}\pic{VLine-j}=\D\delta_{j0}\pic{VLine-j}.
\ee
In order to prove this we have used property \eqref{S through}, the fact that $S_{0j}=v_j^2/\D$, and in the last step the unitarity of the $S$-matrix\footnote{Since in the case of $\SU(2)_\k$ the $S$-matrix is real and symmetric, this reduces to the orthogonality relation $\sum_kS_{ik}S_{kj}=\delta_{ij}$.}. Note that the last equation, as well as the identity \eqref{vacuum F-move}, therefore hold only in modular fusion categories (such as $\SU(2)_\k$). This annihilation property justifies the name ``vacuum loop''. Finally, by combining this relation with an $F$-move we obtain
\be\label{vacuum F-move}
\pic{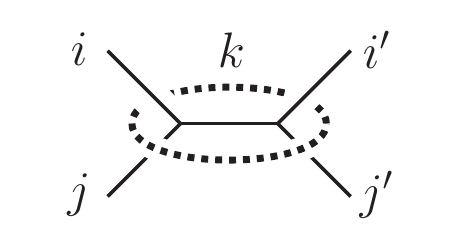}=\sum_lF^{ijk}_{j'i'l}\pic{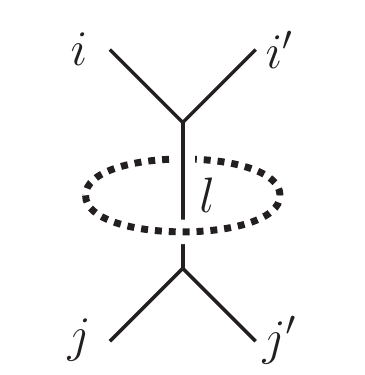}=\D\delta_{ii'}\delta_{jj'}\f{v_k}{v_iv_j}\pic{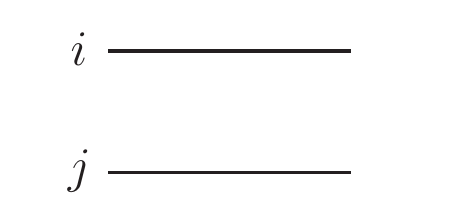}.
\ee
Note that this relation is compatible with \eqref{F split} because of the equality
\be
\pic{HLine-ij}=\f{1}{\D}\sum_k\f{v_k}{v_iv_j}\pic{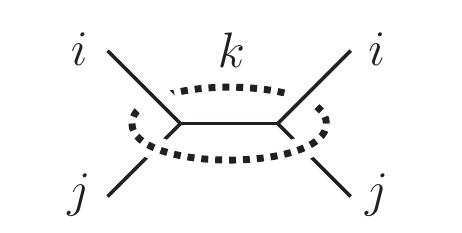}=\sum_k\f{v_k^2}{v_i^2v_j^2}\delta_{ijk}\pic{HLine-ij}\hs=\pic{HLine-ij},
\ee
which holds because of \eqref{vFusion}.

With all these ingredients of graphical calculus and their algebraic interpretation, we are now ready to define the graph Hilbert space and the vacuum.

\section{Graph Hilbert space}
\label{sec:graph hilbert space}

\noindent In this section, we are going to use the rules of graphical calculus introduced above in order to assign a graph Hilbert space to two-dimensional manifolds with defects \cite{KKR,Kir}. We will then see in which sense certain states in this Hilbert space describe the TV vacuum, while others represent excited states carrying curvature and torsion (i.e. violations of the flatness and Gauss constraints). In the next section we will then focus more specifically on these excitations, and explain how they are related to ribbon operators and the structure of a Drinfeld center.

Following the idea of extended TQFTs, the excitations on top of the vacuum should be carried by (or located around) defects in the manifold. In the construction of the $(2+1)$-dimensional $\SU(2)$ BF vacuum \cite{DG1,DG2,BDG}, we have realized this in a geometrical setting where triangulations of the spatial manifold play an important role. There, indeed, the defects were located on the embedded vertices of the triangulation. The refinement operations were given by the Alexander moves, which add triangles and vertices (i.e. defects) in the flat vacuum state. Here, instead, we are going to start with a manifold containing embedded punctures without any a priori reference to a specific triangulation. Let us therefore consider a two-dimensional compact and orientable Riemann surface $\Sigma_p$ with an arbitrary number $p$ of embedded punctures. For reasons which will become clear later on, we should think of the punctures as being obtained by removing discs and placing an embedded marked point on each corresponding boundary component. One can see $\Sigma_p$ as arising from placing $p$ punctures in an initial manifold $\Sigma$.

As a remark, notice that instead of representing the punctures as removed disks with a marked point on the boundary, one could also use an embedded point together with a vector attached to it \cite{Kir}. The strands (representing objects of the fusion category $\SU(2)_\k$) and later the ribbons (representing objects of the Drinfeld centre of $\SU(2)_\k$) then have to arrive to the point representing the puncture by being tangential to this vector. This information (or equivalently the marked point on the boundary of the removed disk) is important for the proper identification of the excitations. In particular, it will make a difference if a strand goes ``straight'' into the puncture, or instead first winds around the puncture and then enters it. The difference between these two situations is a Dehn twist of the cylinder-like region around the puncture. 

Note that in our graphical representation we will suppress the marked points (or the vectors) on the punctures for the clarity of the drawings. Punctures will be depicted by thick black dots and the (suppressed) vectors will be always horizontal, either pointing to the right (if the puncture is located on the left part of the figure) or to the left (if the puncture is located on the right part of the figure).

On the manifold with embedded punctures, we can now define the vector space of graphs.

\begin{Definition}[Space $\V_{\Sigma_p}$ of graphs]\label{Strand-space-def}
First, consider a trivalent graphs embedded in $\Sigma_p$, and allow for the possibility of having, for each puncture, a single strand ending at the marked point. Then, given a fixed level $\k$ and the corresponding set $\mathcal{J}$ of spins, consider colorings of this graph with spins such that the admissibility conditions \eqref{Admissibility} are satisfied at each trivalent node. The space $\V_{\Sigma_p}$ is then defined as the $\mathbb{C}$-linear span of such embedded colored graphs modulo the local equivalence relations \eqref{StrandDeformation}, \eqref{F-move}, \eqref{braiding}, and \eqref{bubble move}. Thus, $\V_{\Sigma_p}$ has the structure of a vector space.
\end{Definition}

In order to turn $\V_{\Sigma_p}$ into a Hilbert space $\H_{\Sigma_p}$, we need to understand graphs in $\Sigma_p$ as states, and in particular specify a basis and an inner product for these states. As we will now see, this construction depends on the number of punctures, on the topology, and on the level $\k$. For the sake of clarity, we are going to focus in the rest of this article solely on the two-sphere, and refer to it simply as the sphere. The construction of the basis states for the higher genus case is explained in appendix \ref{appendix:higher-genus}.

\subsection{Basis states}

\begin{figure}[h]
$\pic{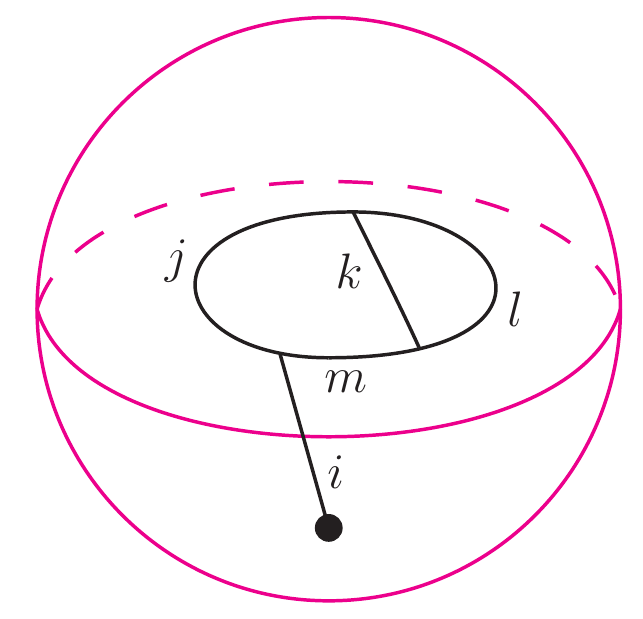}=v_jv_kv_l\delta_{jm}\delta_{i0}\delta_{jkl}\pic{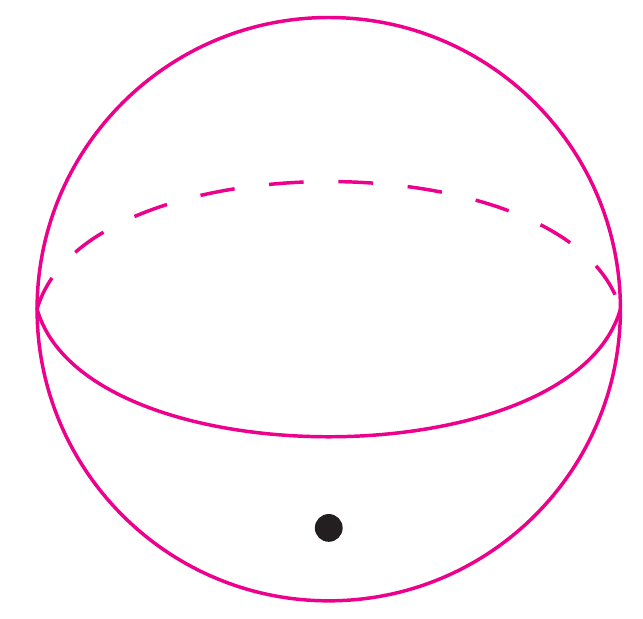}$
\caption{Example of a state on the sphere with one puncture. Using the rules of graphical calculus, one can show that any such state is proportional to the empty graph.}
\label{2S1p}
\end{figure}

\noindent Let us assume that the punctured manifold $\Sigma_p$ is a $p$-punctured sphere $\mathbb{S}_p$. First, if there are no punctures, it is clear that because of the trivial topology any graph on $\mathbb{S}_0$ can be evaluated to a $\mathbb{C}$-number using the rules of graphical calculus introduced in the previous section. We therefore have that\footnote{Although we are only going to define the inner product in the next subsection, we already denote the space of graphs by $\H_{\Sigma_p}$, call it a Hilbert space, and refer to the graphs as states.} $\dim\H_{\mathbb{S}_0}=1$, and the unique basis state can be chosen to be the empty graph. If there is a single puncture, it follows from \eqref{bubble move} that $\dim\H_{\mathbb{S}_1}=1$ as well. An example of a state on $\mathbb{S}_1$ is represented in figure \ref{2S1p}.

\begin{figure}[h]
$\pic{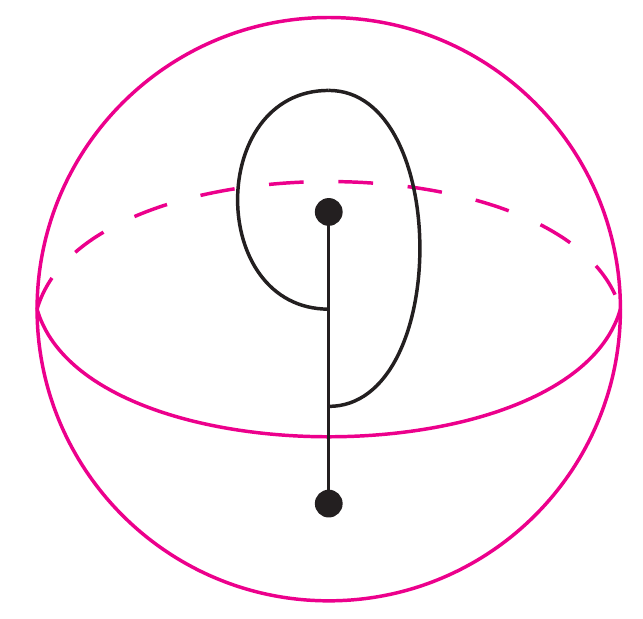}$
\caption{Minimal graph on the sphere with two punctures. The basis states for $\H_{\mathbb{S}_2}$ are given by the admissible spin colorings of this minimal graph.}
\label{2S2p}
\end{figure}

\begin{figure}[h]
$\pic{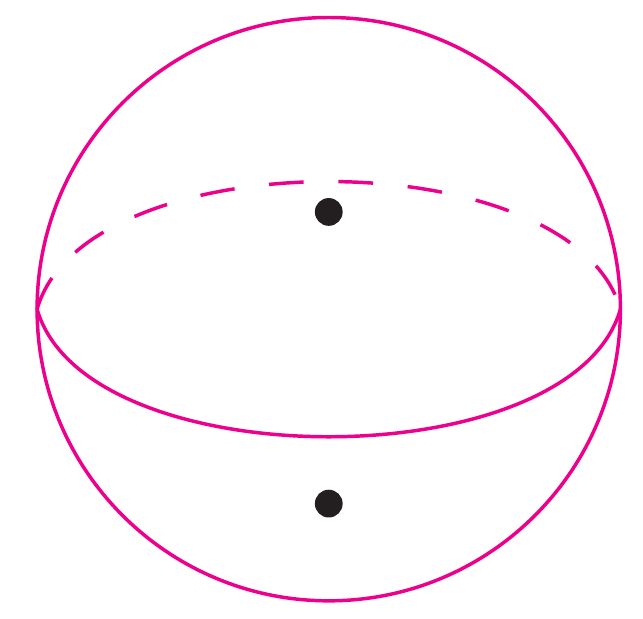}\pic{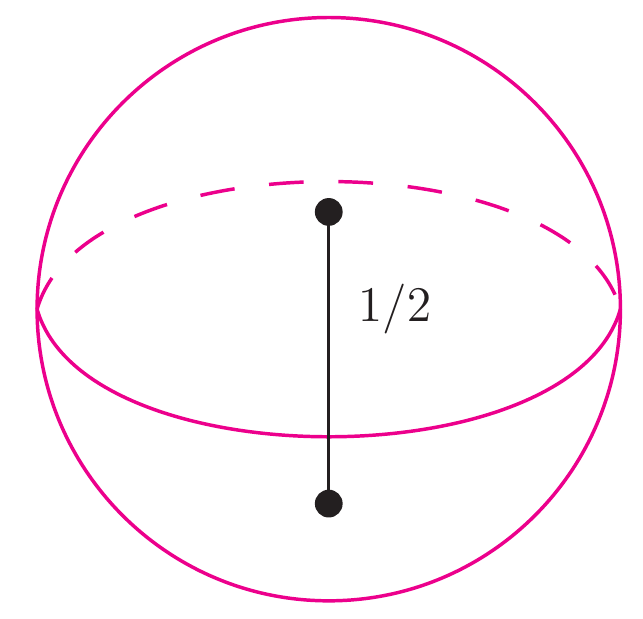}\pic{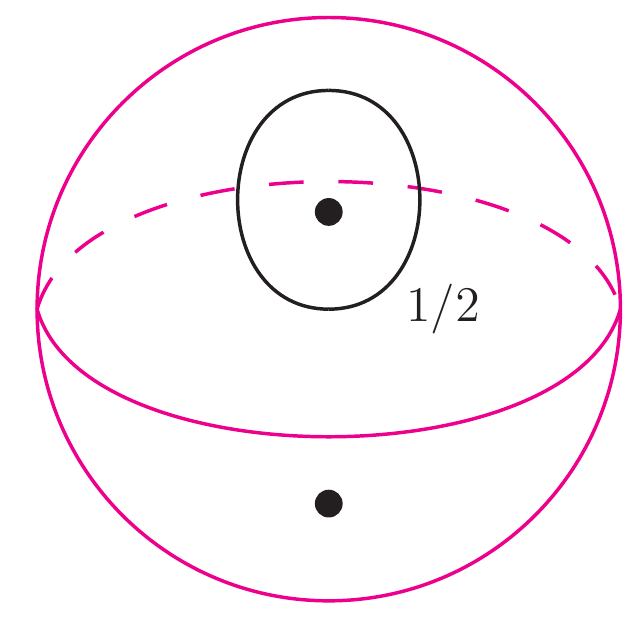}\pic{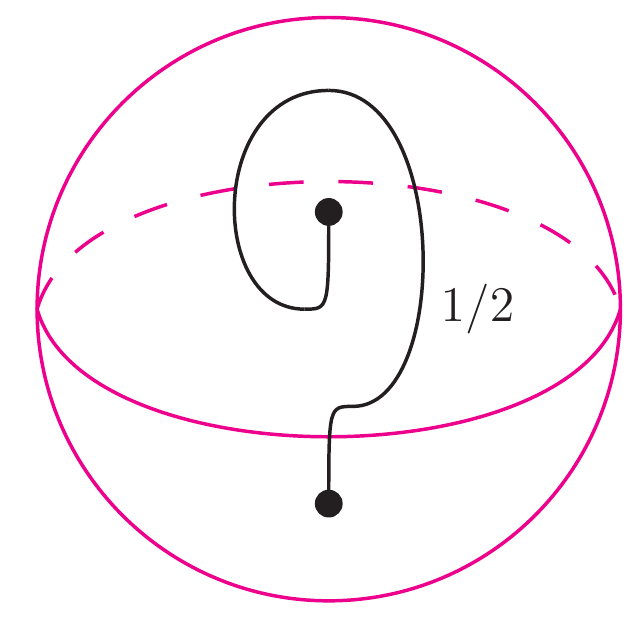}$
\caption{Basis of states for the sphere with two punctures and level $\k=1$.}
\label{2S2pk=1}
\end{figure}

The first non-trivial case is that of the sphere with two punctures, $\mathbb{S}_2$, which is topologically a cylinder. Because of this non-trivial topology there is a non-contractible cycle around which strands can wind, thereby preventing certain graphs from being reduced to a number (i.e. to a certain coefficient multiplying the empty graph on $\mathbb{S}_2$). One can describe a basis of $\H_{\mathbb{S}_2}$ by considering the minimal graph represented in figure \ref{2S2p}. The basis states are given by the admissible spin colorings of this minimal graph, and in what follows we will represent\footnote{When using these planar representations one should always remember that the graphs are actually defined on the sphere.} and denote them by
\be\label{Qijrs}
\Q^{ij}_{rs}\coloneqq\pic{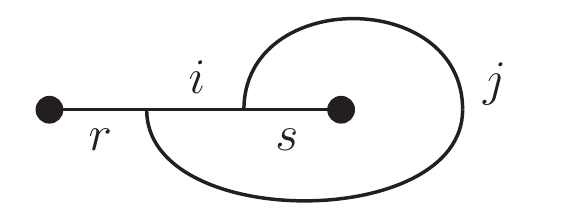}\hs.
\ee
Because this graph has two trivalent nodes, the dimension of the graph Hilbert space (i.e. the number of allowed states) on the cylinder is given by $\dim\H_{\mathbb{S}_2}=\sum_{ijrs}\delta_{ijr}\delta_{ijs}$, and therefore depends on the level $\k$. For example, if $\k=1$, there are only two allowed spin labels, $\mathcal{J}=\{0,1/2\}$, and four basis states. These states are represented in figure \ref{2S2pk=1}. In particular, one can see that the second and fourth states cannot be identified due to the nature of the punctures. Indeed, the embedded marked point prevents us from unwinding the strand which goes around the upper puncture and from smoothly deforming the fourth state into the second one. It turns out that states on the cylinder will play a very important role in what follows. We will study their properties in detail in the next section.

We can now easily generalize this construction to the sphere $\mathbb{S}_p$ with $p$ punctures. In this case, a minimal graph has a tree-like structure and basis states can be written in the form
\be\label{p-punctured basis}
\textbf{Q}^{\vec{\imath}\,\vec{\jmath}}_{\vec{n}\,\vec{r}\,\vec{s}}\coloneqq\pic{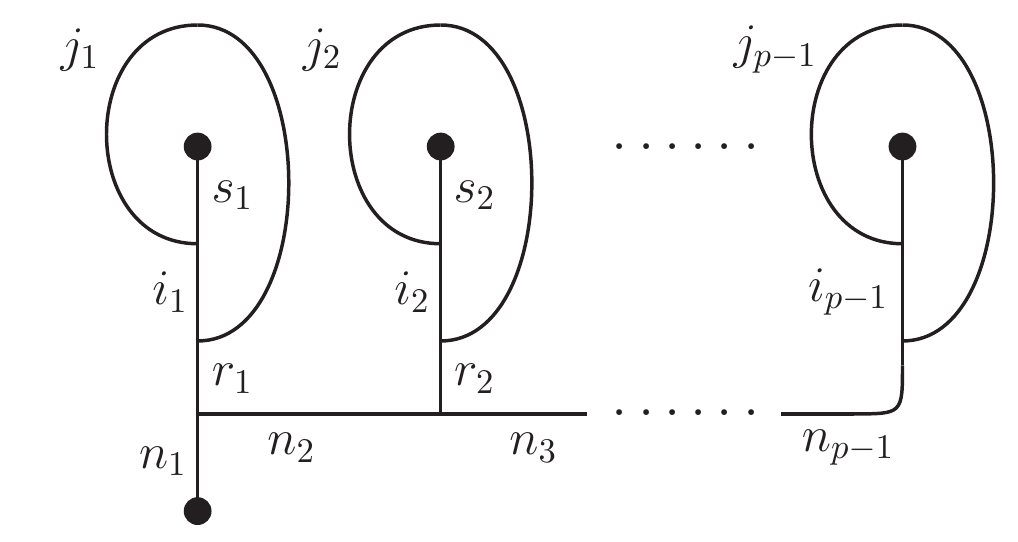}.
\ee
Such a graph has $(5p-6)$ strands for $p\geq2$ punctures, and, as explained above, the basis states are given by the admissible spin colorings of the strands. The dimension of the Hilbert space for $p\geq2$ is given by
\be
\dim\H_{\mathbb{S}_p}=\sum_{\vec{\imath},\vec{\jmath},\vec{n},\vec{r},\vec{s}}\delta_{n_p0}\prod_{\alpha=1}^{p-1}\delta_{i_\alpha j_\alpha r_\alpha}\delta_{i_\alpha j_\alpha s_\alpha}\delta_{n_\alpha n_{\alpha+1}r_\alpha},
\ee
where the vector label denotes a set of spins to be summed over, i.e. $\vec{\jmath}=\{j_1,\dots,j_{p-1}\}$, and each sum should range over the set $\mathcal{J}$ of spin values determined by $\k$.

Let us end this subsection by a comment on triangulations. Since the graphs which we are considering are trivalent, they are dual to (possibly degenerate) triangulations. These triangulations of punctured manifolds can be built in a systematic manner as follows: First, recall that a puncture is represented by a marked point on the boundary of a hole obtained by removing a disk. In order to obtain a triangulation of a punctured manifold, it is important to draw explicitly these holes and to place the marked points on their boundaries. One can then place a circular triangulation edge on the boundary of each hole, i.e. an edge whose source and target nodes are the same, and choose this node to be opposite to the marked point (with respect to the center of the hole). By doing so, one obtains a triangulation of the boundary of the hole on which are sitting the marked point and a node. Once a circular edge has been drawn for every hole, one can connect the nodes with additional edges so as to obtain a triangulation of the punctured manifold. By triangulation, we mean in particular that every face has to have three boundary edges.

\begin{figure}[h]
$\pic{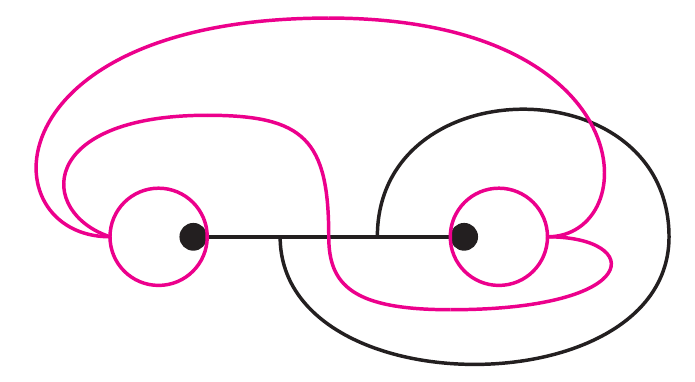}$
\caption{Minimal triangulation (in pink) of the two-punctured sphere and its dual graph $\Q$. One can see that the circular edge tangent to a marked point is the triangulation boundary of the hole defining the puncture and on which the marked point has been placed.}
\label{2pSphereTriangulatedMin}
\end{figure}

A particular class of triangulations of the punctured sphere $\mathbb{S}_p$ are so-called minimal triangulations. These have the defining property of having only $p$ nodes, which are the nodes of the circular edges surrounding the puncture holes. Such a minimal triangulation for the two-punctured sphere is represented on figure \ref{2pSphereTriangulatedMin}.
The graph dual to such a minimal triangulation is of the form described in definition \ref{Strand-space-def}. The equivalence relations \eqref{StrandDeformation}, \eqref{F-move}, and \eqref{bubble move} allow to deform, refine, and change the graph, and the corresponding changes of triangulations includes the $2-2$, $3-1$, and $1-3$ Pachner moves, which are ergodic on the space of two-dimensional triangulations with fixed topology. Notice also that vertices of the triangulation (if this latter is not minimal) which are not situated on the boundaries of holes can freely be moved, thereby realizing (spatial) diffeomorphism symmetry \cite{LouapreFreidel,Noui:2004,BD08} for all vertices not associated to punctures. For the punctures themselves, one would need to perform a group averaging over spatial diffeomorphisms, as is done in the AL case \cite{lqg3}.

\subsection{Inner product}
\label{subsec:innerproduct}

\noindent We have so far obtained a systematic way of writing down basis states for the space of equivalence classes of graphs on the punctured sphere. An inner product on $\H_{\mathbb{S}_p}$ can now simply be obtained by declaring these basis states to be orthonormal. However, for this definition to be consistent, one has to ensure that it is independent of the choice of the basis states themselves. Part of the arbitrariness in choosing these states has already been fixed by restricting ourselves to the minimal graphs with the general tree-like structure of \eqref{p-punctured basis}. However, we are still left with the freedom of choosing these minimal graphs up to local moves which preserve the number of strands, i.e. up to the $F$-moves of equation \eqref{F-move}. In order to see explicitly that the inner product is indeed independent of the choice of basis states, let us focus for simplicity on the case with two punctures. Two choices of basis states which differ by an $F$-move are given by
\be\label{Q and Q tilde basis}
\Q^{ij}_{rs}=\pic{Qijrs}\hs,
\q\q
\widetilde{\Q}^{ij}_{rs}=\pic{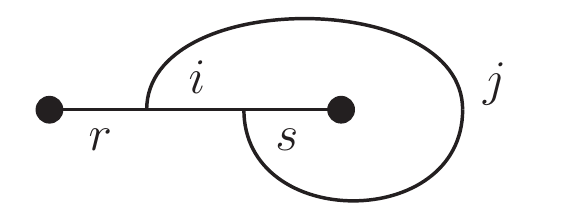}\hs.
\ee
By definition, the inner product constructed from the basis $\Q$ is
\be
\Big\la\Q^{ij}_{rs}\Big|\Q^{i'j'}_{r's'}\Big\ra_\Q=\delta_{ii'}\delta_{jj'}\delta_{rr'}\delta_{ss'}.
\ee
Using the orthogonality relation \eqref{F-orthogonality}, we can check that the inner product defined by the basis $\widetilde{\Q}$ gives the same result, i.e.
\be
\Big\la\Q^{ij}_{rs}\Big|\Q^{i'j'}_{r's'}\Big\ra_{\widetilde{\Q}}=\Big\la\sum_nF^{rji}_{sjn}\widetilde{\Q}^{ij}_{rs}\Big|\sum_{n'}F^{r'j'i'}_{s'j'n'}\widetilde{\Q}^{i'j'}_{r's'}\Big\ra_{\widetilde{\Q}}=\delta_{jj'}\delta_{rr'}\delta_{ss'}\sum_nF^{rji}_{sjn}F^{rji'}_{sjn}=\delta_{ii'}\delta_{jj'}\delta_{rr'}\delta_{ss'}.
\ee
This calculation can easily be generalized to any number of $F$-moves affecting the tree-like minimal graph of the basis \eqref{p-punctured basis}, thereby showing that the inner product is indeed independent of the choice of basis states, and providing us with a consistent definition.

Although this inner product has a simple definition in terms of a basis on $\mathbb{S}_p$, it can be cumbersome to use if one wants to compute the norm of ``complicated states'' (i.e. with braiding for example) since these then have to be expanded in the basis itself. For this reason, it will be useful later on to consider an alternative inner product, which in \cite{KKR} is called the the trace inner product.

The trace inner product $\la\Psi_\Gamma|\Psi_{\Gamma'}\ra_\tr$ between two states on graphs $\Gamma$ and $\Gamma'$ is defined\footnote{Our definition is slightly different and more systematic than the one given in \cite{KKR}.} by reflecting the graph $\Gamma'$, connecting the open strands of the two graphs (i.e. those ending at the punctures), reducing the resulting closed graph to a number, and dividing by a factor of $v_j$ for each pair of strands being connected. In this definition, we have to be precise about the way in which the open strands should be connected, since there can be obstructions to doing so without an additional step. First, notice that it is well-defined and unambiguous to connect the open strands ending at punctures. This is because the punctures are embedded (i.e. posses a specific location) and can therefore be identified and matched two-by-two between states $\Psi_\Gamma$ and $\Psi_{\Gamma'}$ defined on the same punctured manifold $\Sigma_p$. Now, for graphs with no closed strands going around the punctures, the definition of the trace inner product is immediate to apply, and for example one has that
\be
\Big\la\pic{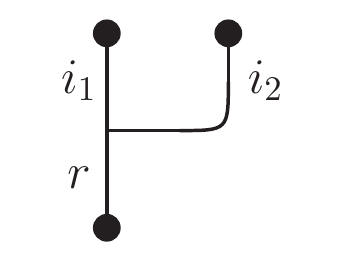}\Big|\pic{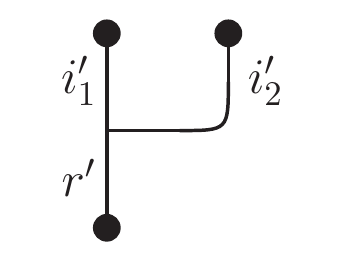}\Big\ra_\tr=\f{1}{v_rv_{i_1}v_{i_2}}\delta_{i_1i_1'}\delta_{i_2i_2'}\delta_{rr'}\pic{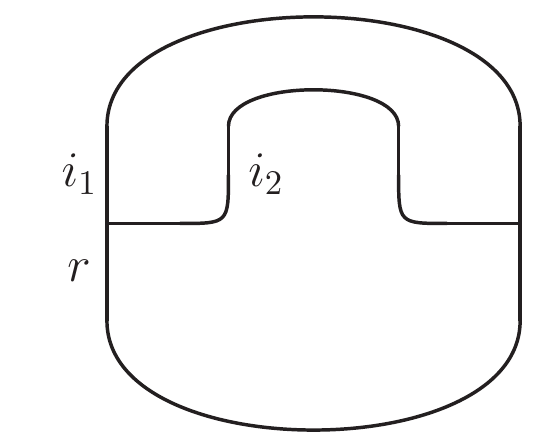}=\delta_{i_1i_1'}\delta_{i_2i_2'}\delta_{rr'},
\ee
in agreement with the orthonormality of the basis states for $\mathbb{S}_3$. However, for states with closed (standard or vacuum) loops going around the punctures, the open strands ending at these punctures in $\Gamma$ and $\Gamma'$ cannot a priori be connected. In order to define the trace inner product for such states, one has to add the following additional step. For each pair of punctures in $\Gamma$ and $\Gamma'$ whose open strands can a priori not be connected, divide by a factor of $\D$, and add a closed vacuum loop passing through and linking the two closed loops around the punctures. Taking the three-punctured sphere as an example, we prove in calculation \eqref{Proof of orthonormality of p-punctured Q} of appendix \ref{appendix2} that this prescription for the trace inner product does indeed produce the result
\be\label{orthonormality of p-punctured Q}
\Big\la\textbf{Q}^{\vec{\imath}\,\vec{\jmath}}_{\vec{n}\,\vec{r}\,\vec{s}}\Big|\textbf{Q}^{\vec{\imath}\,'\,\vec{\jmath}\,'}_{\vec{n}\,'\,\vec{r}\,'\,\vec{s}\,'}\Big\ra_\tr=\delta_{\vec{\imath}\,\vec{\imath}\,'}\delta_{\vec{\jmath}\,\vec{\jmath}\,'}\delta_{\vec{n}\,\vec{n}\,'}\delta_{\vec{r}\,\vec{r}\,'}\delta_{\vec{s}\,\vec{s}\,'},
\ee
as expected.

The trace inner product does in fact posses a geometrical interpretation. Once possible closed strands have been removed from around punctures by inserting the appropriate number of vacuum loops and factors of $\D^{-1}$, the trace inner product amounts to reversing the orientation of the manifold $\Sigma_p$ on which the state $\Psi_{\Gamma'}$ is defined, and then gluing two-by-two the marked points of the punctures between $\Psi_\Gamma$ and $\Psi_{\Gamma'}$ by identity-connected diffeomorphisms. Then, the holes of the resulting higher genus manifold have to be filled, which results in a topologically trivial manifold on which the glued graph can be evaluated to a number. This schematic explanation is the reason for which, when computing the trace inner product, the punctures disappear once their open strands have been connected. In particular, it also means that two naked punctures in $\Psi_\Gamma$ and $\Psi_{\Gamma'}$ will disappear (or be connected by strands of spin $j=0$) under the trace inner product, and therefore not contribute to it.

\subsection{Vacuum}
\label{subsec:vacuum}

\noindent We have so far obtained a description of the graph Hilbert space $\H_{\Sigma_p}$ and of its basis states. As explained above, a generic state in $\H_{\Sigma_p}$ can have open strands ending at the punctures as well as strands winding around the punctures. In fact, a puncture has two special properties. First, it allows for open strands to end, thereby representing violations of the Gauss constraint. Second, it prevents strands from being freely deformed or moved on $\Sigma_p$, thereby representing violations of the flatness constraint\footnote{We call ``flatness'' the property of being able to deform strands over a region of the manifold, as a flat connection does indeed allow to arbitrarily deform paths of holonomies.}. In this sense, the punctures can carry torsion and curvature excitations. By definition, states which do not present such excitations correspond to the vacuum. In the vacuum, there are no open strands ending at the punctures, and strands can freely be pulled over the punctures. States in the vacuum can therefore be seen as living effectively on a manifold with trivial topology, and thus can always be reduced to a coefficient multiplying a minimal vacuum state. In this sense, the vacuum of the cylinder is given by a single state equivalent to the empty graph where the punctures are made ``invisible'' by being surrounded by vacuum loops. On punctured manifolds of arbitrary topology however, the vacuum can be degenerate and described by several states. In order to understand this in more details, one can introduce the Gauss and flatness projection operators.

As is well-known from the spin network representation of lattice gauge theory, the Gauss constraint is defined at the three-valent nodes by the requirement that the triple of spins labeling the incoming strands be admissible in the sense of \eqref{Admissibility}. Let us schematically denote by $n(i,j,k)$ a three-valent node connecting strands with spins $(i,j,k)$. With this notation, we can then define the action of the Gauss projection operator as
\be
B_n\triangleright n(i,j,k)\coloneqq\delta_{ijk}n(i,j,k).
\ee
Because $\SU(2)_\k$ has no fusion multiplicities, this action is simply equal to zero or one depending on whether the given fixed triple $(i,j,k)$ of spins is admissible or not, and therefore does indeed define a projector. When acting on generic basis states with free spin labels, as for example \eqref{Qijrs}, $B_n$ should be understood as a function of the spins which multiplies the state it acts on by a factor of $\delta_{ijk}$. This implements in turn the relations $B_n\triangleright n(i,j,0)=\delta_{ij}n(i,i,0)$ and $B_n\triangleright n(i,0,0)=\delta_{i0}n(0,0,0)$. In particular, we should now see the marked point of a puncture as a three-valent node with spins $(i,0,0)$, where $i$ is the spin labeling the incoming open strand. The operator $B_n$ can therefore act on a marked point, and does so by removing the open strand and inserting a factor of $\delta_{i0}$. This factor of $\delta_{i0}$ then constrains the node $n(i,r,s)$ from which the open strand was departing, and together with $B_n\triangleright n(i,r,s)=\delta_{irs}n(i,r,s)$ consistently enforces that $n(i,r,s)=\delta_{i0}\delta_{rs}n(0,r,r)$. Our definition of the Gauss projection operator therefore guarantees that, when acting on a generic state in $\H_{\Sigma_p}$, the operator $\prod_nB_n$ returns a state in $\H_{\Sigma_p}$. It is clear that since we have defined the graph Hilbert space $\H_{\Sigma_p}$ as being spanned by graphs with admissible spin colorings at the three-valent nodes which are not marked points, only the punctures can potentially carry violations of the Gauss constraint.

Similarly, the local equivalence relations which define the graph Hilbert space implicitly guarantee that, on the sphere, there is no curvature located ``away from the punctures''. The punctures represent the only obstruction to deforming graphs and evaluating them on the sphere, and in this sense can be thought of as carrying the curvature. In order to understand the vacuum as a state with no curvature, let us introduce the flatness projection operator acting on punctures as
\be\label{flatness-projector}
B_p\triangleright\pic{Puncture}\coloneqq\f{1}{\D}\pic{VacuumStacking4}.
\ee
In this definition, we have included a factor of $\D^{-1}$ in order to ensure that $B_p^2=B_p$. Because of the fundamental sliding property \eqref{vacuum sliding} between strands and vacuum loops, we see that the operator $B_p$ renders the punctures ``invisible'' by freely allowing strands to slide over them. In this sense, $B_p$ removes the curvature located at the punctures. This property can also be understood algebraically by going back to the classical group picture. In the limit $q\rightarrow1$, if we interpret the spin label $j$ as being the Fourier component of a group element, we see from the definition of the vacuum loops that\footnote{Note that we have written here the limit of $\D^2B_p$ in order to cancel divergent factors of $\D$.}
\be
\D^2B_p\rightarrow\sum_{j}d_j\chi_j(g_p)=\delta_{\SU(2)}(g_p),
\ee
where $g_p$ is the holonomy around the puncture. The flatness operator is therefore imposing that the holonomy $g_p$ around the puncture be trivial. For $\SU(2)_\k$ at root of unity, since there is no Fourier transform defining a group representation, the only way to define a flatness projection operator is via the spin representation \eqref{flatness-projector}. In the next two subsection we will explain the relationship between $B_p$, the embedding maps, and the TV partition function on a three-manifold.

We can now define the vacuum as the space of states in the original graph Hilbert space $\H_{\Sigma_p}$ which are invariant under the action of the Gauss and flatness projection operators for all nodes and punctures. Explicitly, this is
\be\label{vacuum-space}
\H_{\Sigma_p}^0\coloneqq\text{Im}\left(\prod_{p\in{\Sigma_p}}B_p\prod_{n\in\Gamma}B_n\right)=\left\{\Psi_\Gamma\in\H_{\Sigma_p}\ \Bigg|\ \left(\prod_{p\in{\Sigma_p}}B_p\prod_{n\in\Gamma}B_n\right)\triangleright\Psi_\Gamma=\Psi_\Gamma\right\},
\ee
where $\Psi_\Gamma$ denotes a state in $\H_{\Sigma_p}$ defined by a graph $\Gamma$. In the case of the $p$-punctured sphere, this space is one dimensional and a basis is given by the unique vacuum state corresponding to the empty graph on $\mathbb{S}_p$. Since the operators $B_n$ and $B_p$ commute (both with each other and together, and independently of the location of the punctures), there is no ordering ambiguity and the definition \eqref{vacuum-space} is meaningful. We explicitly show this on the example of the basis $\Q$ for $\H_{\mathbb{S}_2}$ in appendix \ref{appendix:BnBp}. Notice that there is a natural isomorphism $\H_{\Sigma_p}^0\simeq\H_\Sigma$ between the space \eqref{vacuum-space} of vacuum states and the graph Hilbert space on the manifold $\Sigma$ without any punctures.

Finally, let us conclude this subsection by introducing the Wilson loop operator. This operator is defined by inserting a normalized Wilson loop in $\Sigma_p$, i.e.
\be\label{Wilson-operator}
W_l\coloneqq\f{1}{v_l^2}\pic{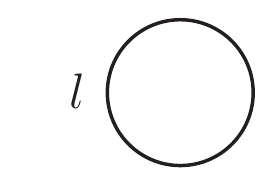}.
\ee
This is nothing but a measure of curvature. Indeed, if $W_l$ acts ``away'' from punctures, the loop can be evaluated and, together with the normalization coefficient, one gets the identity. This is because the local equivalence relations defining the graphs amount to having flatness away from the punctures. Similarly, if $W_l$ acts on a puncture surrounded by a vacuum loop, i.e. a puncture in the flat vacuum state, the $l$ loop can be pulled over the puncture and one gets the identity again. Therefore, there is curvature located in a region whenever the action of the operator $W_l$ is, for any $l$, different from the identity. This is the case in particular around a naked puncture (i.e. a puncture not in the vacuum state).

\subsection{Embedding maps}
\label{sec:embedding}

\noindent Now that we have obtained a description of the vacuum on the graph Hilbert space $\H_{\Sigma_p}$, we are able to define refinement operations and embedding maps. These are the operation which will enable us, in particular, to extend the inner product of section \ref{subsec:innerproduct} to states which live on the same underlying manifold but with different numbers of punctures. As we have seen, a state carries degrees of freedom in the form of torsion and curvature excitations located at the punctures. When refining a state, one allows for the possibility of describing more degrees of freedom by adding a puncture in the vacuum configuration, and this puncture can then be excited with the creation operator which we will define later on in section \ref{sec:creation-operators}. More precisely, the embedding map defined by the vacuum is the mathematical structure which enables us to refine a given state and embed it in a continuum Hilbert space defined as an inductive limit over the punctures.

The embedding maps are given by the operation of adding embedded punctures in the vacuum state. Remember that a puncture in the vacuum state is simply represented graphically by a puncture surrounded by a vacuum loop. If we denote by $\Sigma_{p+q}$ the punctured manifold obtained by adding $q$ punctures to the manifold $\Sigma_p$, the embedding maps are given by
\be\label{refinement-map}
\begin{array}{cccl}
\E_{p,q}:&\H_{\Sigma_p}&\longrightarrow&\H_{\Sigma_{p+q}}\\
&\Psi_\Gamma&\longmapsto&\Psi_\Gamma\times_q\pic{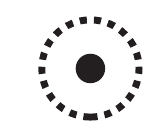},
\end{array}
\ee
where the symbol $\times_q$ simply means that we are attaching $q$ vacuum punctures to the state $\Psi_\Gamma$. As a technical remark, note that we are carefully using the notation $\Sigma_{p+q}$ to indicate that the embeddings of the $p$ common punctures of $\Sigma_p$ and $\Sigma_{p+q}$ have to agree. Similarly, the embedding maps in the $\SU(2)$ BF vacuum \cite{BDG} are defined between a coarse triangulation $\Delta$ and a finer triangulation $\Delta'$ obtained by refining $\Delta$, which means that all the embedded vertices of $\Delta$ are contained in $\Delta'$. Here a similar requirement is necessary for the punctures.

With these embedding maps, we are now able to define the inner product between states on manifolds $\Sigma_p$ and $\Sigma_q$ where the punctures are not required to have the same location. For this, one has to use the map $\E_{p,p+q}$ to add the punctures of $\Sigma_q$ to $\Sigma_p$, and, likewise, the map $\E_{q,q+p}$ to add the punctures of $\Sigma_p$ to $\Sigma_q$. In case one wishes to embed a new puncture at a position already occupied by a strand or a node of the graph, one first has to slightly deform the graph using the equivalences relations \eqref{StrandDeformation}, \eqref{F-move}, and \eqref{bubble move}. By doing this operation, one embeds the states $\Psi_{\Gamma}\in\H_{\Sigma_p}$ and $\Psi_{\Gamma'}\in\H_{\Sigma_q}$ into the common larger Hilbert space $\H_{\Sigma_{p+q}}$ on which the inner product can then be computed. In fact, the punctured manifold $\Sigma_{p+q}$ on which this larger Hilbert space is defined can be thought of, in the terms of \cite{DG1,DG2,BDG}, as the common refinement of the manifolds $\Sigma_p$ and $\Sigma_q$. It is necessary to resort to this common refinement in order to compare states which have support on manifolds with different number or location of the punctures.

As an illustrative example, one can consider the states
\be\label{StatesOn2p3p}
\H_{\mathbb{S}_3}\ni\Psi_\Gamma=\pic{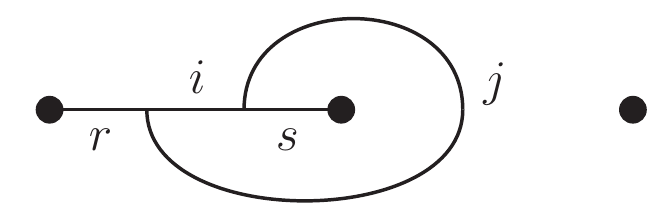},
\q\q
\H_{\mathbb{S}_2}\ni\Psi_{\Gamma'}=\pic{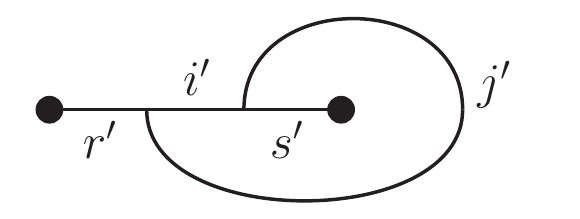}\hs,
\ee
whose inner product can be computed by embedding $\Psi_{\Gamma'}$ in $\H_{\mathbb{S}_3}$. One then finds that
\be
\big\la\Psi_\Gamma\big|\Psi_{\Gamma'}\big\ra_\tr=\Big\la\pic{Qijrs+p}\Big|\pic{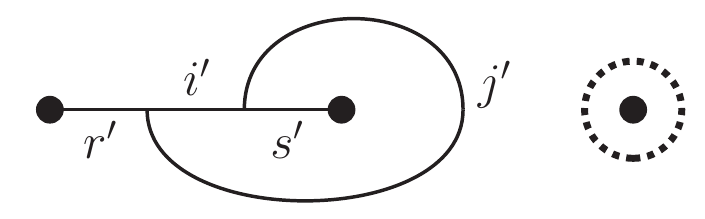}\Big\ra_\tr=\f{1}{\D}\delta_{ii'}\delta_{jj'}\delta_{rr'}\delta_{ss'}.
\ee
To compute this trace inner product we have simply applied the definition given above. In particular, we have divided by a factor of $\D$ and linked an additional vacuum loop through the vacuum loop encircling the new puncture in $\Psi_{\Gamma'}$. Then, using the fact that
\be
\pic{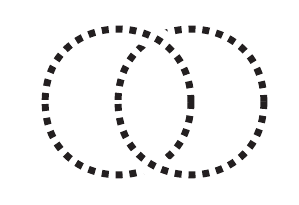}=1
\ee
removes the vacuum loop encircling the new puncture in $\Psi_{\Gamma'}$, we have connected the two naked punctures with strands of spin $j=0$ and evaluated the rest of the trace inner product graph. In particular, this calculation shows that the two states in \eqref{StatesOn2p3p} are not orthonormal because of the factor of $\D^{-1}$ in their inner product. This is to be expected given the fact that a naked puncture is different from a puncture in the vacuum state, i.e. a puncture surrounded by a vacuum line. The former carries curvature while the latter does not.

Let us now show that the inner product is cylindrically consistent with respect to the embedding maps. This crucial property, which also holds in the AL and $\SU(2)$ BF representations (although in a different technical setting), ensures that the inner product between two states is independent of the choice of refined punctured manifold on which it is evaluated. If we denote by $\Psi_\Gamma$ and $\Psi_{\Gamma'}$ two states in $\H_{\Sigma_p}$, the property of cylindrical consistency translates into the following equation:
\be
\big\la\Psi_\Gamma\big|\Psi_{\Gamma'}\big\ra_\tr=\big\la\E_{p,q}\Psi_\Gamma\big|\E_{p,q}\Psi_{\Gamma'}\big\ra_\tr.
\ee
Using the definition of the trace inner product and of the embedding maps, it is straightforward to show that this relation holds. In fact, it is sufficient to show that it holds for the case $q=1$ since then the result can be extended recursively. From the definition of the trace inner product, one can see that the inner product between disconnected components of the graphs can be computed separately and then multiplied back together. In particular, this means that
\be
\big\la\E_{p,1}\Psi_\Gamma\big|\E_{p,1}\Psi_{\Gamma'}\big\ra_\tr=\big\la\Psi_\Gamma\pic{VacuumPuncture}\big|\Psi_{\Gamma'}\pic{VacuumPuncture}\big\ra_\tr=\big\la\Psi_\Gamma\big|\Psi_{\Gamma'}\big\ra_\tr\times\big\la\pic{VacuumPuncture}\;\big|\pic{VacuumPuncture}\;\big\ra_\tr,
\ee
which shows that cylindrical consistency amounts to the requirement that a puncture surrounded by a vacuum loop be a state of unit norm. This is indeed the case since
\be\label{UnitNormPunctures}
\big\la\pic{VacuumPuncture}\;\big|\pic{VacuumPuncture}\;\big\ra_\tr=\f{1}{\D}\pic{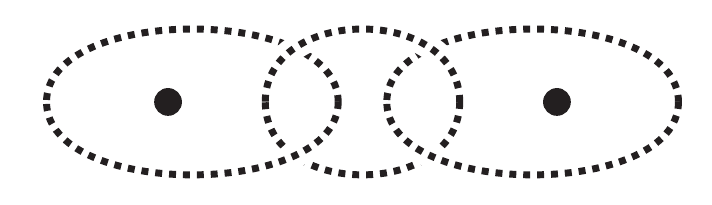}\stackrel{\tr}{=}\f{1}{\D}\pic{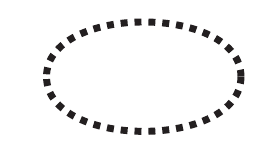}=1.
\ee
For this last calculation, we have used the definition of the trace inner product, i.e. inserted a factor of $\D^{-1}$ together with a vacuum loop, and then used the identity
\be
\pic{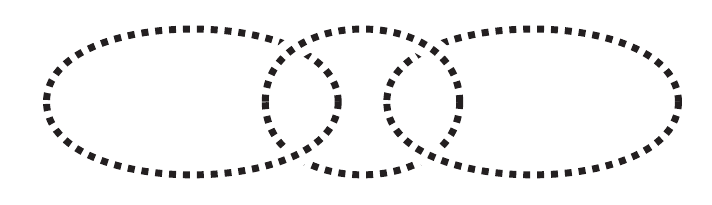}=\pic{VacuumLoop}
\ee
together with the fact that a vacuum loop away from a puncture is in fact equal to the number $\D$.

Finally, let us comment on the following important subtlety: so far, with our definitions, two Hilbert spaces $\H_{\Sigma_p}$ and $\H_{\Sigma_q}$ might not have a common refinement. This lack of a common refinement leads to superselection sections in an inductive limit Hilbert space based on the embedding maps \eqref{refinement-map}.

This possible lack of a common refinement is due to the marked points: consider the case that two surfaces $\Sigma_p$ and $\Sigma_q$ agree in the positioning of the punctures, to the extend that the same small disks have been removed from $\Sigma$. However the punctures do disagree in their marked points. Since we allowed only one marked point per puncture (correspondingly, one open strand possibly ending at this puncture), we cannot construct a common refinement for this case. 

One possibility to deal with this situation is to prescribe a certain way for choosing the marked points at the punctures (e.g. by fixing a coordinate chart and using the coordinates to this end), so that this situation does not arise. Another possibility is to allow for an arbitrary number of marked points (and correspondingly open strands possibly ending at the punctures). We will briefly discuss this possibility in the appendix \ref{App_Last}.

\subsection{Relationship with the Turaev--Viro state sum}

\noindent One of the numerous motivations for working with the BF representation and its $\Lambda\neq0$ generalization to the TV representation, is that it connects the kinematical structure of the canonical theory together with the spin foam amplitudes of the covariant approach. As already explained in \cite{DG1} for the BF representation, the reason for this is that the embedding maps describe refinements of the triangulation which can seen as the gluing of flat tetrahedra (and their associated spin foam amplitudes) onto the triangles being refined. In this section, we explain how this is also realized in the TV representation, and explain the relationship between the kinematical vacuum of section \ref{subsec:vacuum} and the TV state sum.

The TV state sum, whose definition is recalled in appendix \ref{appendix:TV}, is a prescription for computing a bulk topological invariant for a triangulated three-dimensional manifold with possible boundary components. It can be understood as taking as input some boundary data, which consists of a triangulation $\Delta$ of the boundary $\Sigma$, and a coloring $\psi$ of its edges with spins so that the coupling rules at the dual nodes (or equivalently for spins associated to one triangle) are satisfied. These data live in state spaces $\K_{\Sigma,\Delta}$. This already bears a resemblance with the TV representation, since this latter is based on the graph Hilbert space $\H_{\Sigma_p}$ whose elements are colored three-valent graphs, and three-valent graphs are dual to triangulations. Here we assume that the coloured graphs in $\H_{\Sigma_p}$ satisfy the Gau\ss~constraints (thus the coupling rules at the nodes of the dual graph hold). More precisely, one can always map by duality a state in $\K_{\Sigma,\Delta}$ to a state in $\H_{\Sigma\backslash\Delta_0}$, where $\Sigma\backslash\Delta_0$ is the punctured manifold obtained by removing from $\Sigma$ the vertices of its triangulation.

Now, just like the space of ground states $\H_{\Sigma_p}^0$ is defined in \eqref{vacuum-space} as the image of a projection operator, the TV state sum assigns to two-dimensional manifolds a vector space
\be
\K^0_\Sigma\coloneqq\text{Im}\big(\Z_\text{TV}(\Sigma\times[0,1],\Delta,\psi)\big),
\ee
which is the image of the projector defined by the state sum itself, and which can be shown not to depend on a particular choice of triangulation $\Delta$ \cite{BalKir}. Then the precise relationship between the TV state sum and the graph Hilbert space, which has been proven in \cite{BalKir,Kir}, is that there is an isomorphism $\K^0_\Sigma\simeq\H^0_\Sigma$. This result, which means in essence that the TV state sum is a projector onto the TV vacuum state, is an isomorphism between spaces which are characterized by the manifold $\Sigma$ alone, and by no other auxiliary structure (such as triangulations or punctures) thereon. 

This result can also be understood by looking at the explicit algebraic relation between the projection operator $B_p$ and the TV state sum. Consider for instance that the initial and final boundary triangulations are the same. Such boundary triangulations can then be connected by a series of so-called tent moves \cite{Sorkinetal,bonzom,bdhoehn2}, which have the property of not changing the connectivity of the triangulation. Starting from the initial triangulation, in order to ensure that the bulk triangulation has everywhere some ``thickness'', one has to apply a tent move to every vertex of the triangulation. It turns out that the TV state sum for a tent move on a vertex can be seen as an operator acting on $\H_{\Sigma\backslash\Delta_0}$. This operator agrees with the action of the operator $B_p$. In order to see this, one can first map  the initial (colored boundary triangulation) state in $\K_{\Sigma,\Delta}$ to a state in $\H_{\Sigma\backslash\Delta_0}$. This is done by considering the dual graph $\Gamma_\Delta$ and the spin coloring of its strands inherited by duality from the edges of $\Delta$. In this duality, the vertices $\Delta_0$ can be seen as punctures at the centre of the faces of $\Gamma_\Delta$. One can then compute the action of the projector $B_p$ on such a face/puncture. We recall in \eqref{Proof of Bp on face} (in the case $s=6$) the proof that
\be\label{Bp on face}
B_\text{face}\triangleright\pic{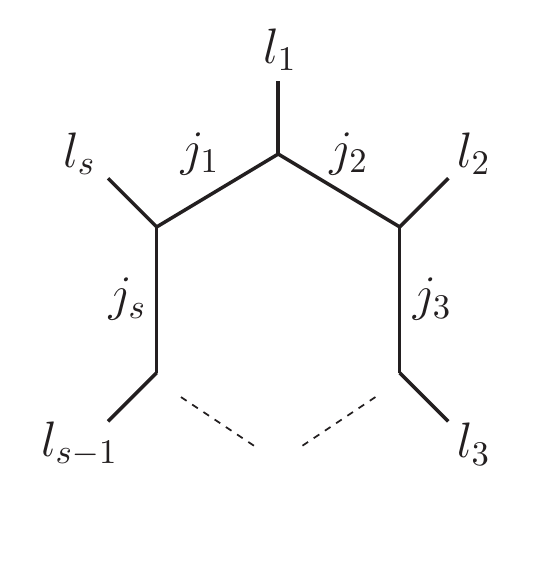}=\f{1}{\D^2}\sum_{kn_1\dots n_s}v_k^2\,\delta_{s+1,1}\prod_{i=1}^sv_{n_i}v_{j_i}G^{n_{i+1}l_in_i}_{j_1kj_{i+1}}\pic{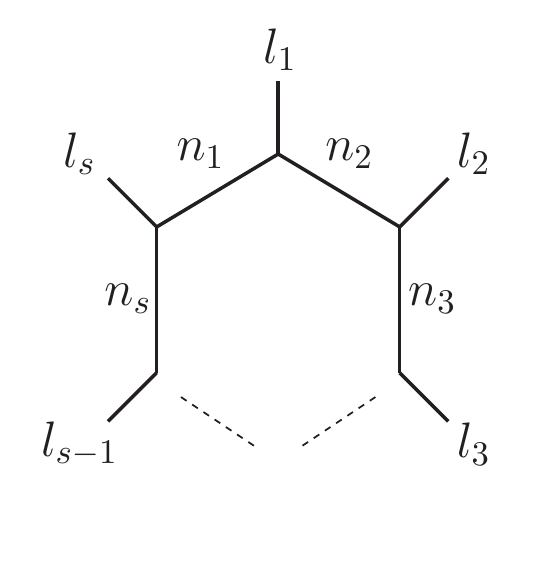},
\ee
where we have used the totally tetrahedral-symmetric symbol $G^{ijm}_{kln}\coloneqq F^{ijm}_{kln}/(v_mv_n)$, and omitted for clarity the puncture in the middle of the face. This formula is the algebraic expression for the tent move, which, seen from the point of view of the triangulation, corresponds to the following gluing of tetrahedra:
\be\label{tent move}
\pic{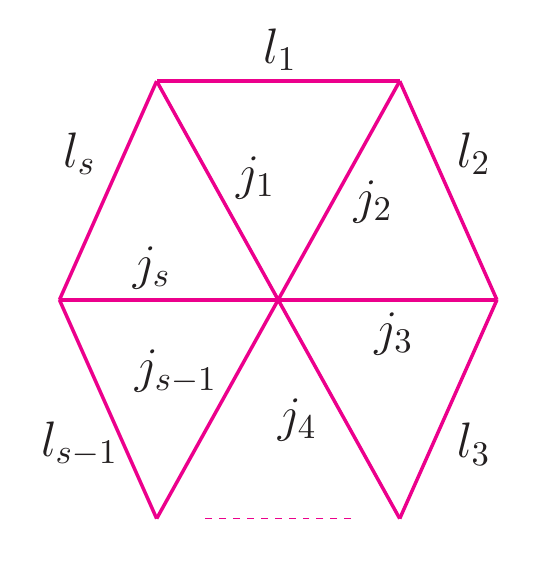}\stackrel{\text{tent move\vphantom{$\f{1}{2}$}}}{\longrightarrow}\pic{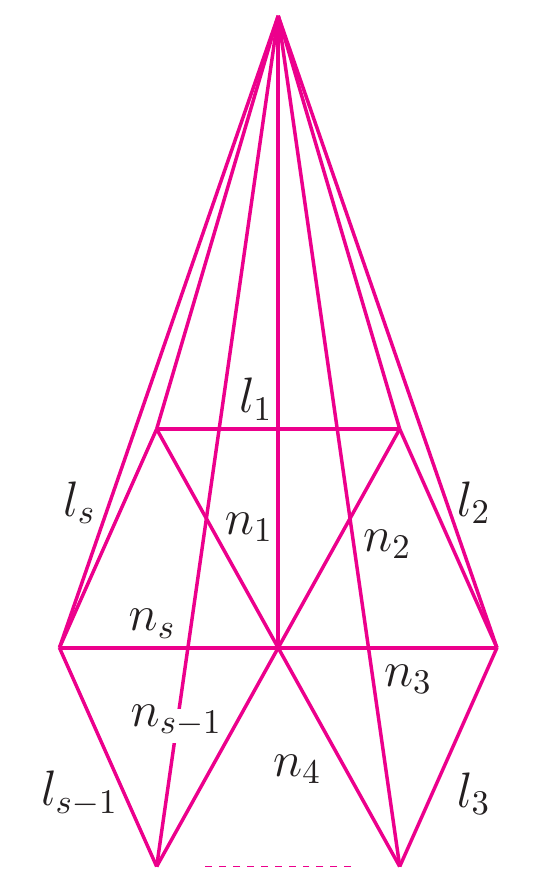}\text{with labels e.g.}\pic{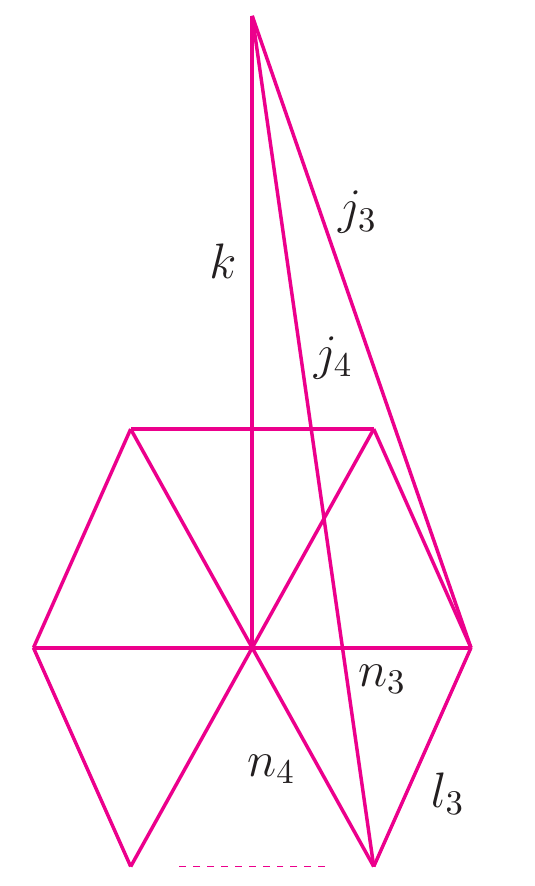}.
\ee
Each of the $G$-symbols in \eqref{Bp on face} labels a tetrahedron with the pattern indicated above on the example of $G^{n_4l_3n_3}_{j_3kj_4}$. The gluing of tetrahedra in \eqref{tent move} represents a triangulation of a ball $\mathbb{B}$, whose TV state sum  is related to \eqref{Bp on face} via the following formula:
\be\label{tentmoveIP}
\big\la\Gamma_\Delta,\vec{\jmath},\vec{l}\;\big|B_\text{face}\big|\Gamma_\Delta,\vec{\jmath}\;',\vec{l}\;\big\ra=\f{\D^s}{\prod_lv_l}\Z_\text{TV}\big(\mathbb{B},\Delta(\partial\mathbb{B}),\vec{\jmath},\vec{\jmath}\;',\vec{l}\;\big).
\ee
Note that the prefactors which appear on the right-hand side of this expression can be reabsorbed in the definition of the map $\K_{\Sigma,\Delta}\rightarrow\H_{\Sigma\backslash\Delta_0}$ (as is done in \cite{Kir}).

Equation \eqref{tentmoveIP} gives a relationship between the inner product in $\H_{\Sigma_p}^0$ with insertion of a projector, and the TV partition function for the example of a tent move. This relation can be generalized in two ways. First, one can consider the partition function between initial and final triangulations which differ from each other. Since the partition function is independent from the choice of bulk triangulation, we can choose any arbitrary one. We can first preform a tent move on each vertex of the initial triangulation, which does not change the connectivity of the initial triangulation but projects out any curvature from the state associated with the initial spin coloring. We can then connect the initial and final triangulations by some bulk triangulation. This can be interpreted as gluing tetrahedra to a series of hypersurfaces, starting with the initial triangulation and ending with the final triangulation. The gluing of tetrahedra to the hypersurface corresponds to a change of the two-dimensional triangulation via $2-2$, $3-1$ and $1-3$ Pachner moves \cite{bdhoehn2}. These Pachner moves are implemented by the state equivalence relations \eqref{StrandDeformation}, \eqref{F-move}, and \eqref{bubble move}. Now, one can check that the equivalences are defined in such a way that each gluing of a tetrahedron comes with a $G$-symbol attached to it, which is also consistent with the gluing rule for the TV partition function. The inner product between two states in $\H_{\Sigma_p}$ is defined by using the state equivalences \eqref{StrandDeformation}, \eqref{F-move}, and \eqref{bubble move} to transform the two states to a common computational basis (in particular a common underlying graph). One can then evaluate the inner product using the orthonormality of this basis. These steps can again be understood through the relation of the equivalences to gluing tetrahedra as including a further refinement of the bulk triangulation, which leaves the TV partition function unchanged.
 
In summary, \eqref{tentmoveIP} generalizes to different initial and final boundary triangulations with respective spin colorings $\psi$ and $\psi'$, and one has that
\be
\big\la\Gamma_\Delta,\psi\big|\prod_\text{faces}B_\text{face}\big|\Gamma_{\Delta'},\psi'\big\ra=\Z_\text{TV}\big(\Sigma\times[0,1],\Delta,\Delta',\psi,\psi'\big).
\ee

A further generalization of this result is to allow for punctures with curvature and possible torsion excitations as well. This requires the generalization of the TV state sum model to the extended TV model \cite{BalKir}. In the $(2+1)$-dimensional picture, the two-dimensional punctures are ``evolved'' into three-dimensional tubes which have to be excised from the three-dimensional manifold. One can then show that the extended TV state sum computes the inner product in $\H_p$ between boundary kinematical states with excitations.

\section{The two-punctured sphere}
\label{section:2p}

\noindent We have so far developed a general understanding of the graph Hilbert space and the vacuum states on the punctured sphere. As we will now see, in the case of the cylinder the Hilbert space and its basis states do actually posses some extra mathematical structure, the understanding of which will play an important role in the construction of the ribbon excitation operators. We devote this section to the study of these properties.

On punctured manifolds, there exists a gluing operation $\pound$ along the punctures. If we denote by $\Sigma_p^g$ a $p$-punctured manifold of genus $g$, leaving aside properties other than topological, this operation is such that $\Sigma_{p_1}^{g_1}\pound\Sigma_{p_2}^{g_2}=\Sigma_{p_1+p_2-2}^{g_1+g_2}$. Two-punctured spheres therefore play a special role from the point of view of this operation, since two of them can be glued along a puncture to again obtain a two-punctured sphere, i.e. $\Sigma_2^0\pound\Sigma_2^0=\Sigma_2^0$. Moreover a two-punctured sphere can always be glued along a puncture to another punctured surface of arbitrary topology without changing this topology or the number of punctures, i.e. $\Sigma_p^g\pound\Sigma_2^0=\Sigma_p^g$. In the previous section, we have constructed a Hilbert space of states by considering graphs on punctured manifolds. Since each puncture carries a marked point where open strands are allowed to end, when gluing punctured manifolds we can also consistently glue their graphs by matching the strands at the marked points. In this sense, we can understand the gluing of a two-punctured sphere onto a punctured manifold $\Sigma_p$ as the action on $\H_{\Sigma_p}$ of an operator, where the details of this action depends on the graph state on the two-punctured sphere being glued. Let us explain in more details how this comes about.

\begin{figure}[h]
$\pic{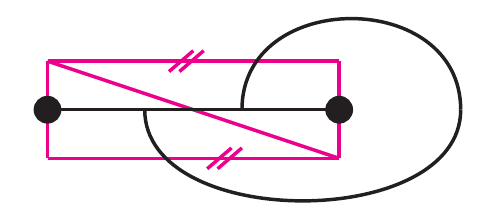}$
\caption{Triangulation of a cylinder and its dual graph. For each of the two circle boundaries of the cylinder, the dual graph has an open strand ending at a marked point. These circle boundaries and their marked point are the punctures of the two-punctured sphere.}
\label{2pSphereTriangulated}
\end{figure}

Recall that the two-punctured sphere is topologically equivalent to a cylinder. Since a cylinder can be seen as a quadrangle with two opposite sides identified, the two-punctured sphere can be minimally triangulated by two triangles, as represented on figure \ref{2pSphereTriangulated}.
This is just another way of understanding the result of the previous section, namely the fact that an orthonormal basis on the cylinder is given by the $\Q$ states defined in \eqref{Qijrs}. Since these $\Q$ states represent a basis, any other state in $\H_{\mathbb{S}_2}$ can always be written as some linear combination $f(\Q)$. The gluing of a cylinder state can therefore be understood as an operator\footnote{In order to lighten the notation we do not write hats on operators.}
\be
\begin{array}{cccl}
f(\Q):&\H_{\Sigma_p}&\longrightarrow&\H_{\Sigma_p}\\
&\Psi_\Gamma&\longmapsto&\Psi'_{\Gamma'}\coloneqq f(\Q)\times\Psi_\Gamma.
\end{array}
\ee
We are going to encounter below examples of states expanded in the $\Q$ basis and explicit computations of their action as operators. In particular, among all the operators defined by states on the cylinder, operators that act as projectors will play an important role in what follows. They are the operators which will enable us to characterize via certain stability conditions the quasi-particle excitations (i.e. the excited states of the theory) and to construct the excitation operators. In order to understand the role of states acting as projectors and to see in which sense they satisfy a certain stability property, let us first focus on states which only carry curvature excitations.

\subsection{Pure curvature states}\label{pure curv}

\noindent Pure curvature states are states which are invariant under the action of the operator $B_n$ for every node, i.e. which have admissible colorings of the strands at the three valent nodes and no strands ending at the punctures. Such states are in general not invariant under the action of the projection operators $B_p$, and can therefore carry curvature. On the cylinder, a basis for such states is given by
\be
\left(\prod_nB_n\right)\triangleright\Q^{ij}_{rs}=\delta_{ijr}\delta_{ijs}\delta_{r00}\delta_{s00}\pic{Qijrs}\hs=\delta_{ij}\delta_{r0}\delta_{s0}\Q^{ij}_{rs},
\ee
which therefore simply reduces to
\be
\Q^{jj}_{00}=\pic{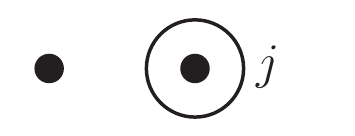}\hs.
\ee
These states can carry curvature at the punctures. Indeed, using the normalized Wilson loop operator introduced in \eqref{Wilson-operator} as a measure of curvature, one finds the non-trivial action
\be\label{wilson on curvature state}
W_l\triangleright\Q^{jj}_{00}=\f{1}{v_l^2}\pic{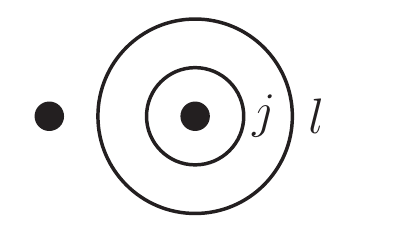}\hs=\f{1}{v_l^2}\sum_m\delta_{jlm}\pic{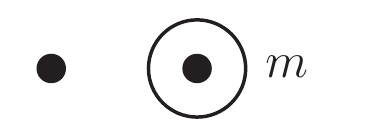}\hs.
\ee
Moreover, this action is also non-trivial on the state $\Q^{00}_{00}$ with a loop of spin $j=0$, in which case one finds $W_l\triangleright\Q^{00}_{00}=v_l^{-2}\Q^{ll}_{00}$. This is just a manifestation of the fact that naked punctures carry curvature. We see from the simple calculation \eqref{wilson on curvature state} that the  states $\Q^{jj}_{00}$ are not eigenstates of the Wilson loop operator. Instead, the action of this latter leads to a superposition of these states. In this sense, the states $\Q^{jj}_{00}$ do not satisfy any particular stability property under the action of the Wilson loop operator.

We are therefore naturally brought to look for  states which are eigenstates of the Wilson loop operator. In terms of group variables, such states would be analogous to $\psi([g])=\int\delta(hgh^{-1})\de h$, which is indeed a gauge-invariant eigenstate of the Wilson loop operator. Here, in the quantum group picture with spin labels, such states are given by the graphical representation
\be
\O^{jj}_{00}\coloneqq\pic{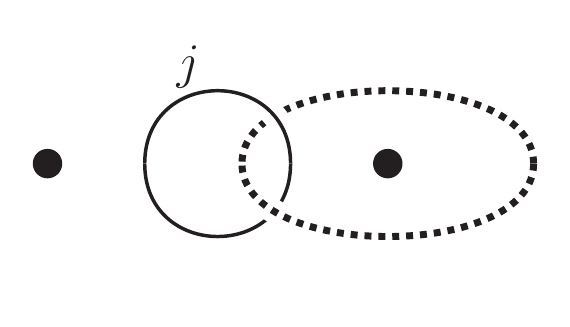},
\ee
which is a strand of spin $j$ linked with non-trivial braiding to a vacuum loop. These states have a rather complicated expansion in terms of the $\Q$ basis given by
\be\label{OjtoQ}
\O^{jj}_{00}=\f{1}{\D}\sum_{km}v_m^2\big(R^{jk}_m\big)^2\Q^{kk}_{00}=\sum_kS_{jk}\Q^{kk}_{00}, 
\ee
where a proof of the first equality is given in \eqref{Proof of OjtoQ}, and for the second equality we have used the definitions \eqref{SM1} and \eqref{s matrix in terms of R} for the $S$-matrix. This is an example of an expansion $\O=f(\Q)$ of a cylinder state into the original $\Q$ basis. Now, one can indeed check that these states have the desired property, namely that of being eigenstates of the Wilson loop operator. The action $W_l\triangleright\O^{jj}_{00}$ of this latter is given by
\be\label{lWilsonAroundOii00}
\f{1}{v_l^2}\pic{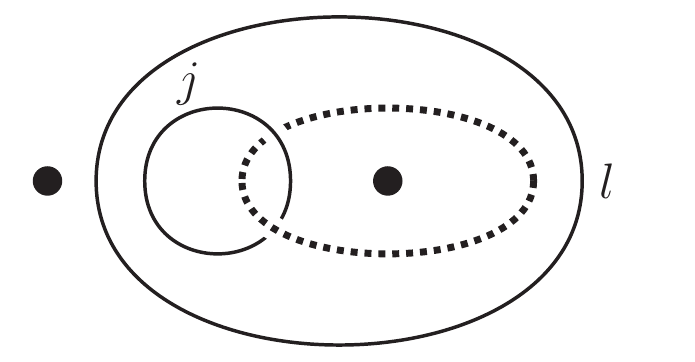}\hs=\f{1}{v_l^2}\pic{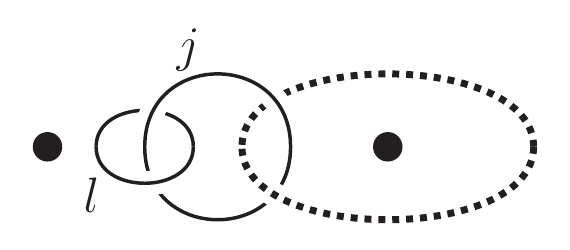}=\f{1}{v_j^2v_l^2}s_{jl}\pic{Oii00},
\ee
where we have  used \eqref{S through} to obtain the non-normalized $s$-matrix $s_{ij}=\D S_{ij}$.

We therefore see that the states $\O^{jj}_{00}$ are eigenstates of the normalized Wilson loop operator, and, using the explicit expression for the $s$-matrix in  \eqref{s matrix in terms of R} and the quantum dimensions in \eqref{qNumber}, one finds that the eigenvalues are given by
\be\label{Wilson q eigenvalue}
W_l\triangleright\O^{jj}_{00}=\f{\displaystyle\sin\left(\f{\pi}{\k+2}\right)\sin\left(\f{\pi}{\k+2}(2l+1)(2j+1)\right)}{\displaystyle\sin\left(\f{\pi}{\k+2}(2l+1)\right)\sin\left(\f{\pi}{\k+2}(2j+1)\right)}\O^{jj}_{00}.
\ee
This should be compared with the eigenvalue of the (normalized) Wilson loop operator in the undeformed $\SU(2)$ case, which is given by
\be
\f{\sin\big((2l+1)\theta\big)}{(2 l+1)\sin(\theta)}
\ee
for a state $\int\delta(hgh^{-1})\de h$ peaked on a class angle $\theta(g)\in[0,\pi]$. This does indeed agree with the large $\k$ (i.e. ``classical'' group) limit of \eqref{Wilson q eigenvalue}, where the class angle is given by $\theta\sim\pi(2j+1)/(\k+2)$, and we can conclude that the spin label $j$ in $\O^{jj}_{00}$ is a measure of curvature.

Note that the calculation \eqref{lWilsonAroundOii00} is a special case of the following more general result:
\be\label{pulling over O}
\pic{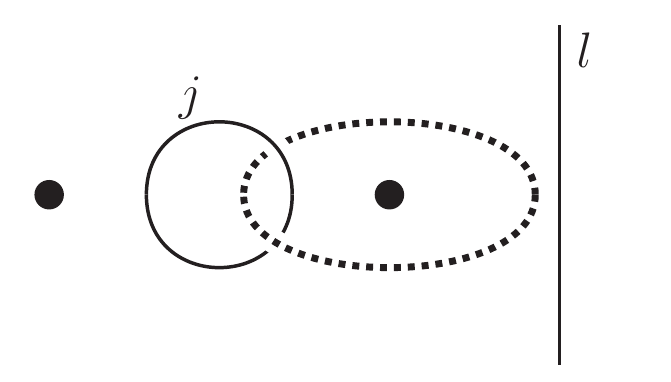}\hs=\sum_{mp}\f{v_m}{v_jv_l}\big(R^{jl}_m\big)^2F^{jlm}_{ljp}\pic{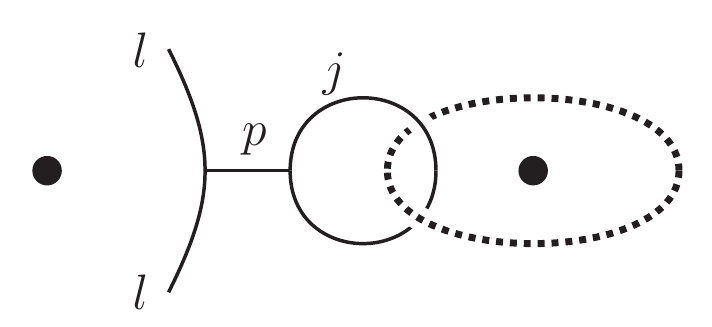},
\ee
which tells us how the state $\O^{jj}_{00}$ changes when it crosses a strand of spin $l$. A proof of this relation is given in \eqref{proof of pulling over O}. One can see that by closing the strand $l$ (while avoiding the left puncture) and dividing by $v_l^2$ we indeed recover \eqref{lWilsonAroundOii00}. Moreover, one can recognize from \eqref{pulling over O} that the set of states $\{\O^{jj}_{s0}\}_s$ are stable under the operation which consists in crossing a strand. We will see shortly that this property characterizes quasi-particles excitations. 

The appearance of the $S$-matrix in \eqref{OjtoQ} and \eqref{lWilsonAroundOii00} is not a coincidence. We were looking for eigenstates of the (here non-normalized) Wilson loop operators $\widetilde{W}_j=v_j^2W_j$. As can be seen from \eqref{wilson on curvature state}, the Wilson loop operators themselves form an algebra
\be\label{W algebra}
\widetilde{W}_j\triangleright\widetilde{W}_k=\sum_m\delta_{jkm}\widetilde{W}_m,
\ee
where the multiplication coefficients are given by the fusion matrices $\delta_{jkm}=(\mathbf{N}_j)_{km}$. Therefore, a diagonalization
\be
(\mathbf{N}_j)_{km}=\sum_p\big(U^{(j)}\big)_{kp}\lambda^{(j)}_p\big(U^{(j)}\big)^*_{pk}
\ee
of $(\mathbf{N}_j)_{km}$ leads to a diagonalization for the Wilson loop operator $\widetilde{W}_j$. As stated in \eqref{diagN}, such a diagonalization is provided by the $S$-matrix, i.e.
\be
\big(U^{(j)}\big)_{kp}=S_{kp},
\ee
the matrices diagonalizing $\mathbf{N}_j$ are independent of $j$, and the eigenvalues are given by $\lambda^{(j)}_p=S_{jp}/S_{0p}=s_{jp}/v_p^2$. The eigenstates $\sum S_{pk}\widetilde{W}_k$ are thus also defining modules of the algebra \eqref{W algebra}, i.e. these states are invariant under the action of $\widetilde{W_j}$ for all admissible spins $j$. 

In the following subsection, we are going to generalize this kind of reasoning to more generic states.

\subsection{General states and the tube algebra}
\label{section:2pg}

\noindent Let us now consider the general states on the cylinder given by the orthonormal $\Q$ basis \eqref{Qijrs}. As we have explained above, cylinders can naturally be glued together along punctures, and states on the cylinder can be seen as operators under this operation. Since the gluing of two cylinders gives back another cylinder, it is clear that the state obtained after the gluing can be again expanded in the cylinder $\Q$ basis. In particular, it is natural to ask what happens to the cylinder basis states themselves under this gluing. In other words, we would like to understand the action of $\Q$, when seen as an operator, on the $\Q$ basis states of $\H_{\mathbb{S}_2}$. For the sake of definiteness, when gluing two cylinder states by matching their left and right punctures respectively, we require the spins labeling the strands to match, and otherwise define the result of the gluing to be zero. By using the local equivalence relations, we can then compute the expansion in the $\Q$ basis of the gluing of two such basis states. This amounts to computing the multiplication rule of the algebra defined by this gluing of states. The calculation, which is given in \eqref{proof of QAlgebra}, leads to
\be\label{QAlgebra}
\Q^{ij}_{rs}\times\Q^{kl}_{tu}=\delta_{st}\sum_{mn}v_s\f{v_mv_n}{v_rv_u}F^{jsi}_{lmk}F^{mnr}_{jil}F^{mnu}_{lkj}\Q^{mn}_{ru}.
\ee
If we now introduce the following linear combinations of $\Q$ basis states:
\be
X\coloneqq\sum_{ijrs}X^{rs}_{ij}\Q^{ij}_{rs},
\q\q
Y\coloneqq\sum_{kltu}Y^{tu}_{kl}\Q^{kl}_{tu},
\ee
we find the product rule
\be\label{tube algebra multiplication}
X\times Y=\sum_{\substack{ijrs\\kltu}}X^{rs}_{ij}Y^{tu}_{kl}\left(\Q^{ij}_{rs}\times\Q^{kl}_{tu}\right)=\sum_{mnru}Z^{ru}_{mn}\Q^{mn}_{ru}\eqqcolon Z,
\ee
with
\be\label{z components definition}
Z^{ru}_{mn}=\sum_{ijkls}X^{rs}_{ij}Y^{su}_{kl}v_s\f{v_mv_n}{v_rv_u}F^{jsi}_{lmk}F^{mnr}_{jil}F^{mnu}_{lkj}.
\ee
This multiplication rule for linear combinations of $\Q$ basis states defines an algebra known in the literature as Ocneanu's tube algebra \cite{Ocneanu1,Ocneanu2,Mueger1,Mueger2}.

We have therefore shown that the generic $\Q$ basis states define at the same time a vector space and an algebra. This means that the space $\V_{\mathbb{S}_2}$ of graphs on the cylinder is a representation space for the tube algebra. Now, this representation space can be decomposed into indecomposable components, which by definition are modules over the tube algebra. Modules are by definition invariant under the action of cylinder basis states, and this stability property justifies the identification of the modules with quasi-particles \cite{Lan}. In fact, given the multiplication rule \eqref{tube algebra multiplication} for the tube algebra, we can write down the condition for the $\Q$ basis coefficients to define a projection operator. This condition is that $Z\times Z=Z$, which is a complicated equation when written in terms of the coefficients \eqref{z components definition}. We are therefore naturally brought to look for states which act as projectors.

For the case of $\SU(2)_\k$, a set of basis states satisfying this projection property is known \cite{Ocneanu1,Ocneanu2,KKR}, and graphically given by\footnote{Notice that there is a certain arbitrariness in the choice of these basis states, since one is free to choose the opposite crossings between the vacuum loop and the strands $i$ and $j$. This is analogous to the freedom in choosing the basis $\Q$ or $\widetilde{\Q}$ in \eqref{Q and Q tilde basis}. More precisely, the two choices of $\O$ basis states are related by the transformation \eqref{O upside down transformation}.}
\be\label{O basis state}
\O^{ij}_{rs}\coloneqq\pic{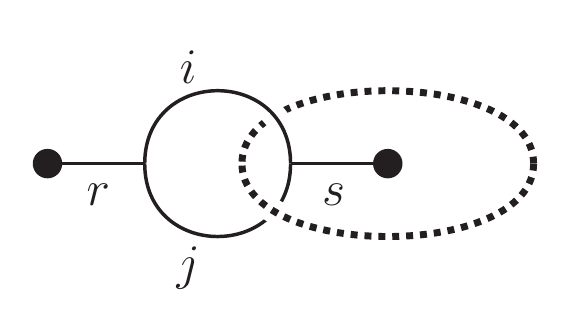}.
\ee
As shown in the calculation \eqref{proof of OtoQ}, these new basis states can be expanded in the $\Q$ basis as
\be\label{OtoQ}
\O^{ij}_{rs}=\sum_{kl}\widetilde{\Omega}^{rs}_{kl,ij}\Q^{kl}_{rs},
\ee
where we have introduced the tensors
\be\label{Omega definition}
\widetilde{\Omega}^{rs}_{kl,ij}\coloneqq\f{1}{\D}\f{v_iv_j}{v_s}v_l^2\Omega^{rs}_{kl,ij},
\q\q
\Omega^{rs}_{kl,ij}\coloneqq\sum_{mn}\f{v_mv_n}{v_rv_l^2}R^{il}_mR^{lj}_nF^{nmr}_{ijl}F^{kls}_{ijm}F^{rlk}_{jmn}.
\ee
As we shall see shortly, the tensor $\Omega$ will play a very important role in our construction. The projection property for the $\O$ basis states, which is proved in \eqref{proof of O times O}, is that
\be\label{O times O}
\O^{ij}_{rs}\times\O^{i'j'}_{s'u}=\D\delta_{ii'}\delta_{jj'}\delta_{ss'}\f{v_s}{v_iv_j}\O^{ij}_{ru}.
\ee
In addition, these new states $\O$ are also orthonormal, as can be seen from the trace inner product computation
\be\label{orthonormality of Os}
\Big\la\pic{Oijrs}\Big|\pic{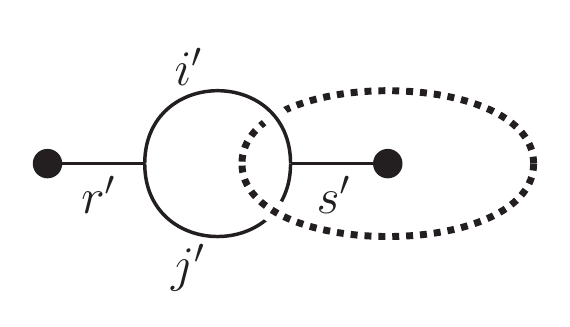}\Big\ra_\tr=\delta_{ii'}\delta_{jj'}\delta_{rr'}\delta_{ss'}
\ee
given in \eqref{proof of orthonormality of Os}.

All these properties of the $\O$ states become crucial when considering the crossing of a strand. In this situation, one can compute as in \eqref{proof of pulling through Oijrs} the following generalization of formula \eqref{pulling over O}, which shows the stability of the $\O$ states:
\be\label{pulling through Oijrs}
\pic{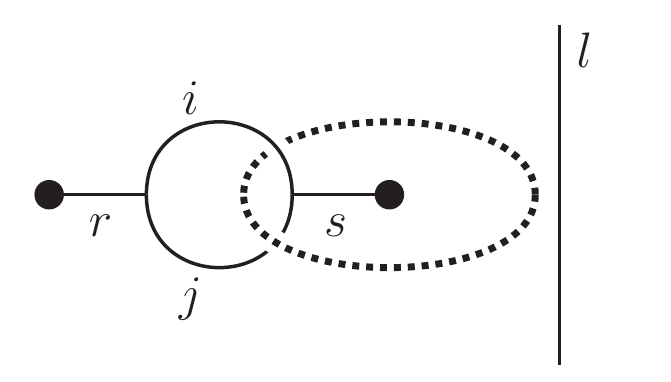}\hs
&=\sum_p\f{v_p}{v_iv_j}\pic{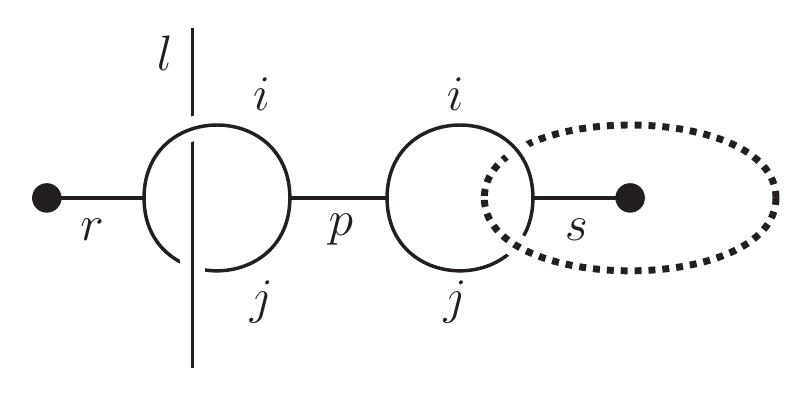}\nn\\
&=\sum_{pq}\Omega^{rp}_{ql,ij}\pic{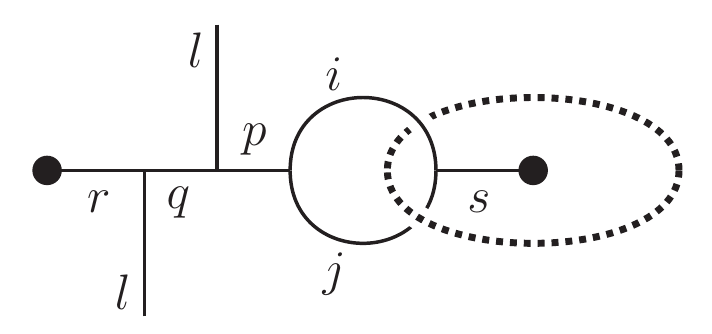}.
\ee
From this calculation one can derive the following graphical representation for the tensor $\Omega$:
\be
\Omega^{rp}_{ql,ij}=\f{1}{v_iv_jv_rv_l^2}\pic{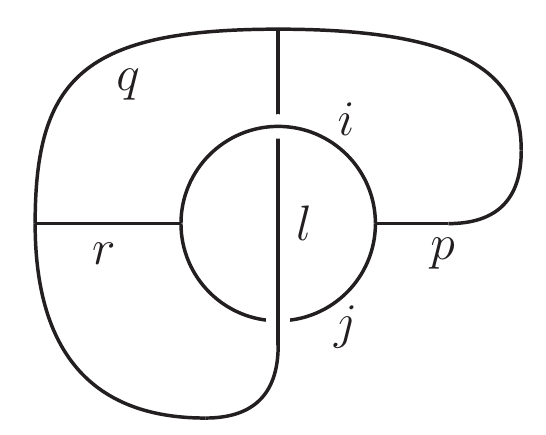}.
\ee
This graphical representation can be used to easily determine some special cases (with vanishing spins) for the coefficients of the half-braiding tensor.

Although the tensors $\Omega$ have already been introduced above, it is this stability property of the $\O$ states when passing through a strand which should be understood as their definition. We have already seen in the previous subsection that the $\Q$ states do not posses this nice stability property when crossing a strand. Instead, in this case, as shown in \eqref{proof of pulling through Qijrs}, one obtains a stacking of two $\Q$ states:
\be\label{pulling through Qijrs}
\pic{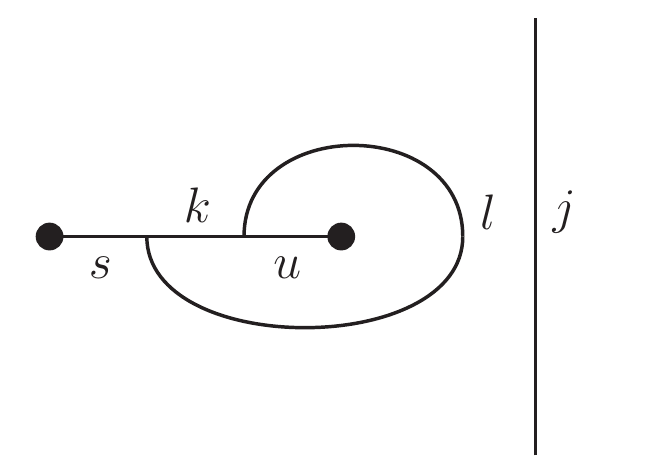}\hs=\sum_{imr}\f{v_r}{v_j^2v_s}F^{jsm}_{jri}\pic{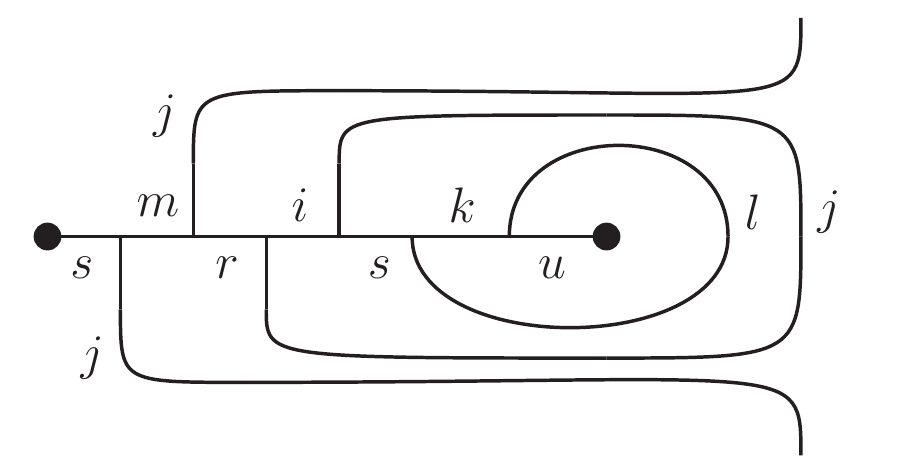}\hs.
\ee
Equation \eqref{pulling through Qijrs} is the reason why modules of the tube algebra, which by definition are stable under stacking, are also automatically stable when they cross a strand. Since these modules are spanned by $\{\O^{ij}_{kl}\}_{kl}$, we prefer to work with the $\O$ states instead of the $\Q$ ones. Indeed, we will see later on that the $\O$ states appear as eigenstates of so-called closed ribbon operators, which give a measure of both curvature and torsion.

\subsection{Half-braiding and Drinfeld centre}

\noindent We have just seen that the projection property of the $\O$ states ensures their stability when they cross a strand in the sense of \eqref{pulling through Oijrs}. This motivates their identification as quasi-particle excitations, and the introduction of the following graphical notation:
\be\label{half-braiding definition}
\pic{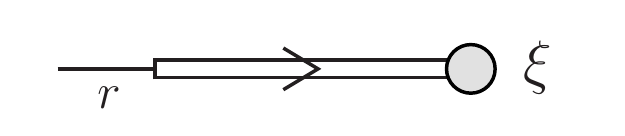}\hs\coloneqq\pic{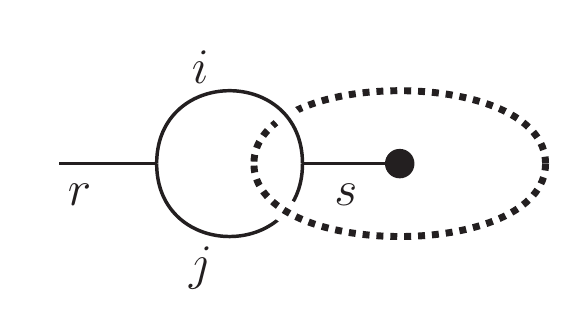},
\ee
where the quasi-particle label $\xi$ is a combined notation for the labels $(i,j,s)$ of $\O$. In the literature, it is also customary to call $\xi$ a quasi-particle of type $i\bar{j}$, so depending on the context we will use $\xi$ to denote either the particle type $i\bar{j}$ or the labels $(i,j,s)$. Notice that these new doubled strands now posses an orientation. With this new graphical notation, the property \eqref{pulling through Oijrs} tells us how the doubled strands cross and braid with standard strands. In particular, when going over a strand we have that
\be\label{pulling through half-braiding}
\pic{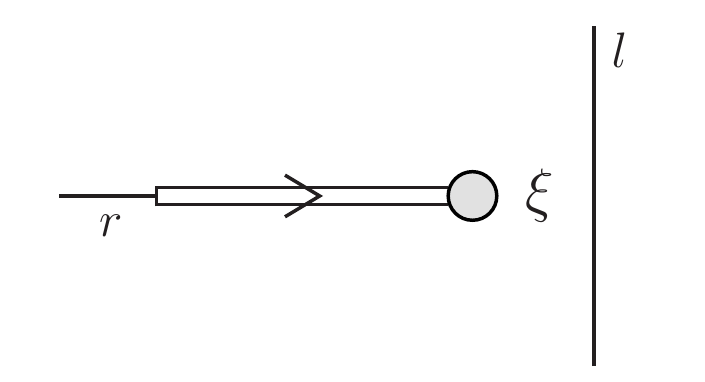}\hs
&=\sum_p\f{v_p}{v_iv_j}\pic{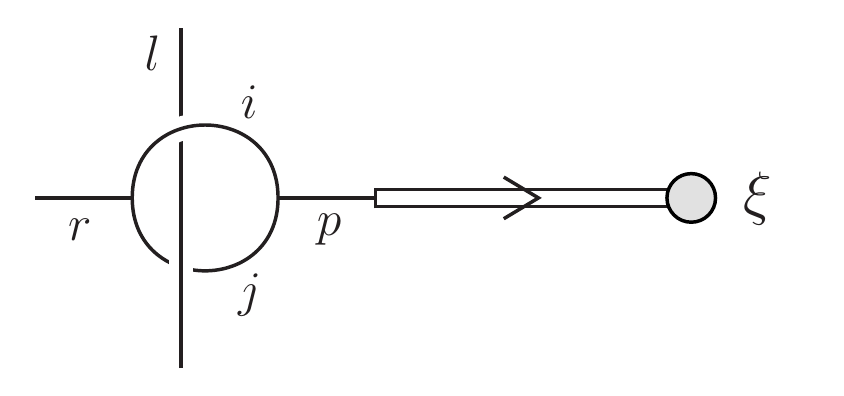}\nn\\
&=\sum_{pq}\Omega^{rp}_{ql,\xi}\pic{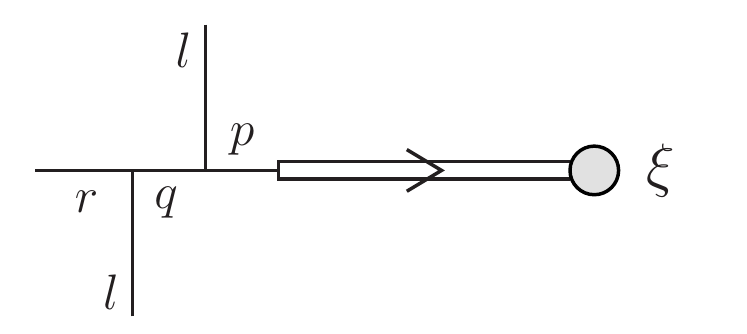}\nn\\
&\eqqcolon\pic{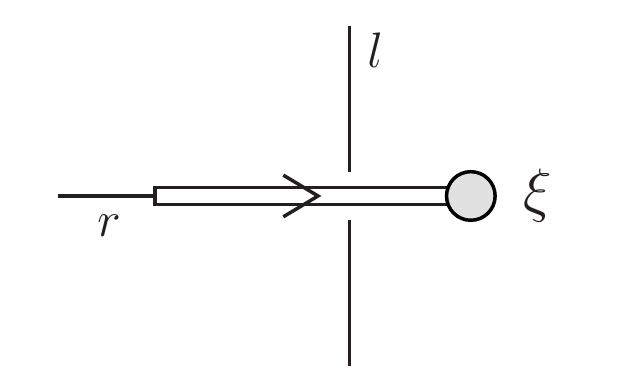}\hs,
\ee
where the tensor $\Omega^{rp}_{ql,\xi}\coloneqq\Omega^{rp}_{ql,ij}$ is given in \eqref{Omega definition}. This should be compared with the standard resolution \eqref{resolution of crossing} of a crossing between two strands. Here, instead of being resolved by a braiding $R$-matrix, the crossing is resolved by the tensor $\Omega$. Mathematically speaking, the map $\omega:V_l\otimes V_\xi\rightarrow V_\xi\otimes V_l$ is known as half-braiding, and is essential for the definition of the Drinfeld centre category\footnote{In the case of a modular fusion category $\C$ one can also use the term (categorical) Drinfeld double, since in this case the Drinfeld centre is isomorphic to the direct product of $\C$ and $\C^*$, where $\C^*$ is the fusion category $\C$ with opposite braiding (which here is the complex conjugated braiding tensor) \cite{Mueger1,Mueger2}.}. 

This half-braiding has to satisfy a so-called naturality condition. Graphically, this condition amounts to being allowed to pull a double strand over a node, and is therefore given by
\be\label{half-braiding sliding}
\pic{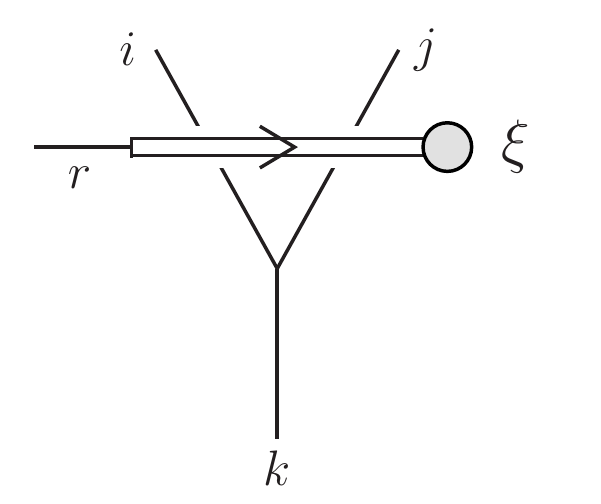}=\pic{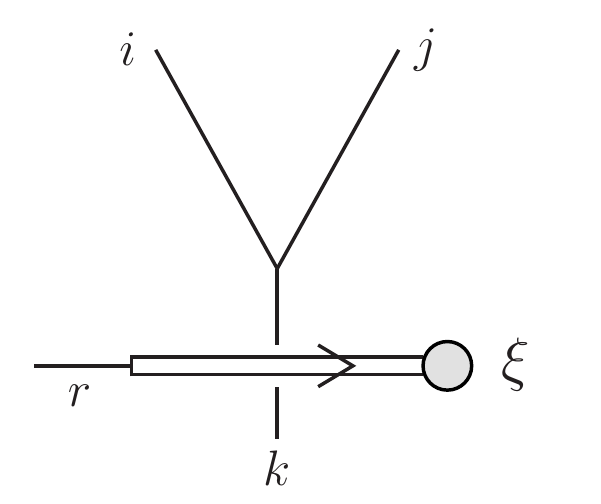}.
\ee
One can recognize that this graphical relation is simply the generalization to the half-braiding (or to the double strands) of the (Yang--Baxter) relation \eqref{strand sliding}. This equation \eqref{strand sliding} holds as a consequence of the definition and of the axioms for the $F$-symbols and the $R$-matrix. The condition \eqref{half-braiding sliding} represents the same kind of naturality condition as \eqref{strand sliding}, but now for the half-braiding tensor. In terms of components, this condition is given  by
\be\label{sscomp}
\sum_{iks}\Omega^{rs}_{ij}\Omega^{su}_{kl}v_s\f{v_mv_n}{v_rv_u}F^{jsi}_{lmk}F^{mbr}_{jil}F^{mnu}_{lkj}=\delta_{bn}\Omega^{ru}_{mn},
\ee
where we have dropped the label $\xi$ for clarity. Since we have defined the half-braiding tensors $\Omega$ via the strand sliding property \eqref{pulling through Oijrs} for the $\O$ basis states, which uses only the local equivalence relations (and in particular \eqref{strand sliding} which follows from these local equivalence relations), the naturality condition for the half-braiding is automatically satisfied\footnote{Indeed, the graphical definition of the half-braiding tensor in \eqref{pulling through half-braiding} shows that the crossing of a ribbon labelled by $i\bar{j}$ with strands $a,b,\dots$ can be resolved by replacing the crossing part of the ribbon with a strand $i$ which over-crosses the strands $a,b,\dots$, and a parallel strand $j$ which under-crosses the strands $a,b,\dots$. Therefore, the naturality condition for the half-braiding reduces to the naturality condition for the $R$ matrix.}.

Notice, that the relation \eqref{sscomp} is exactly the one we would have obtained in \eqref{z components definition} by looking at the projection condition $Z\times Z=Z$ for the multiplication rule \eqref{tube algebra multiplication} of the tube algebra. Therefore, the solutions to the projection condition are given by the half-braiding tensors $\Omega$ used in \eqref{OtoQ} and \eqref{Omega definition}.

In summary, when defining the half-braiding tensor via the sliding property \eqref{pulling through Oijrs}, we do satisfy the naturality condition \eqref{sscomp}. In order for this to hold, it is essential that the $\O$ states are stable under the operation of passing through an edge. From the naturality condition follows also the projection condition for the tube algebra, i.e. the stacking property \eqref{O times O} for the $\O$ states.

\subsection{Cylinder ribbon operator}

\noindent With all these ingredients, we are now able to give a preliminary definition of the ribbon excitation operator and to study its action on the vacuum of the cylinder. As a generalization of \eqref{half-braiding definition}, let us define the following oriented ribbons, which start at a quasi-particle of type $\bar{\xi}=\bar{i}j$ and end at a quasi-particle of type $\xi=i\bar{j}$:
\be\label{ribbon operator}
\pic{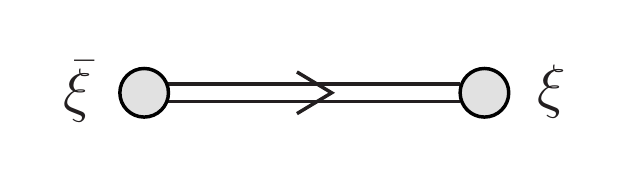}\hs
&\coloneqq\sum_r\pic{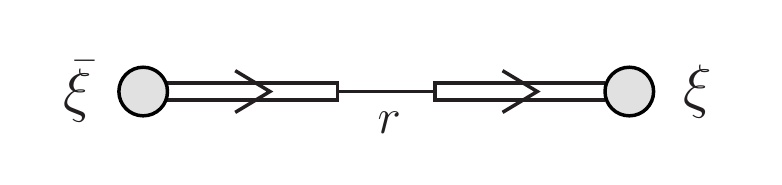}\nn\\
&=\sum_r\f{v_r}{v_iv_j}\pic{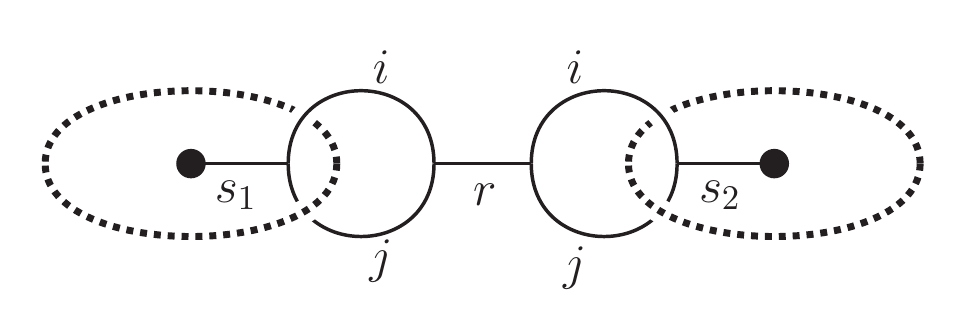},
\ee
where, as an important and useful convention, we have absorbed the factor $v_r/(v_iv_j)$ into the graphical representation for the source quasi-particle as a filled disk labelled by $\bar{\xi}$ (this does however not apply to the target quasi-particle of the ribbon, which does not posses any weight in the graphical representation, and is therefore just represented by \eqref{half-braiding definition}). In what follows we will denote these ribbons simply by $\R_{\bar{\xi}\xi}$.

The important thing to notice about this definition is that, when seen as a state on the cylinder, this ribbon operator reduces to
\be\label{elementary ribbon operator on cylinder}
\R_{\bar{\xi}\xi}\stackrel{\mathbb{S}_2}{=}\D\pic{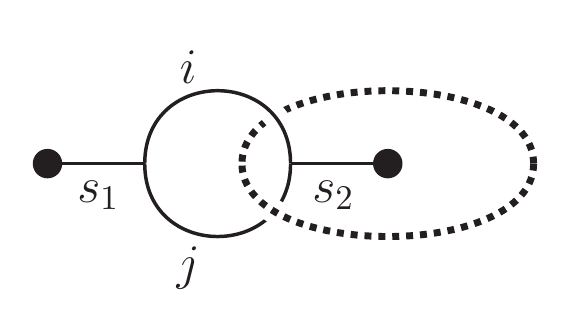}=\D\O^{ij}_{s_1s_2}.
\ee
This can easily be seen by using \eqref{F split}, then wrapping the left vacuum loop around the sphere so as to encircle the puncture on the right, and finally using the stacking property \eqref{stacking vacuum loops}.

The ribbon operator \eqref{ribbon operator} carries conjugated quasi-particles excitations at its ends. Intuitively, the action of this ribbon operator on a pair of disjoint punctures should therefore amount to placing these quasi-particle excitations at the punctures, i.e. to exciting the punctures. Since these punctures on which we wish to act need not be in the vacuum state, but can already carry their own curvature or torsion excitations, the operation of placing the ribbon quasi-particle excitations at the punctures should in fact be understood as a fusion of excitations. If the punctures are in the vacuum state, this fusion is trivial. This gives us an intuitive idea of how the ribbon operator should be used. One has to start with a ribbon operator where the quasi-particles $\bar{\xi}$ and $\xi$ are side-by-side in a same neighborhood of $\Sigma_p$ (i.e. not separated by strands). Then, one can move the ends of the ribbon operator to the punctures where we want the quasi-particle excitations to act. By doing so, the ribbon operator has to cross over several strands, but, importantly, the exact path of the ribbon does not matter because of the naturality condition and the sliding property \eqref{half-braiding sliding}. Then, one has to evaluate these crossings using the half-braiding tensors, and finally fuse the quasi-particles at both ends with the states at the punctures where they are acting. We are going to illustrate this procedure and in particular define the fusion process in the next section.

By analogy with \eqref{pulling through half-braiding}, we can write the behavior of the ribbon operator \eqref{ribbon operator} when it crosses a strand. This defines the braiding of a ribbon operator with a strand as 
\be\label{ribbon braiding strand}
\pic{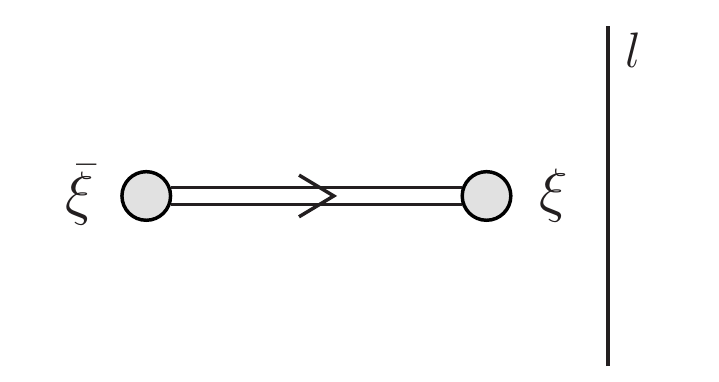}\hs
&=\sum_{pr}\f{v_p}{v_iv_j}\pic{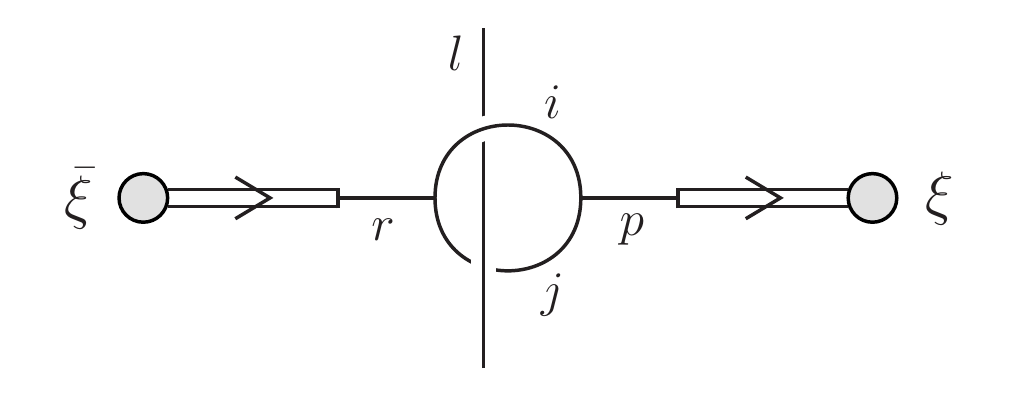}\nn\\
&=\sum_{pqr}\Omega^{rp}_{ql,\xi}\pic{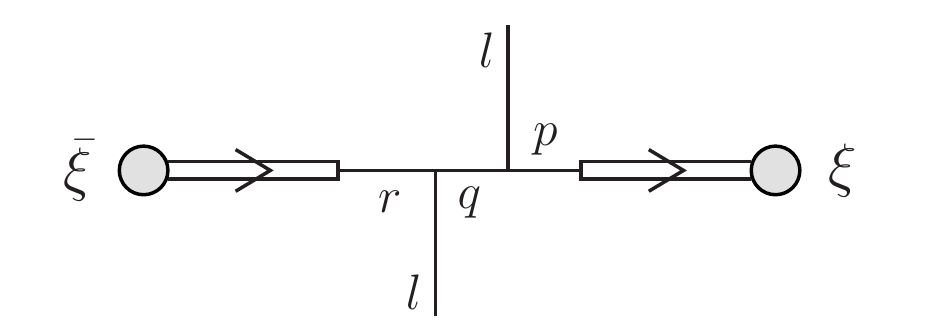}\nn\\
&\eqqcolon\pic{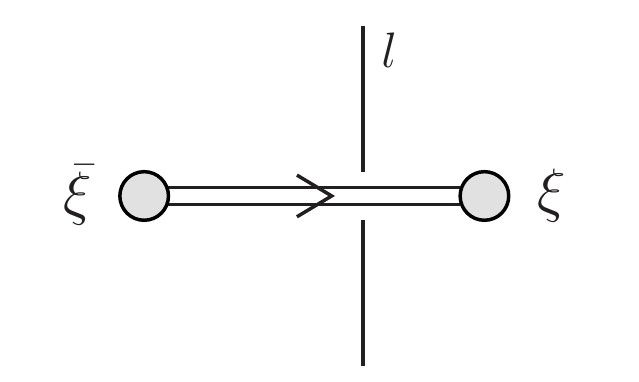}\hs.
\ee
These are the crossing identities which have to be used when acting with a ribbon operator on punctures which are ``far away'', i.e. separated by strands. Again, notice that the path along which we choose to connect the two punctures with the ribbon operator does not matter because of the naturality condition \eqref{half-braiding sliding} for the half-braiding tensor.

Notice that the half-braiding tensor for the ribbon $i\bar{j}$ crossing a strand $l$ can be resolved into a strand $i$ over-crossing $l$ and a parallel strand $j$ under-crossing $l$. This generalizes if the ribbon crosses several strands $l_1,\dots,l_n$, and all the half-braiding tensors can be resolved by drawing a strand $i$ over-crossing the strands $l_1,\dots,l_n$ and a parallel strand $j$ under-crossing these strands $l_1,\dots,l_n$ (see \eqref{resolution of half-braiding}).

Without knowing the details of the fusion process (which we will describe in the next section) or having to resolve any crossing, we can already guess the action of the ribbon operator on the cylinder vacuum state. For this, consider $\R_{00}$, which is the ribbon operator with vanishing quasi-particle excitations. This should correspond to the unit element of the Drinfeld center fusion algebra, and therefore under the fusion product satisfy\footnote{We do not worry for the moment about whether the ribbons act from the left or from the right, and will come back to this point in the next section.} $\R_{\bar{\xi}\xi}\triangleright\R_{00}=\R_{00}\triangleleft\R_{\bar{\xi}\xi}=\R_{\bar{\xi}\xi}$. Now, by noticing that $\R_{00}$ is proportional to the vacuum on the cylinder, i.e.
\be
\R_{00}=\D\pic{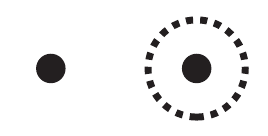}=\D\O^{00}_{00},
\ee
we see that the natural requirement $\R_{\bar{\xi}\xi}\triangleright\R_{00}=\R_{\bar{\xi}\xi}$ leads to $\R_{\bar{\xi}\xi}\triangleright\O^{00}_{00}=\O^{ij}_{s_1s_2}$. This simple argument shows that the definition \eqref{ribbon operator} of the ribbon operator has the desired property, namely that it creates an orthonormal basis state by acting on the vacuum. In the next section we are going to make this statement more precise by defining the action of the ribbon operators via the fusion of quasi-particle excitations.

\section{The three-punctured sphere}
\label{tpsphere}

\noindent We have so far focused on the cylinder in order to understand the nature of the quasi-particle excitations and the construction of the ribbon operators. Moreover, we have been able to derive the action of a ribbon operator on the simplest state which is the vacuum for the cylinder. However, we eventually wish to define the action of ribbon operators on general states and not only on the vacuum. This means in particular that we have to consider arbitrary states on the sphere with more than two punctures. Intuitively, if we imagine two nearby punctures carrying arbitrary excitations, we would like to replace these two punctures by a single puncture in a new fused state. For this, we should use the local equivalence relations to define a new boundary going around the two punctures and traversed by only one strand. In addition, we should require that the fused state behaves in the same way as the original state on the two punctures, as long as we probe the region outside of the new boundary. This applies in particular to the behavior under the operation of crossing a strand. Topologically, one way of replacing two punctures by a single puncture is equivalent to gluing a three-punctured sphere and filling-in the resulting hole. An ``inverse'' procedure for replacing two punctures by a single puncture is to cut the manifold $\Sigma_p$ in a region going around the two punctures. This has the effect of removing a three-punctured sphere and of leaving the manifold $\Sigma_p$ with a new puncture in place of the two old ones. This reasoning shows that it is sufficient to focus on the three-punctured sphere in order to understand the fusion process.

One possible choice of basis for the graph Hilbert space on the three-punctured sphere is given by the generalization \eqref{p-punctured basis} of the $\Q$ basis. However, just like for the $\Q$ basis itself, the states \eqref{p-punctured basis} do not transform with the half-braiding tensor when they cross a strand, and therefore do not posses a stability property. For this reason, one should consider the generalization of the $\O$ basis to a higher number of punctures. A possible choice of orthonormal basis would be to consider a tree-like basis formed by $\O$ states. But this would also not transform nicely when the quasi-particle ends cross a strand, since it would lead to two half-braiding tensors (in the three-punctured case). Therefore, following \cite{KKR}, one should consider the so-called fusion basis given by
\be\label{sphere fusion basis choice}
\bO^{a\,b\,\vec{\imath}\,\vec{\jmath}}_{r\,\vec{s}}\coloneqq\pic{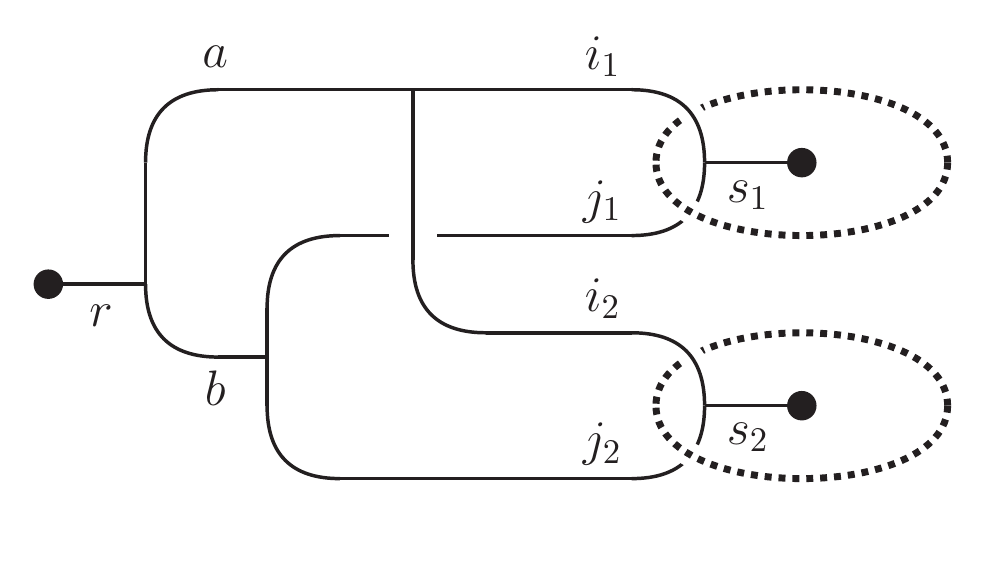}.
\ee
This can of course easily be generalized to the case of $p\geq3$ punctures. As shown in the trace inner product calculation \eqref{proof of orthonormality of p-punctured O}, these states are indeed orthonormal. Using a calculation similar to \eqref{pulling through Oijrs}, it is also straightforward to determine the behavior of these basis states when they cross a strand. This is given by
\be
\pic{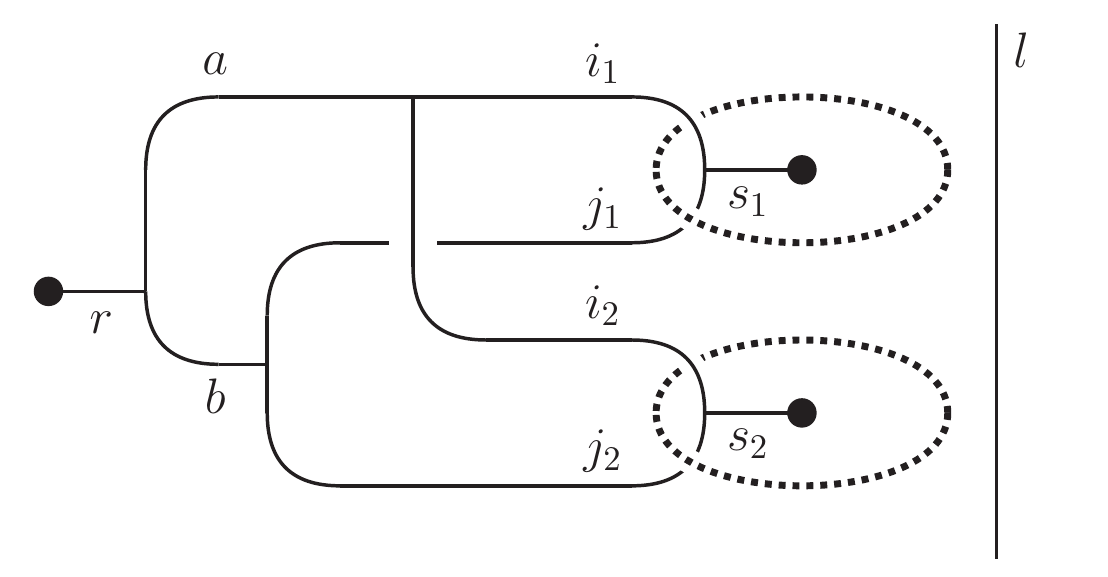}=\sum_{pq}\Omega^{rp}_{ql,ab}\pic{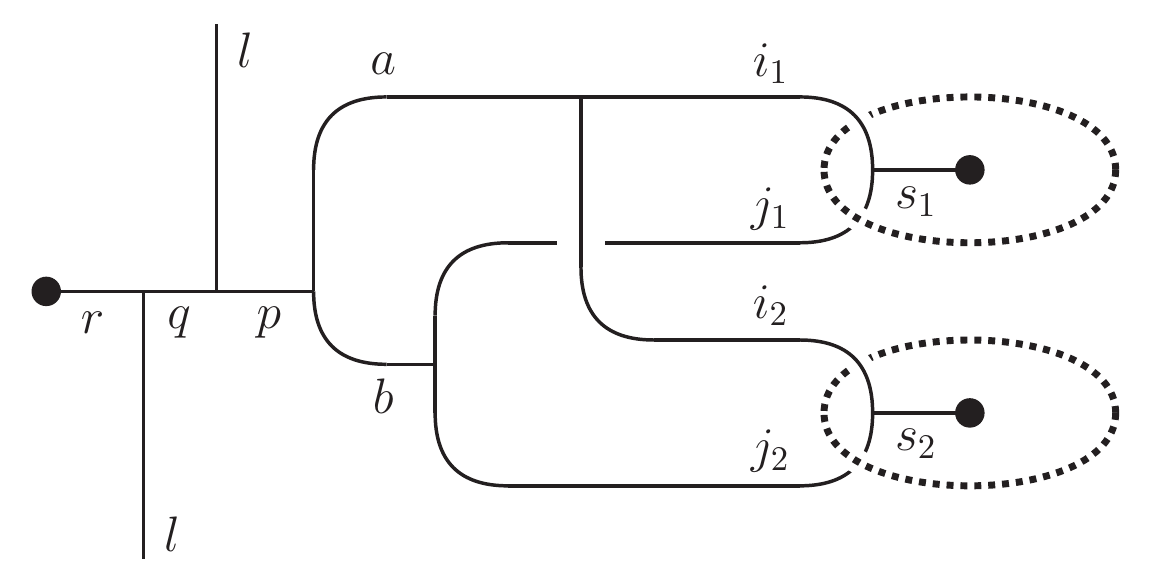},
\ee
which indicates that, from the point of view of the half-braiding, the two quasi-particle ends on the right behave like one fused quasi-particle. This justifies the name ``fusion basis'' and the use of these states to define the fusion of two puncture quasi-particles into one. This is what we are now going to do explicitly.

As discussed above, we would like to describe the fusion process as the cutting of a three-punctured sphere away from the manifold, or equivalently as the gluing of a three-punctured sphere followed by a filling of the resulting hole. To make this concrete, let us consider a state with two neighboring punctures carrying excitations. In order to describe the fusion as the removal of a three-punctured sphere by cutting around these two punctures, we should first locally rewrite the state in the $\mathbb{S}_3$ fusion basis $\bO$. The choice \eqref{sphere fusion basis choice} is however not unique, and there exist different fusion basis given by arbitrary orientation choices for the crossings. For our purposes, it will be convenient to use the following rewriting formula to a particular fusion basis:
\be\label{fused O states 1}
\pic{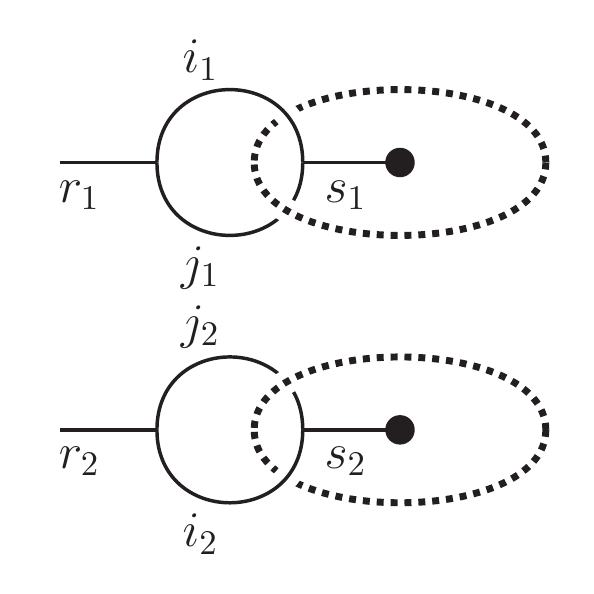}
&=\sum_{abmt}\f{v_a}{v_{i_1}v_{i_2}}\big(R^{i_2j_2}_{r_2}\big)^*R^{i_2j_1}_mF^{ar_1m}_{j_1i_2i_1}F^{j_1mi_2}_{r_2j_2b}F^{abt}_{r_2r_1m}\nn\\
&\q\q\pic{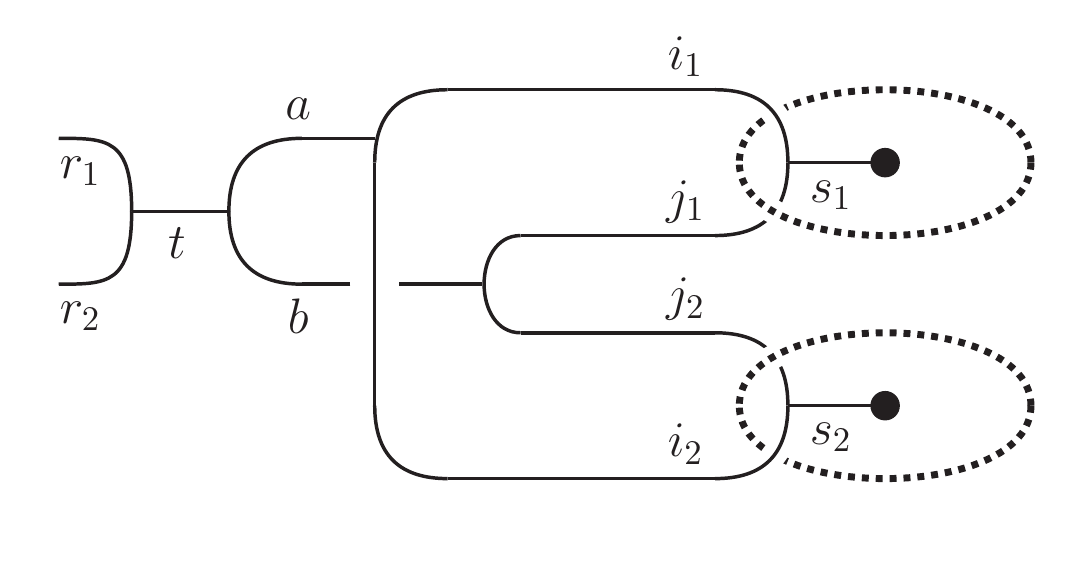}.
\ee
Other possible choices are given in appendix \ref{appendix:other fusion basis}, and will lead to different Clebsch--Gordan coefficients for the fusion. Since we are only interested in what happens in the neighborhood of the two punctures, we have here deliberately left the strands $r_1$ and $r_2$ open in order to signify that there could be an arbitrary complicated graph continuing in this direction. Now, using formula \eqref{vacuum F-move} from right to left and sliding the vacuum loop over the graph, we can further rewrite the state \eqref{fused O states 1} as
\be\label{O states rewritten}
\pic{TwoOStatesMirrorLower}=\sum_{abmt}\f{v_a}{v_{i_1}v_{i_2}}\big(R^{i_2j_2}_{r_2}\big)^*R^{i_2j_1}_mF^{ar_1m}_{j_1i_2i_1}F^{j_1mi_2}_{r_2j_2b}F^{abt}_{r_2r_1m}\times\G,
\ee
where
\be\label{fused O states 2}
\G=\f{1}{\D}\f{v_av_b}{v_u}\pic{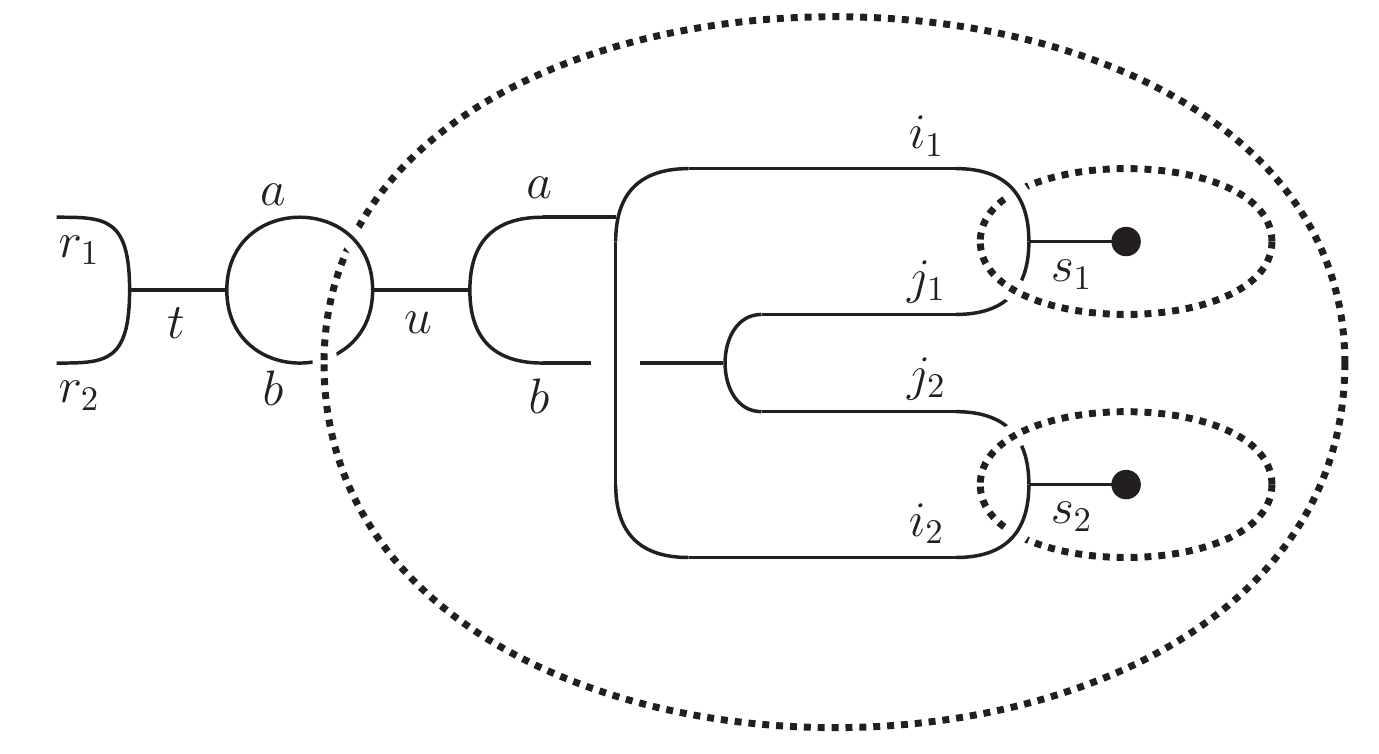}.
\ee
For the moment, we have simply used the local graphical equivalence relations to obtain a strict equality between the left- and right-hand sides of \eqref{O states rewritten}. With this rewriting, we are now ready to define the fusion operation.

First of all, one can verify with the trace inner product that $\G$ is a state of unit norm. It is therefore meaningful to try to replace it by another state of unit norm, namely
\be\label{fused O states 3}
\G\rightarrow\pic{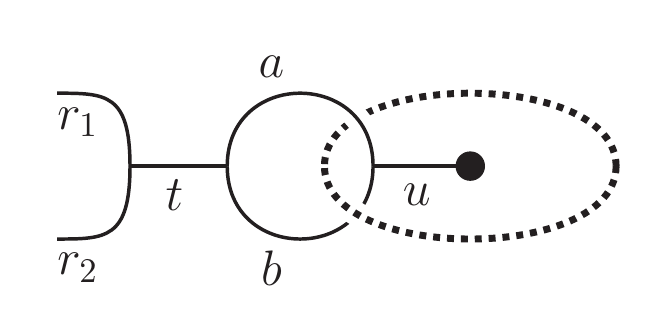},
\ee
which can be seen as being obtained by cutting away a three-punctured sphere in \eqref{fused O states 2} at the level of the strand $u$. Furthermore, this fused state has the same behavior (given by the half-braiding tensor) as $\G$ when crossing a strand. However, notice that the label $u$ becomes free and unspecified in \eqref{fused O states 3}, while it does initially not appear in \eqref{fused O states 2} by virtue of \eqref{vacuum F-move}. Therefore, the cutting of the three-punctured sphere at the level of the strand $u$ should be done in all possible channels which are compatible with the strands $s_1$ and $s_2$ arriving at the two punctures being fused. We thus wish to sum over the label $u$ and to insert a fusion coefficient $\delta_{s_1s_2u}$, i.e. to consider
\be
\sum_u\delta_{s_1s_2u}\G\rightarrow\sum_u\delta_{s_1s_2u}\pic{TwoOStatesFusion3}.
\ee
However, if we want to define the fusion by cutting away and replacing several states (because of the sum over $u$), an additional natural requirement is that the total norm of the states which are removed should be one. Because of the fusion rule, we have that
\be
\sum_u\f{v_u^2}{v_{s_1}^2v_{s_2}^2}\delta_{s_1s_2u}=1,
\ee
which shows that we should rather replace
\be
\sum_u\f{v_u}{v_{s_1}v_{s_2}}\delta_{s_1s_2u}\G\rightarrow\sum_u\delta_{s_1s_2u}\pic{TwoOStatesFusion3}.
\ee
This argument motivates the definition of the following fusion operation:
\be\label{fusion rule}
\pic{TwoOStatesMirrorLower}\stackrel{\otimes}{=}\sum_{abtu}\C[a,b,t,u|i_1,i_2,j_1,j_2,r_1,r_2,s_1,s_2]\pic{TwoOStatesFusion3},
\ee
where
\be\label{fusion coefficient}
\C\coloneqq\sum_m\f{v_{s_1}v_{s_2}v_a}{v_{i_1}v_{i_2}v_u}\big(R^{i_2j_2}_{r_2}\big)^*R^{i_2j_1}_mF^{ar_1m}_{j_1i_2i_1}F^{j_1mi_2}_{r_2j_2b}F^{abt}_{r_2r_1m}\delta_{i_1j_1s_1}\delta_{i_2j_2s_2}\delta_{abu}\delta_{s_1s_2u},
\ee
and where $\stackrel{\otimes}{=}$ denotes the fact that the equality comes from the fusion of the two puncture states. In this algebraic expression we have made explicit the coupling rules which would otherwise only be seen graphically on \eqref{fused O states 2}. As we are going to see in what follows, this definition of the fusion operation leads to a consistent gluing of the ribbons.

Before going on, notice that \eqref{fusion rule} is sufficient to describe all possible fusion processes. Indeed, even if one of the quasi-particle ends being fused has a different orientation, one can use the transformation
\be\label{O upside down transformation}
\pic{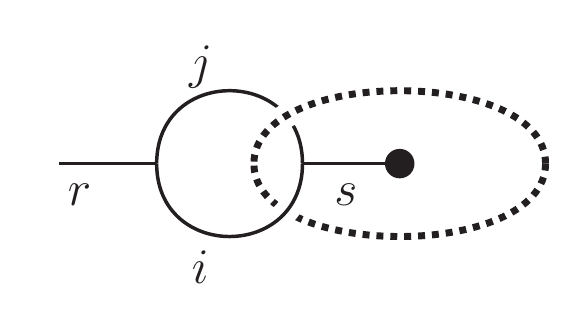}=\pic{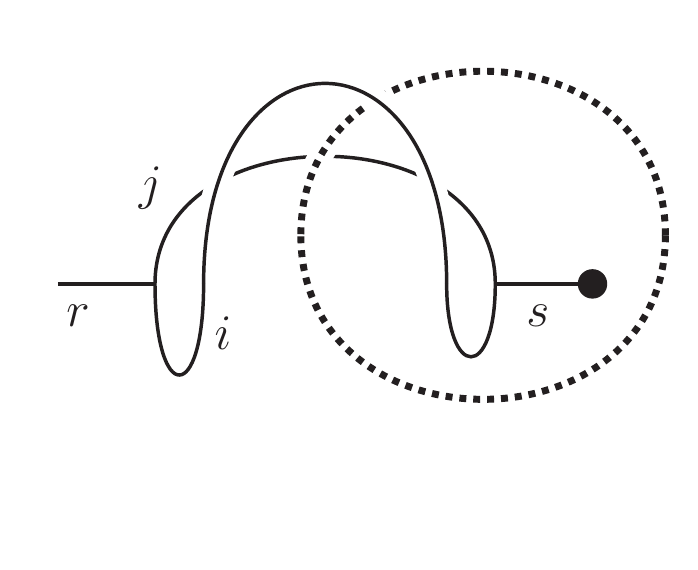}=R^{ij}_s\big(R^{ij}_r\big)^*\pic{OijrsOnePunctureOnly}
\ee
or its inverse to go back to the left-hand side of \eqref{fusion rule}. We are now going to give examples of this fusion rule by computing the action of ribbon operators.

\section{Ribbon operators}
\label{sec:creation-operators}

\noindent This section is devoted to the precise definition and study of the action of the ribbon operators \eqref{ribbon operator} on general states. First, we will define the action of the open ribbon operators. Then, we will compute this action on vacuum states and show that it does create excitations, similarly to the action of exponentiated fluxes of the BF representation. Here we will have in addition torsion excitations generated by open Wilson line operators. Next, we will define the gluing of open ribbon operators, and use this to show that the action of closed ribbons generalizes that of Wilson loops. Finally, we will comment on the interplay between curvature and torsion excitations. This section does contain the main result of our work, namely the explanation of why  holonomy and exponentiated flux operators are represented  by closed and open ribbon operators.

\subsection{Definition of (left) ribbon operators}

\noindent We want to use the fusion operation introduced above in order to define the action of a ribbon operator as the fusion of its quasi-particle ends with the excitations of the punctures it is acting on.

First, a choice of convention has to be addressed. The fusion which we are considering here is the fusion in the Drinfeld centre category, and the order of the factors plays a very subtle role \cite{Kir}. We therefore have to decide whether the left or right factor is given by the ribbon's quasi-particle ends or the excitations of the punctures. First, note that since a ribbon $\R_{\bar{\xi}\xi}$ has an orientation, it is meaningful to refer to its source and target quasi-particle ends $\bar{\xi}$ and $\xi$. Referring to this orientation, i.e. with the ribbon pointing upwards, we will define the action of a ribbon from the left as follows: In the fusion process of the target quasi-particle end $\xi$ with a puncture excitation $\chi$, we will use $\xi\otimes\chi$ with the ribbon target $\xi$ providing the left factor. For the fusion process of the source quasi-particle $\bar{\xi}$ with a puncture excitation $\gamma$, we have to rotate our viewpoint by $180^\circ$, and consider an ordering $\gamma\otimes\bar{\xi}$. In this way, the ribbon with its ends is situated on the left of an auxiliary strand connecting the two punctures on which the ribbon will act.

The precise definition of the action of a left ribbon operator is as follows:
\begin{itemize}
\item If one wants to use a ribbon operator to create an excitation at a location in $\Sigma_p$ where there is initially no puncture, one first has to embed the state under consideration into the Hilbert space $\H_{\Sigma_{p+1}}$ using \eqref{refinement-map}, i.e. to add a puncture in the vacuum configuration at the desired position.

\item 
Use the representation \eqref{ribbon operator} for the ribbon as a state around two auxiliary punctures. Place the two auxiliary punctures of the ribbon near the intended source puncture in $\Sigma_p$. If the marked point of the source puncture with excitation $\gamma$ in $\Sigma_p$ is pointing \textit{upwards}, the source puncture of the ribbon should be placed to the left side of the marked point such that the marked point of the source puncture also points upwards. The state $\gamma$ around the source puncture of $\Sigma_p$ and the source puncture $\bar{\xi}$ of the ribbon should be brought into the form depicted on the left-hand side of \eqref{fusion rule}, with the upper puncture there representing the excitation $\gamma$ of $\Sigma_p$ and the lower puncture the (source) ribbon puncture $\bar{\xi}$.

\item One moves the target puncture of the ribbon along the intended ribbon path to the intended target puncture in $\Sigma_p$. For each strand being crossed over, one applies the sliding rule \eqref{pulling through half-braiding} which involves the half-braiding tensor. We can graphically resolve the half-braiding tensors using \eqref{ribbon braiding strand}: the ribbon $i\bar{j}$ crossing strands $a,b,c,\ldots$ can be replaced by a double strand labelled by $i$ and $j$, where the strand $i$ \textit{over-crosses} the strands $a,b,c,\ldots$, and the strand $j$ \textit{under-crosses} the same strands. Graphically, this is given by
\be\label{resolution of half-braiding}
&\pic{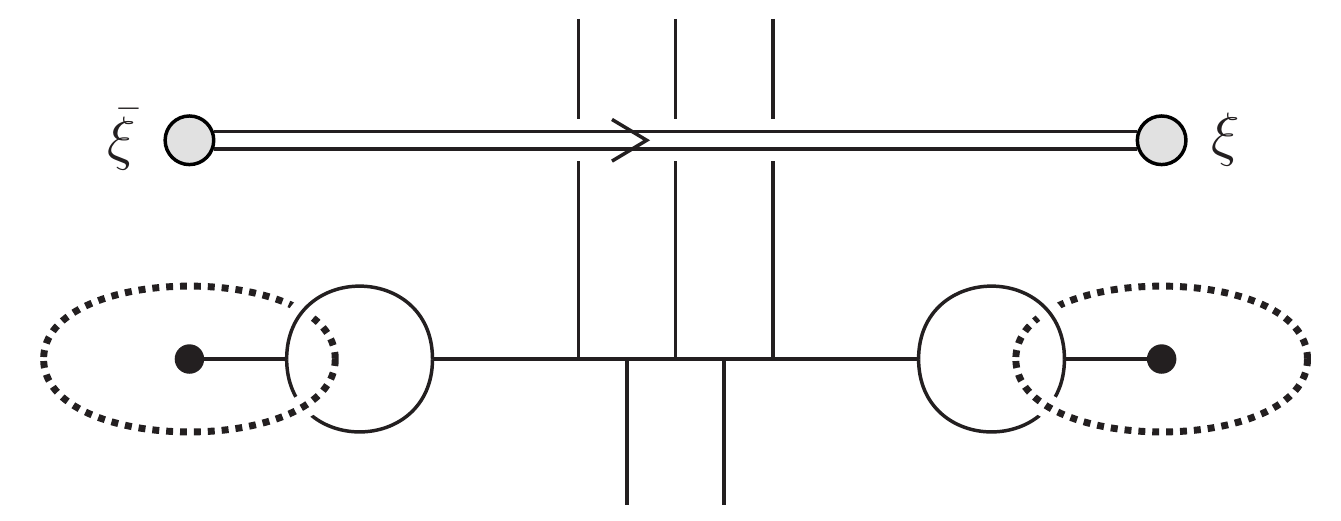}\nn\\
~\nn\\
=\sum_{pr}\f{v_r}{v_iv_j}\f{v_p}{v_iv_j}&\pic{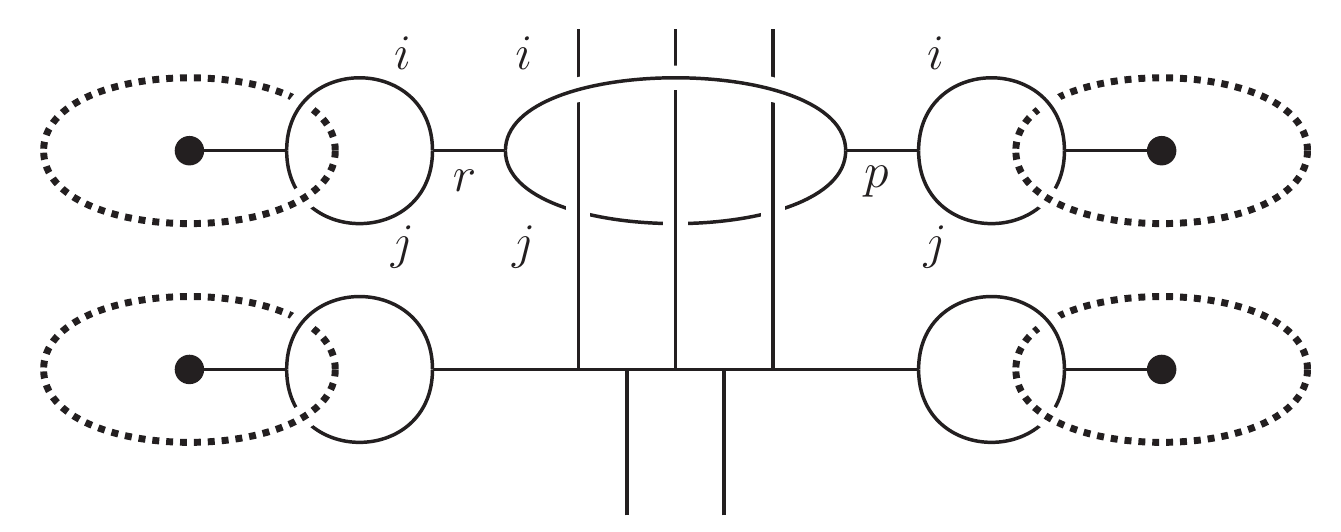}.
\ee

\item For the fusion of the target puncture $\xi$ of the ribbon and the target puncture $\chi$ in $\Sigma_p$, we move the ribbon target puncture to the left (with respect to an upward orientation of the ribbon) of the puncture $\chi$ in $\Sigma_p$. The marked points of both punctures should now be pointing \textit{downwards}. Again, we then bring the state around the target puncture in $\Sigma_p$ and the target puncture of the ribbon into the form depicted on the left-hand side of \eqref{fusion rule}. This time, the upper puncture in \eqref{fusion rule} is representing the target puncture of the ribbon, and the lower puncture is representing the target puncture in $\Sigma_p$.

\item Finally, one applies the replacement rule \eqref{fusion rule} to fuse the ribbon's quasi-particle ends with the punctures they are meant to act on. One then obtains new fused punctures whose location agree with that of the initial punctures in $\Sigma_p$ which have been acted on.
\end{itemize}

Notice that for a given $\H_{\Sigma_p}$ the action of an open ribbon operator depends on three different data, namely:
\begin{itemize}
\item The quasi-particle excitations $\bar{\xi}$ and $\xi$ carried by the ribbon;
\item The quasi-particle excitations $\gamma$ and $\chi$ carried by the punctures with which $\bar{\xi}$ and $\xi$ are fusing;
\item An equivalence class of paths connecting the two quasi-particles $\gamma$ and $\chi$ and the state in a neighborhood of this path. The state-dependent equivalence class of paths is defined by the sliding property \eqref{half-braiding sliding} for the ribbons, which allows to deform the ribbon's path as long as it does not cross a non-vacuum puncture.
\end{itemize}

\subsection{Open ribbon operators}

\noindent The first interesting case to consider is the action of a ribbon operator on the cylinder vacuum state. We have already argued above that this should reproduce an $\O$ basis state, but this result can now be proven using the fusion rule. Graphically, the action of a ribbon on the cylinder vacuum state is given by
\be
\R_{\bar{\xi}\xi}\triangleright\O^{00}_{00}=\f{1}{\D}\pic{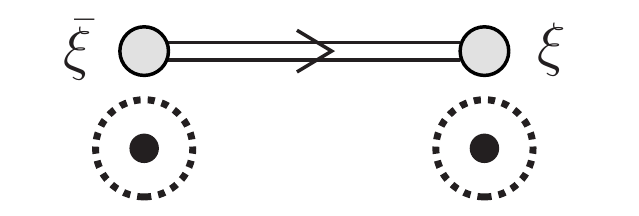}\hs=\f{1}{\D}\sum_r\f{v_r}{v_iv_j}\pic{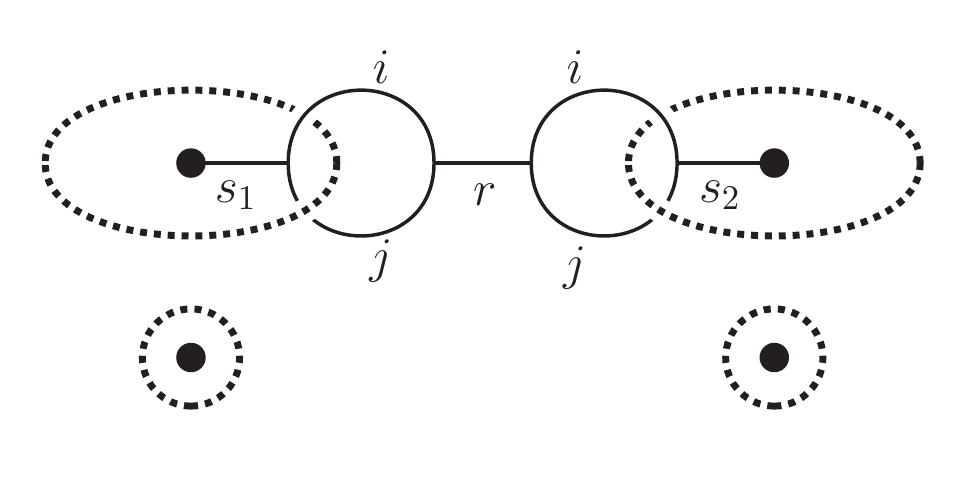},
\ee
where we should use the fusion rule \eqref{fusion rule} to fuse $\xi$ from above with the right vacuum puncture, and then \eqref{fusion rule} rotated by $180^\circ$ to fuse $\bar{\xi}$ from below with the left vacuum puncture. It is straightforward to compute that, for the fusion of $\xi$ for example, we have
\be
\pic{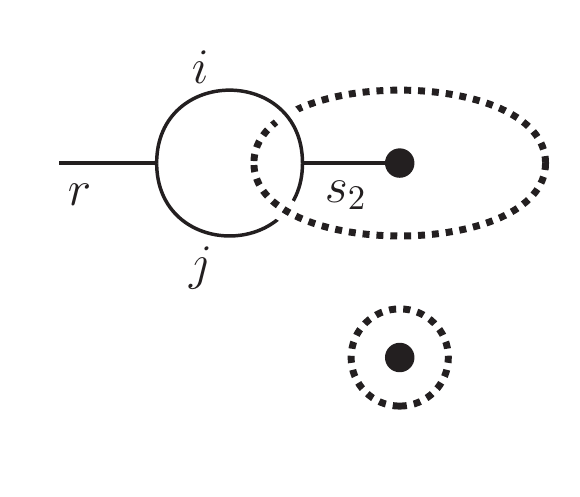}\stackrel{\otimes}{=}\pic{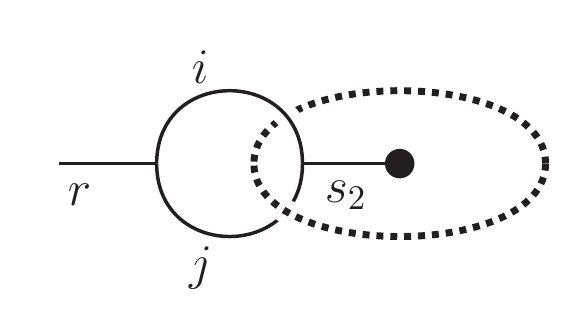}.
\ee
After fusing $\bar{\xi}$, one gets that the final result of the fusion process is
\be
\R_{\bar{\xi}\xi}\triangleright\O^{00}_{00}=\f{1}{\D}\R_{\bar{\xi}\xi}=\O^{ij}_{s_1s_2}.
\ee
This shows that an open ribbon operator does indeed act on the cylinder vacuum state by producing an orthonormal basis state.

This simple result is the counterpart in the TV representation of the action of the exponentiated flux operators in the BF representation. Note that here the open ribbons also include the action of open Wilson lines. These open Wilson lines were not allowed in the $\SU(2)$ BF representation as they lead to (non-normalizable) torsion excitations. The open ribbons are the basic excitation (or creation) operators of the theory. Now, just like the simplicial fluxes in the BF representation can be composed to yield ``longer'' fluxes along so-called co-paths (i.e. paths in the triangulation itself) \cite{DG2,BDG}, open ribbons can also be glued with an operation which we now define.

\subsection{Gluing of ribbons}

\noindent Open ribbon operators can be glued along punctures, which allows us to represent a ``longer'' ribbon as the gluing of ``shorter'' ribbons. Since a ribbon operator leads to excitations only at its end, we have to project the glued ends of the shorter ribbons back to the vacuum state. In fact, the considerations we make here will show the cylindrical consistency of the ribbon operators in the sense explained in the last item of section \ref{BFrep}. The gluing can be represented graphically as
\be\label{graphical gluing}
\pic{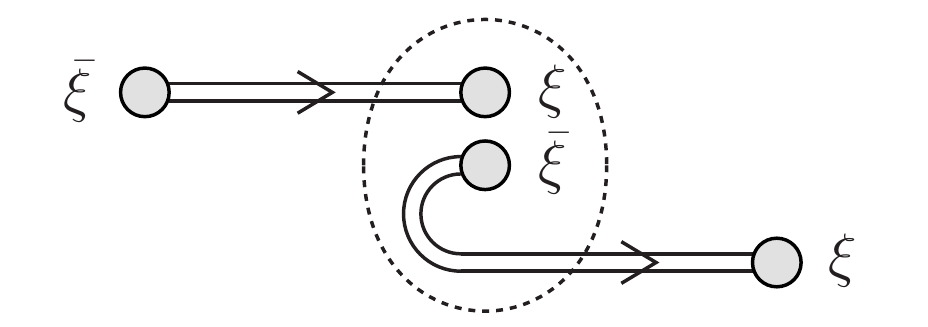}=\pic{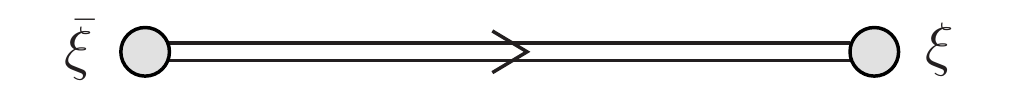},
\ee
and here we are going to describe the operation depicted by the dashed line around the quasi-particle ends $\xi$ and $\bar{\xi}$.

For this, consider the situation in which we first fuse a vacuum puncture with the target quasi-particle $\xi=i\bar{j}$ of a ribbon operator, and then fuse the resulting excited puncture with the source quasi-particle of a second ribbon operator labelled $\bar{\xi}=\bar{i}j$. Looking at the two quasi-particles $\xi$ and $\bar{\xi}$ being fused on the left-hand side of \eqref{graphical gluing}, and following our convention which requires to place the marked points on the same side, we are led to considering the fusion
\be\label{glue1}
\pic{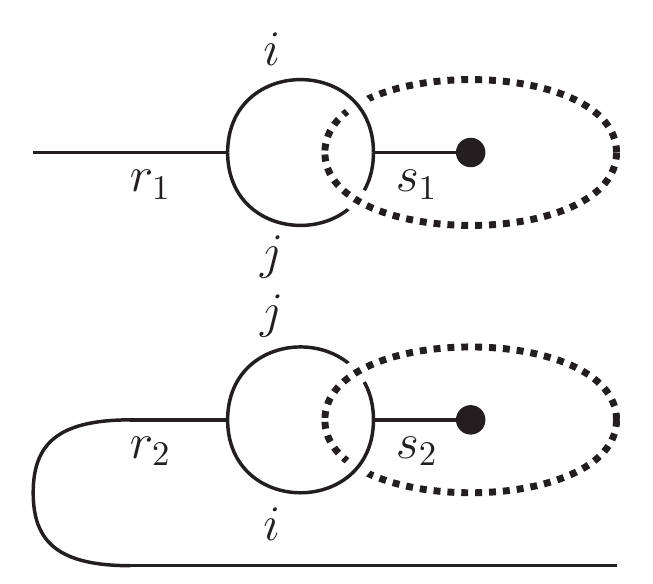}\stackrel{\otimes}{=}\sum_{abtu}\C[a,b,t,u|i,i,j,j,r_1,r_2,s_1,s_2]\pic{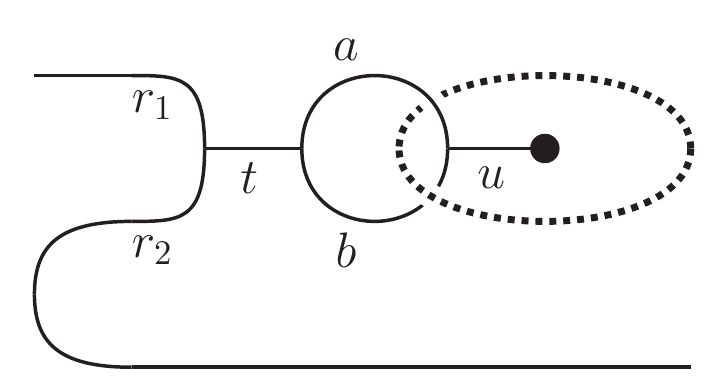}.
\ee
The gluing operation is now defined on this resulting fused puncture by the following additional steps:
\begin{itemize}
\item One first projects the state around the fused puncture onto the Gauss and flatness constraints using the projections $B_n$ and $B_p$ respectively. This amounts to projecting the puncture onto the vacuum state, and therefore the resulting state looks locally like a long ribbon which passes by this vacuum puncture.
\item One then contracts the two ``ribbon tail'' labels $s_1$ and $s_2$ by summing over $s_1=s_2$.
\end{itemize}
The projection onto the Gauss constraint of the right-hand side of \eqref{glue1} forces $u=0$, and therefore $a=b$ and $s_1=s_2$ as well. Graphically, it leads to
\be
\pic{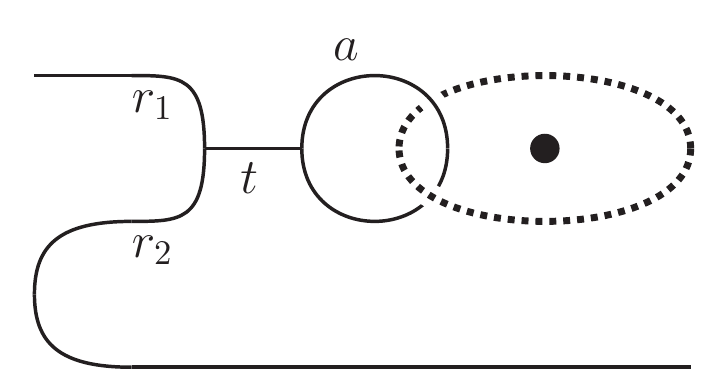}.
\ee
Projecting onto the flatness constraint with \eqref{flatness-projector} then leads to
\be
\f{1}{\D}\pic{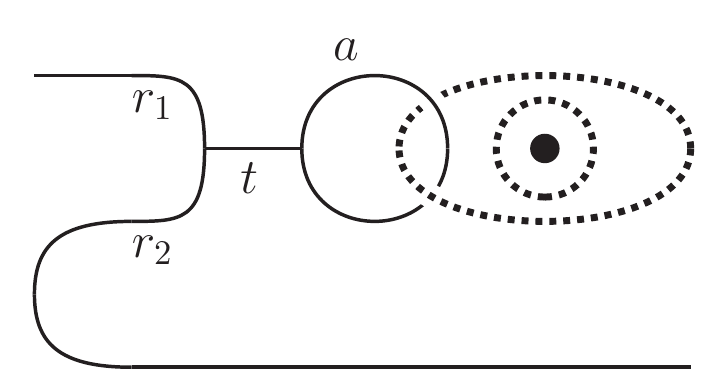}=\f{1}{\D}\pic{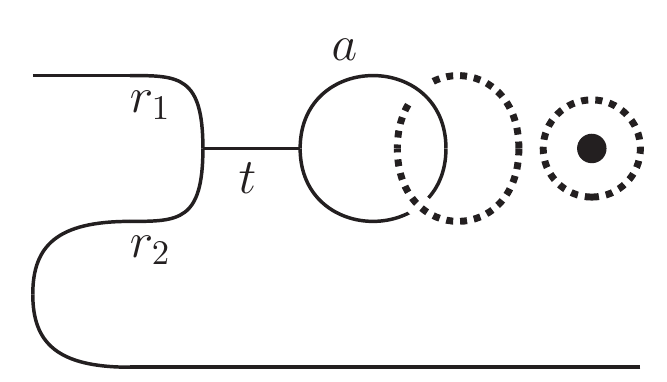}=\delta_{a0}\delta_{t0}\delta_{r_1r_2}\pic{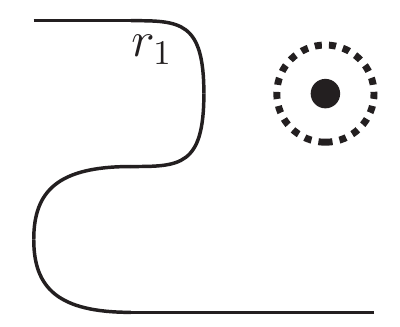},
\ee
and therefore one ends up with a vacuum puncture.

Now, one should remember that the quasi-particles being glued on the right-hand side of \eqref{glue1} come from the ends of the two ribbons in \eqref{graphical gluing}. Since the ribbon operators are defined in \eqref{ribbon operator} with a sum over dimension factors, these have to be taken into account here in order to get the final result. It will be convenient to attach these dimension factors to the source quasi-particles of the ribbons. In the present case, we therefore only write the factor coming from the end $\bar{\xi}$ which is below on the left-hand side of \eqref{glue1}. Also, one should implement the sum over $s_1=s_2$. Putting this together leads to the coefficient
\be
\sum_{s_1s_2}\sum_{r_1r_2}\f{v_{r_2}}{v_iv_j}\sum_{abtu}\C[a,b,t,u|i,i,j,j,r_1,r_2,s_1,s_2]\,\,\delta_{u0}\delta_{ab}\delta_{a0}\delta_{t0}\delta_{r_1r_2},
\ee
which, using
\be
\C[0,0,0,0|i,i,j,j,r_1,r_1,s_1,s_1]=\f{v_{s_1}^2}{v_iv_jv_{r_1}}\delta_{ijr_1}\delta_{ijs_1},
\ee
leads around the puncture to the final glued state
\be\label{glue3}
\sum_{r_1}\sum_{s_1}\f{v_{s_1}^2}{v^2_iv^2_j}\delta_{ijr_1}\delta_{ijs_1}\pic{TwoOppositeOStatesFusionCurvature3}=\sum_{r_1}\delta_{ijr_1}\pic{TwoOppositeOStatesFusionCurvature3}.
\ee
Remembering now that the source quasi-particle attached at the left end of the strand $r_1$ carries the factor $v_{r_1}/(v_iv_j)$, one therefore sees that the gluing of two open ribbons along a vacuum puncture is equivalent to having a long open ribbon passing by the vacuum puncture. This is what is represented in \eqref{graphical gluing}.

The fact that the gluing of two ribbons along a vacuum puncture gives a longer ribbons confirms our definitions for the fusion process which underlies the action of the ribbons. It also allows us to represent a long ribbon, which is a priori defined on some coarse Hilbert space $\H_p$, as a combination of shorter ribbons that are connecting additional vacuum punctures for states in a refined Hilbert space $\H_{p+q}$ resulting from an embedding of states in $\H_p$. This notion is important in order to define a cylindrical consistent family of (ribbon) operators which agrees on states that are connected by the embedding maps. In this way, the action of a (long) ribbon operator coming from some ``coarser'' Hilbert space $\H_p$ is uniquely defined for the set of states in a ``finer'' Hilbert space $\H_{p+q}$ which results from the embedding maps. Note that the ribbon operators can be extended in various ways to the full  ``finer'' Hilbert space $\H_{p+q}$. For example, in the computation above we can choose to represent the refined ribbon as the glued ribbon which passes by the puncture to the left or to the right. This feature of having different possible extensions (without an obvious canonical choice) also appears in the BF representation and is discussed in detail in \cite{DG2,BDG}.

\subsection{Closed ribbon operators}
\label{sec:cribbon}

\noindent The gluing of open ribbon operators now allows us to define closed ribbons. For this we simply have to glue, following the prescription of the previous subsection, the source and target quasi-particles of the same ribbon. Using the result \eqref{glue3}, this can be done by closing the strand labelled $r_1$ (which we can relabel $r$). If the ribbon (and therefore the strand $r$) does not enclose a (non-vacuum) puncture, the loop can be evaluated to $v_r^2$. Therefore, a closed ribbon which does not enclose an excited puncture is given by
\be
\R_\xi\coloneqq\pic{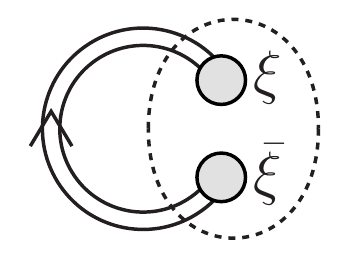}=\pic{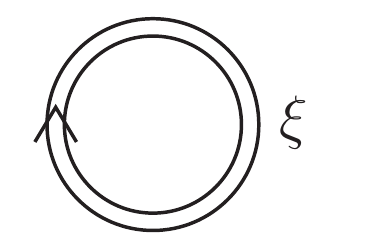}\hs=\sum_rv_r^2\delta_{ijr}=v_i^2 v_j^2\coloneqq\sum_{r\in\,\xi}v_r^2,
\ee
where the dotted line around the two quasi-particle ends indicates that they should be fused and projected with the operators $B_p$ and $B_n$, and where we have then omitted the puncture in a vacuum loop produced by the fusion. This equation is a generalization to ribbons (and therefore to the objects of the Drinfeld centre) of the evaluation \eqref{Dimension} of a closed strand.

Let us now consider the more general case of a closed ribbon operator acting on an excited state $\O$ basis state. Using the half-braiding \eqref{ribbon braiding strand} to resolve the crossing between a ribbon and a strand, we get that
\be\label{cribbon1}
\pic{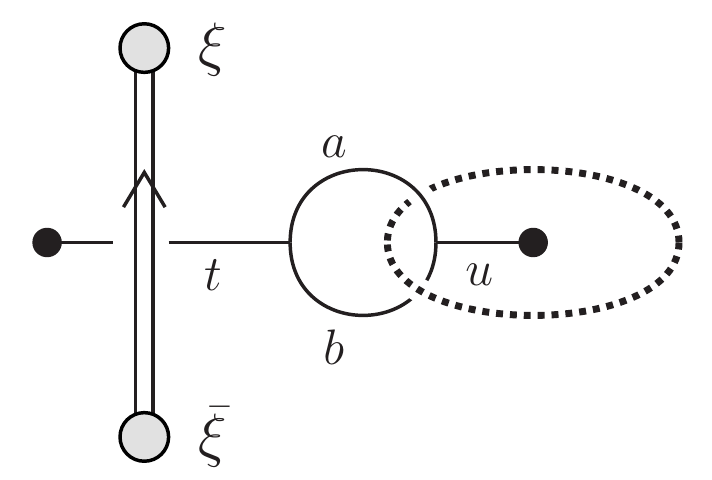}=\sum_{pr}\f{v_p}{v_iv_j}\pic{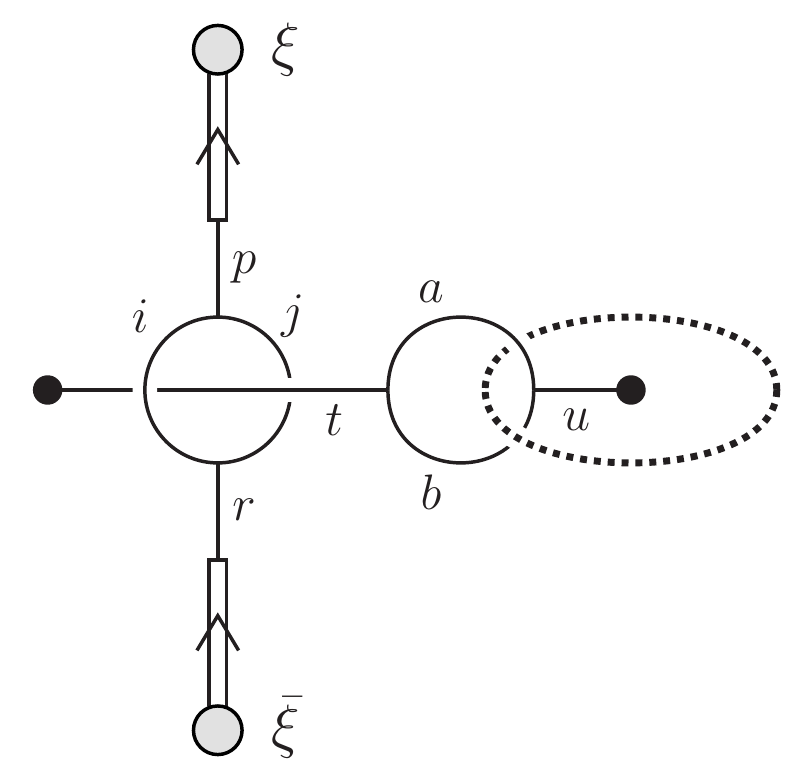}.
\ee
Now, the ribbon quasi-particles $\xi$ and $\bar{\xi}$ can be fused on the right of the puncture, leading to
\be\label{closed ribbon on O state}
\pic{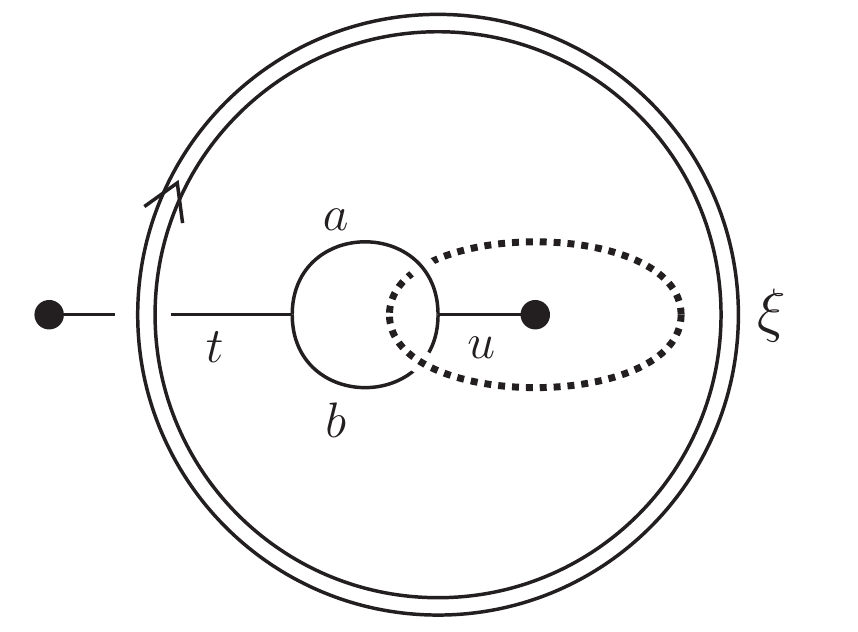}
&=\sum_p\f{v_p}{v_iv_j}\pic{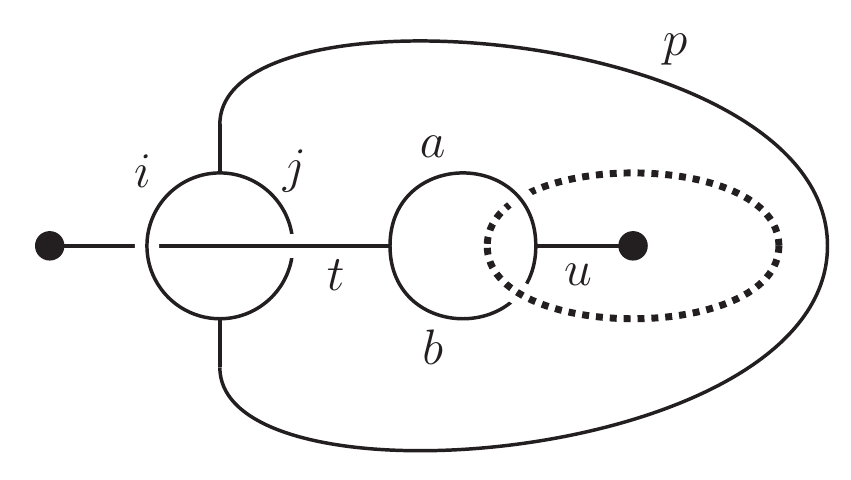}\nn\\
&=\sum_{pr}\f{v_p}{v_iv_j}\f{v_r}{v_av_b}\pic{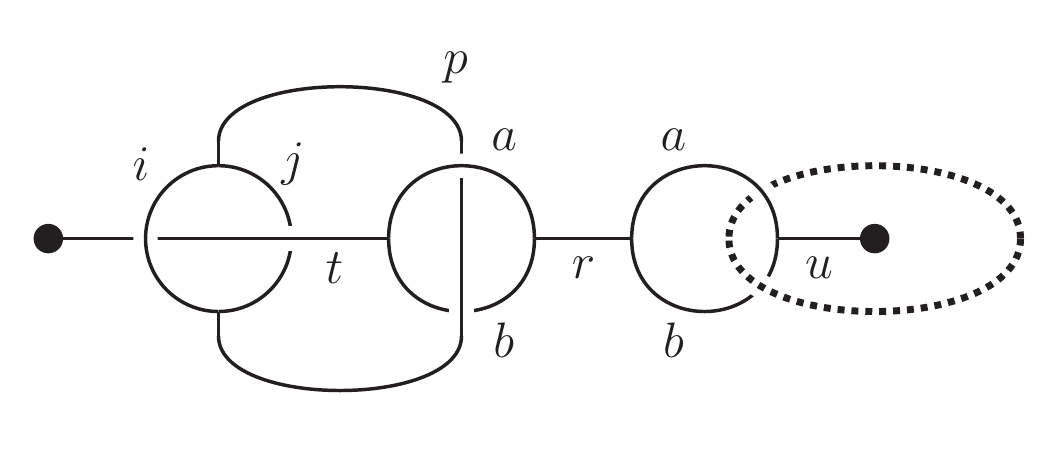}\nn\\
&=\sum_p\f{v_p}{v_iv_j}\f{1}{v_av_bv_t}\pic{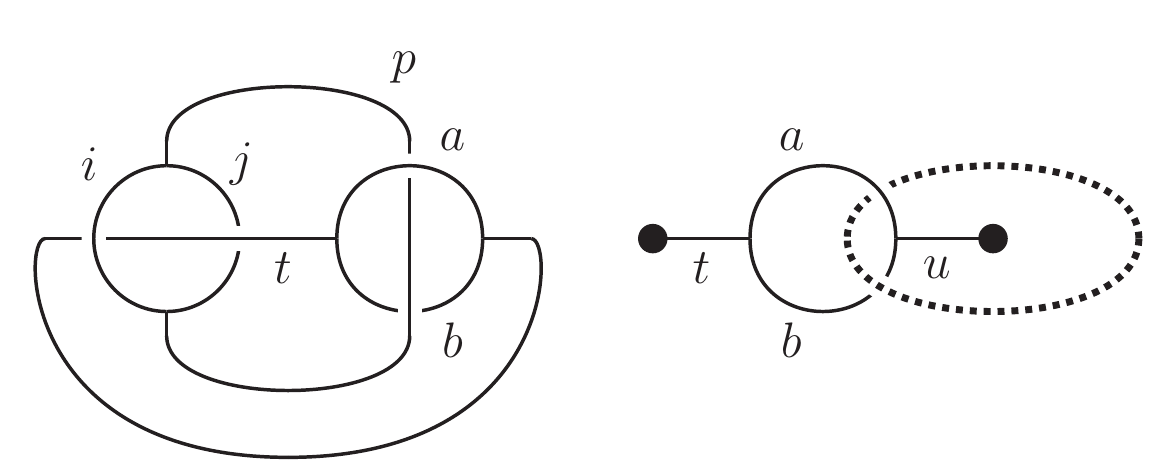}.
\ee
In the last step of this calculation, we have used an argument identical to that below \eqref{bubble move} in order to detach from the tail of the $\O$ basis state the graphical evaluation which appears as a pre-factor.

We have therefore shown that the $\O$ basis states \eqref{O basis state} are eigenstates of closed ribbon operators $\R_{i\bar{j}}$, with eigenvalue given by
\be\label{eigenvalue of closed ribbon}
\lambda_{ij,ab}\coloneqq\sum_p\f{v_p}{v_iv_j}\f{1}{v_av_bv_t}\pic{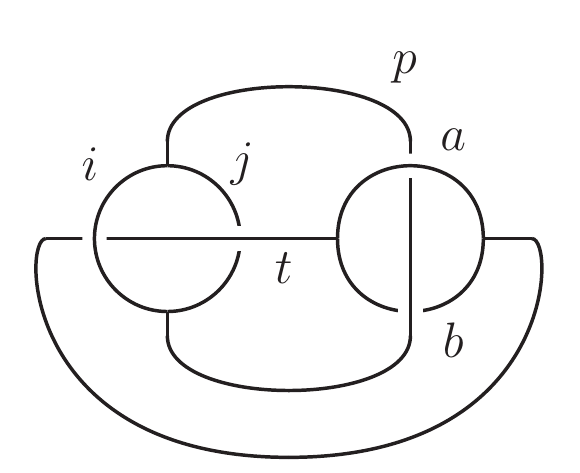}=\f{1}{v_a^2v_b^2}s_{ia}s_{jb}.
\ee
A proof of this graphical evaluation is given in \eqref{proof of eigenvalue of closed ribbon}. This result does also hold for the fusion basis states \eqref{sphere fusion basis choice} which are a generalization of the $\O$ basis states.

Now, one can also consider the braiding of ribbon operators. If a ribbon $a\bar{b}$ is crossing over a ribbon $i\bar{j}$, it means that we first apply the ribbon operator $i\bar{j}$ and then the ribbon operator $a\bar{b}$. By using the gluing of ribbons along vacuum punctures introduced above, we can have different parts of a ribbon acting in different orderings. This allows us to consider two intertwined closed ribbons, and, as shown in \eqref{proof of double S matrix}, we have that
\be\label{double S matrix}
\textbf{s}_{(i\bar{j})(a\bar{b})}\coloneqq\pic{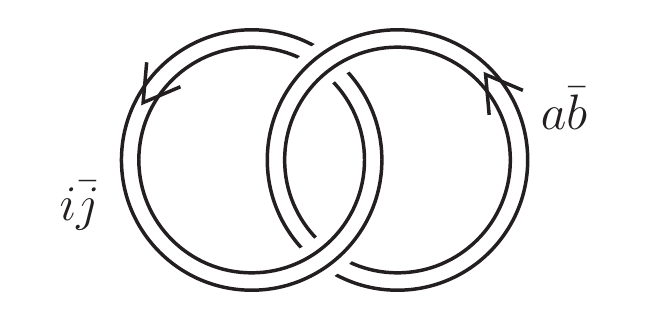}=s_{ia}s_{jb}.
\ee
We have thus defined the $S$-matrix of the Drinfeld centre $\textbf{S}_{(i\bar{j})(a\bar{b})}=\D^2 \textbf{s}_{(i\bar{j})(a\bar{b})}$ and shown that it factorizes as $\textbf{S}_{(i\bar{j})(a\bar{b})}=S_{ia}S_{jb}$ (this holds more generally for modular fusion categories).

As a remark, note that we can use the $S$-matrix of the Drinfeld center in order to define the new closed ribbon operators
\be
\widetilde{\R}_{a\bar{b}}\coloneqq v_a^2v_b^2\sum_{ij}\textbf{s}_{(i\bar{j})(a\bar{b})}^*\R_{i\bar{j}}=\f{1}{\D^4}v_a^2v_b^2\sum_{ij}s_{ia}s_{jb}\R_{i\bar{j}},
\ee
where the second equality comes from the fact that the $S$-matrix of $\SU(2)_\k$ is real and symmetric. With this definition, we obtain that
\be
\widetilde{\R}_{a\bar{b}}\triangleright\O^{kl}_{cd}=\delta_{ac}\delta_{bd}\O^{kl}_{cd}
\ee
and
\be
\widetilde{\R}_{a\bar{b}}\triangleright\widetilde{\R}_{c\bar{d}}=\delta_{ac}\delta_{bd}\widetilde{\R}_{a\bar{b}}.
\ee
These new operators are the so-called projective ribbon operators, whose properties were investigated (although in the group representation and for finite groups) in \cite{ABC}.

As a further remark, note that in the calculation of the eigenvalues of the closed ribbon operators and of the $S$-matrix we could have also used the half-braiding tensor $\Omega$ in order to resolve the crossings between a ribbon and a strand. By comparing the result of this alternative calculation (which is given at the end of appendix \ref{appendix2}) with that obtained above, we find the following contraction identity for the half-braiding tensors:
\be\label{cribbonEV}
\lambda_{ij,ab}=\sum_{pq}v_p^2\Omega^{tt}_{qp,ab}\Omega^{pp}_{qt,ij}=\f{1}{v_a^2v_b^2}s_{ia}s_{jb}.
\ee

Finally, let us point out that in a context where one is interested in the physical Hilbert space for $(2+1)$ dimensional gravity with a cosmological constant, the closed ribbon operators represent the Dirac observables of the theory. The physical Hilbert space contains only the states satisfying all the flatness and Gauss constraints, and it is only non-trivial for surfaces of higher genus or of non-trivial topology because of punctures (for which the flatness and Gauss constraints then do not need to hold). Dirac observables are operators which leave invariant the subspace of states satisfying the constraints.

\subsection{Wilson loop and line operators}

\noindent We have already encountered the Wilson loop operator in section \ref{pure curv}, where it has been applied to a pure curvature state. Since these pure curvature states do not have an outgoing strand, we did not have to decide whether to place the Wilson loop by crossing over or under the strand.

The previous discussions show that the ribbon $\R_{\bar{\xi}\xi}=\R_{(\bar{i}0)(i0)}$ implements an over-crossing Wilson line in the representation $i$. Likewise, with $i=0$ one finds that a ribbon $\R_{(0j)(0\bar{j})}$ agrees with the action of an under-crossing Wilson line in the representation $j$. A ribbon operator $\R_{(\bar{i}0)(i0)}$ acting on two vacuum punctures generates a torsion excitation given by a strand $i$ over-crossing the vacuum loops and all the other strands in-between the two punctures, i.e.  
\be
\pic{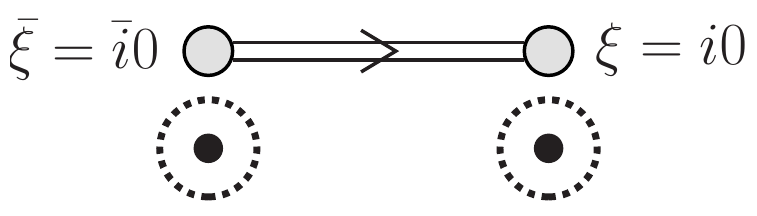}=\pic{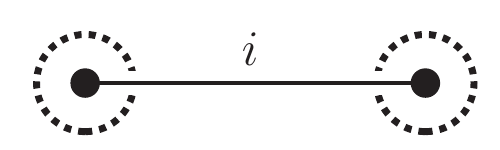}.
\ee
Similarly, the ribbon $\R_{(0j)(0\bar{j})}$ connects the two punctures with a strand $j$ under-crossing the vacuum loops and all the other strands in-between the two punctures. Therefore, the ribbon operators $\R_{(\bar{i}0)(i0)}$ and $\R_{(0j)(0\bar{j})}$ generalize the Wilson line operators by adding the information about whether they cross pre-existing strands from above or from below.

As an example, we can consider the action of a closed ribbon $\R_\xi=\R_{i0}$ on a cylinder basis state of the type $\O^{a0}_{aa}$. According to the general result \eqref{cribbonEV}, the $\O^{a0}_{aa}$ state is an eigenstate with eigenvalue
\be
\lambda_{i0,a0}=\f{1}{v_a^2}s_{ai}.
\ee
This can also be checked with a direct calculation by applying an over-crossing Wilson loop to $\O^{a0}_{aa}$ and using the sliding property \eqref{vacuum sliding} for the vacuum line. Therefore, we see that the $\O^{a0}_{aa}$ basis states have non-trivial curvature when measured with respect to over-crossing Wilson loops. On the other hand, a closed ribbon $\R_{0\bar{j}}$ acting on a cylinder basis state $\O^{a0}_{aa}$ results in an eigenvalue
\be
\lambda_{0j,a0}=v_j^2,
\ee
which again can also be checked by a direct calculation. Hence, the states $\O^{a0}_{aa}$ have vanishing curvature when measured with respect to under-crossing Wilson loops (since with the normalization of \eqref{Wilson-operator} the eigenvalue is one).

\subsection{ Flux operators}

\noindent In the $\SU(2)$ BF representation, the exponentiated flux operators act in the group representation by right or left multiplication, and generate curvature excitations. We have already seen in section \ref{pure curv} that the states $\O^{jj}_{00}$ carry (only) curvature, and that the spin label $j$ measures the amount of curvature (since in the limit of large $\k$ the spin $j$ is proportional to the class angle). On the cylinder, such states are generated by ribbons $\R_{\bar{\xi}\xi}$ with $\bar{\xi}=(j,j,0)$ and $\xi=(j,j,0)$, which we can denote simply by $R_j$. The states $\O^{jj}_{00}$ do not violate the Gauss constraints, and thus correspond to (gauge) group-averaged curvature states $\int\de h\delta(h^{-1}gh)$. We can therefore conclude that the ribbon operators $R_j$ act as exponentiated flux operators\footnote{Generic ribbons $\R_{\bar{\xi}\xi}$ do also generate ``generalized'' curvature, which can be measured with over-crossing and under-crossing Wilson loops. However, excitations generated by such open ribbons are projected to the zero state by the Gauss constraint projector $B_n$, which does not happen in the group case for curvature excitations generated by exponentiated fluxes. Therefore, we can uniquely identify the ribbons $R_j$ as exponentiated fluxes.}.

In the group case, the action of the exponentiated flux operator by multiplication from the right or left does lead to a Gauss constraint violation for the source or target node of the link (on which the flux acts) respectively. The exponentiated flux, and with it the Gauss constraint violation, is however usually parallel transported to a special node, called the root. In the quantum group case, we can also consider ribbons $\R_{\bar{\xi}\xi}$ with $\xi=(j,j,0)$ and $\bar{\xi}=(j,j,s)$, i.e. ribbons which have a vanishing tail for the target puncture but a non-vanishing tail $s\neq 0$ for the source puncture. We therefore also have operators which lead to curvature excitations and a violation of the Gauss constraint at the source puncture. One can also transport this Gauss constraint violation towards another (root) puncture. To this end, one has to apply (a linear combination of) ribbon operators corresponding to open Wilson lines, and finally project the source puncture with the Gauss constraint projector $B_n$.

Besides open ribbon operators $R_j$ corresponding to exponentiated fluxes leading to curvature excitations, we can also consider closed ribbon operators $R_j^\text{cl}$. In the $\SU(2)$ BF representation, these correspond to the following operators: First of all, one can add up flux operators associated to a so-called co-path, which is a connected set  of triangulation edges. The fluxes associated to the edges of the co-path are first parallel transported to a common frame, and then added up. The corresponding operator leads to curvature excitations only at the endpoints of the co-path. Now, by choosing a closed co-path, we could conclude that we have obtained the analogue of a closed ribbon operator. However, the closed co-path operator can still lead to a Gauss constraint violation (at the node which defines the common frame) and to a flatness constraint violation at the face where the co-path starts and ends. To obtain a similar operator as in the quantum group case, we do have to apply the projector to the flatness constraints for the face in question and to also apply a gauge group averaging at the node in question. The resulting operators for the group case will be discussed more deeply in \cite{ABC}.

We have seen in \eqref{closed ribbon on O state} that the quasi-particle excitations $a\bar{b}$ are eigenstates of the closed ribbon operators $\R_\xi=\R_{i\bar{j}}$ with eigenvalues $\lambda_{ij,ab}$. Moreover, the eigenvalues for these closed ribbon operators factorize in two values associated to over- and under-crossing Wilson loops respectively (which can be seen by setting respectively $i$ and $j$ equal to zero in \eqref{closed ribbon on O state}). These over- and under-crossing Wilson loops encode both the curvature and the torsion (or spin) content of the excitation. Torsion is connected to a Gauss constraint violation, and is generated by open ribbons $\R_{(\bar{i}0)(i0)}$ and $\R_{(0j)(0\bar{j})}$ corresponding to open over- and under-crossing Wilson lines. Curvature, on the other hand, is generated by open ribbons $R_{\bar{\xi}\xi}$ (which, depending on the ``tails'', can also lead to a Gauss constraint violation).

The open ribbons $\R_{\bar{\xi}\xi}$ can be also obtained from applying $\R_{(\bar{i}0)(i0)}$ and $\R_{(0j)(0\bar{j})}$ one after the other. We therefore see that flux operators (and more generally all ribbon operators) can be obtained from over-crossing and under-crossing Wilson lines.

On the other hand,  via fusion we can obtain an excitation including torsion out of two ``pure curvature'' excitations. This is encoded in the fusion rules for the Drinfeld centre, which are simply given by the double of the fusion rules for the category of representations of $\SU(2)_\k$, i.e.
\be
\big(i,\bar{j}\big)\otimes\big(i',\bar{j'}\big)=
\;&\big(|i-i'|,\overline{|j-j'|}\big)\oplus\big(|i-i'|+1,\overline{|j-j'|}\big)\oplus\cdots\oplus\nn\\
&\big(i+i',\overline{|j-j'|}\big)\oplus\big(|i-i'|,\overline{|j-j'|+1}\big)\oplus\cdots\oplus\big(i+i',\overline{j+j'}\big),
\ee
where one should understand the implicit imposition of an overall set of admissibility conditions for every triple (two ``ingoing'' and one ``outgoing'' or ``fused'') of spins. These fusion rules can be read from the fusion basis \eqref{sphere fusion basis choice}. From this expression, one can see that the fusion of two excitations $i\bar{i}$ and $i\bar{i}$ includes the components $i0$ and $0\bar{i}$. A closed ribbon encircling these two ``pure'' curvature excitations will therefore detect components corresponding to a torsion excitation.

In the $\SU(2)$ BF case, this \textit{curvature-induced torsion} can be explained by the unavoidable parallel transport involved in the definition of the fluxes in the coarser region, and therefore can be measured by fluxes associated to closed co-paths \cite{DG2}. There is an alternative explanation in terms of quasi-particle excitations, which is that two spinless particles can have an overall non-vanishing spin. Curvature-induced torsion is one of the reasons why it is advantageous to allow also for torsion degrees of freedom at the punctures. Indeed, this leads to a stability of the state space under coarse graining, and the fusion basis makes this stability explicit. Such a stability is rather difficult to implement with (gauge-invariant) spin network states (see the discussion in \cite{ABC,EteraCoarseGrain}).

\section{Conclusion and perspectives}
\label{sec:conclusion}

\noindent In the work \cite{DG1,DG2,BDG}, we have shown that the holonomy-flux algebra of LQG admits a representation based on a kinematical vacuum state which is peaked on flat connections. This leads in turn to a quantum theory which is inequivalent to that based on the AL vacuum, and the properties of this new representation have been recalled in \ref{BFrep}. As argued in the introduction, since the flat curvature vacuum and its excitations can be seen respectively as a physical state of BF topological field theory and curvature defects therein, we are naturally led to wonder whether there exists a deeper relationship between quantum representations of LQG and extended TQFTs. In the present work, we have taken a first step towards this understanding by explaining how the extended TV TQFT, when seen as a curvature vacuum together with its excitations, does lead to a representation of the holonomy and flux operators of LQG.

More precisely, we have focused in this work on $(2+1)$-dimensional Euclidean gravity with a positive cosmological constant, which is a topological field theory. When focussing on the description of the \textit{kinematics} of the theory, i.e. when looking for a representation of the kinematical observable algebra\footnote{The kinematical observable algebra does not commute with the constraints defining the dynamics of the theory. But its representation is usually required as a first step to be able to define the constraints of the theory as well as (Dirac) observables commuting with these constraints.} encoded into the holonomy-flux algebra, we also have to consider excitations on top of the topological field theory. These are incorporated in the framework of so-called extended TQFTs. 

By choosing the vacuum state underlying our representation to be given by the TV TQFT, we are forced to work with the category of representations of $\SU(2)_\k$ and the tools of graphical calculus which we have summarized in section \ref{sec:graphical}. We have shown in section \ref{sec:graph hilbert space} that it is possible to define on punctured manifolds vacuum states which are invariant under the action of projection operators implementing the vanishing of curvature (in excess to the homogeneous curvature forced by the cosmological constant) and torsion. These projection operators implement the same constraints as the path integral defined by the TV model. This defines the kinematical vacuum state peaked on gauge-invariant ``flat'' configurations. Then we have explained in section \ref{section:2p} how to characterize the (quasi-particle) excitations on top of the vacuum state. These are described by modules (or representations) of the tube algebra, and are labelled by elements of the Drinfeld center. We have shown in section \ref{tpsphere} how to define the fusion of the quasi-particle excitations, and explained how this can be used to define the action of ribbon operators. Finally, we have studied in details the properties and the action of open and closed ribbon operators in section \ref{sec:creation-operators}. The open ribbons act as  creation operators which are the analogue of (but generalize) the exponentiated flux operators of the $\SU(2)$ BF representation, and which act on the vacuum state by creating, depending on the spin labels or the ribbon's quasi-particle ends, torsion or curvature excitations.

This results provides a new concrete realization (the first one being the construction of the $\SU(2)$ BF vacuum) of the conjectured general result that the kinematical vacua and their excitations can be understood in terms of TQFTs with defects \cite{DittStein}. Moreover, the framework developed here presents several advantages compared to the previous ($\SU(2)$ BF and AL) representations of the holonomy-flux algebra. We list a few of these important features below.
\begin{itemize}
\item First of all, this work establishes the contact between tools of extended TQFTs and the study of representations of quantum gravity and realizations of quantum geometry.  We believe that this mathematical framework will facilitate the understanding of the structure of the vacua in quantum gravity, as well as the investigation of the continuum limit and the phase structure for spin foams and group field theories.
\item Because the excitations on top of the TV vacuum correspond to local curvature and torsion degrees of freedom, this framework enables for a natural description of the coupling of massive spinning point particles to $(2+1)$-dimensional gravity with a positive cosmological constant. In addition, we have seen that the structure of the Drinfeld center labelling these degrees of freedom does appear naturally from simple stability considerations for the quasi-particles. This opens the possibility of understanding in a deeper manner the inclusion of particles in both the canonical and covariant quantizations of three-dimensional gravity with a cosmological constant.
\item A major advantage of the present construction over the $\SU(2)$ BF representation is that it comes with a built-in finiteness (i.e. one does not need to resort to a Bohr compactification like in the group case). The Hilbert spaces $\H_p$ describing excitations located at a fixed number $p$ of punctures are finite-dimensional because of the finiteness of the category of representations of $\SU(2)_\k$. Thus, the spectra of geometrical operators (which can be restricted to a given $\H_p$) are discrete, as opposed to the continuous spectra which appear in the $\SU(2)$ BF case \cite{BDG} for operators built from the fluxes, or in the AL case for operators built from the holonomies.

\item The fusion basis describes states which do posses both curvature and torsion excitations and the punctures and which can be naturally fused together. The state space spanned by these states is therefore stable from the point of view of the creation of curvature-induced torsion, since it encompasses all possible degrees of freedom from the onset. This is a major advantage for the study of coarse graining of such states, as opposed to the study of usual gauge-invariant spin network states. In the case of a gauge theory with classical groups, one can also construct the fusion basis using the holonomy representation \cite{ABC}. This construction provides also an interpretation of the charge labels appearing in the quantum deformed case as encoding mass and spin.
\end{itemize}

Although the present construction was carried out in the restricted context of Euclidean three-dimensional gravity with a positive cosmological constant, the results which we have obtained and the tools which have been used can potentially be generalized to various other cases of interest. We list below some developments which we postpone to further work.
\begin{itemize}
\item The construction and the study of the vacua and excitations should be extended and made systematic for arbitrary spacetime signature, dimension (three or four), and sign of the cosmological constant. There are two main challenges in this task. First of all, the present formalism has to be adapted to the case of a category which is not finite.  This would for example be the case for the Euclidean theory with a negative cosmological constant, i.e. for $\mathcal{U}_q\big(\su(2)\big)$ with $q\in\mathbb{R}$. Note that the $\SU(2)$ BF representation \cite{DG2,BDG} is in this sense not finite and can give some clues about how to deal with this.

When going to the four-dimensional case, we expect that the $\SU(2)$ BF representation constructed in \cite{DG2,BDG} will provide important hints about how to formulate possible quantum deformations. In the four-dimensional Euclidean theory with a positive cosmological constant, one can expect that the Crane--Yetter TQFT \cite{CY,CKY} will provide the structure of the vacuum, while the excitations will be described by an extended Crane--Yetter TQFT, whose precise definition needs still to be completed. 

One promising possibility to obtain a $(3+1)$-dimensional framework from a $(2+1)$-dimensional one is to use a Heegard splitting to encode a three-dimensional triangulation  through a two-dimensional so-called Heegard surface, or more precisely a Heegard diagram \cite{BC2016}. In \cite{BC2016}, this strategy has been applied to the BF representation with classical gauge group. This work shows that the ribbon operators of the $(2+1)$-dimensional theory do map to surface operators which have been constructed in \cite{DG2,BDG}. Therefore, in order to obtain a quantum deformed $(3+1)$-dimensional representation, one should apply the same strategy to the TV representation constructed here.

\item The representation we built for the $(2+1)$-dimensional quantum geometrical operators can be also useful for the four-dimensional theory. One example is the quantization of the horizon geometry of black holes incorporating the ``isolated horizon condition'' (see \cite{sahlmann1,sahlmann2}). Another example is given by the quantization of holographic screens \cite{SmolinTop}.

\item It would be interesting to compare in more detail the relationship between the present approach to three-dimensional quantum gravity and the other schemes such as the Chern--Simons quantization and the combinatorial quantization \`a la Fock--Rosly \cite{fock-rosly}. At the level of the path integral, the relation between the TV model and Chern--Simon theory has been rigorously established in \cite{TV-WRT1,TV-WRT2} (for the relation between three-dimensional quantum gravity without cosmological constant described by the Ponzano--Regge model and Chern--Simons theory, see \cite{Freidel2}). In the context of canonical quantization, the work \cite{MeusburgerNoui} relates the observable algebra in the Fock--Rosly quantization to the holonomy-flux algebra of loop quantum gravity. This has been worked out in the context of the AL representation, but the relationship is even tighter in the BF representation. It remains to generalize these arguments to the case of a non-vanishing cosmological constant.

\item By comparing the framework presented here to the usual spin network basis, one might wonder about the fate of the magnetic indices (attached to a choice of basis in a given representation space $V_j$) which one would expect to have at the open ends of the strands. In fact, here we were following the setup of so called (non-extended) string-nets \cite{LevinWen,Kir}, in which such magnetic indices do not appear.

For the case of proper groups, so-called \textit{extended} string-nets which include magnetic indices have been defined \cite{Bur1,BurKong}. This setup is needed in order to match the extended string-nets with a description based on group variables (also known as Kitaev models \cite{Kitaev1}), as will be further explored in \cite{ABC}. The additional magnetic indices provide local information, i.e. can be manipulated by local operators. This is different from the quasi-particle quantum numbers $i\bar{j}$ which can be changed by ribbon operators that are rather quasi-local. 

In the case of quantum groups, the definition of extended string-nets has not been completed yet, but it has been argued to exist in \cite{BurKong} based on a so-called weak Frobenius fiber functor \cite{Pfeiffer}. This would then also allow for the definition of a generalized Fourier transform to a (generalized) group picture. The reason why the quantum group case (at root of unity) is much more involved is due to the fact that the tensor product of two representations can lead to an indecomposable part which needs to be factored out. This is mirrored in quantum dimensions which do not need to be natural numbers.

\item The ribbon operators constructed in this work replace the holonomies and fluxes which encode the quantum geometry. It remains to construct out of these ribbons the explicit geometrical operators, e.g. an operator giving the length of a geodesic, or an operator giving the dihedral angle (e.g. the extrinsic curvature) attached to an edge in the triangulation, and to study their spectra (for the BF representation, area and length operators have been constructed in \cite{BDG}.

\item Several works attempt at imposing the dynamics of three-dimensional quantum gravity with a cosmological constant starting from a kinematical set-up, which so far is based on the AL representation and does not incorporate a quantum deformation (see \cite{Improved} for a completion of this program at the classical level). It is hoped that this ``exercise'' will give important insights about how to impose the dynamics in the four-dimensional case. It might however be easier (due to the finiteness of the underlying fusion category) to start with the framework presented here and adapted to a cosmological constant $\Lambda$, and to impose the dynamics defined by a larger cosmological constant $\Lambda'>\Lambda$. This would give interesting lessons on how (curvature) excitations condense to lead to a new vacuum state.

\end{itemize}

\begin{center}
\textbf{Acknowledgements}
\end{center}
The authors would like to thank Cl\'ement Delcamp, Laurent Freidel, Liang Kong, Aldo Riello, Lee Smolin, and Wolfgang Wieland for discussions and comments. This work is supported by Perimeter Institute for Theoretical Physics. Research at Perimeter Institute is supported by the Government of Canada through Industry Canada and by the Province of Ontario through the Ministry of Research and Innovation.

\appendix

\section{Details about SU(2)${}_\text{k}$}
\label{appendix1}

\noindent To construct the algebraic data for $\SU(2)_\k$, let us start by choosing a positive integer $\k$. This latter can be thought of as the coupling constant (the level) of the Chern--Simons formulation of Euclidean three-dimensional gravity with a positive cosmological constant. In this context, $\k$ appears as the ratio
\be
\k=\f{\lc}{\lp}=\f{1}{G\hbar\sqrt{\Lambda}}
\ee
of the cosmological length to the Planck length, and one can show that gauge-invariance requires that $\k$ be an integer. This in turn implies that the deformation parameter
\be
q=e^{2\pi\i/(\k+2)}
\ee
is a root of unity. With this deformation parameter, we can introduce the quantum numbers
\be\label{qNumber}
[n]=\f{q^{n/2}-q^{-n/2}}{q^{1/2}-q^{-1/2}}=\f{\displaystyle\sin\left(\f{\pi}{k+2}n\right)}{\displaystyle\sin\left(\f{\pi}{k+2}\right)},
\q
\forall\,n\in\mathbb{N}-\{0\},
\ee
and define $[0]=1$. This then defines the quantum dimensions $d_j=[2j+1]$, and in particular, using the notation $v_j^2=(-1)^{2j}d_j$, the so-called total quantum dimension
\be
\D\coloneqq\sqrt{\sum_jv_j^4}=\sqrt{\f{\k+2}{2}}\sin^{-1}\left(\f{\pi}{\k+2}\right)=\f{1}{\Z}.
\ee
Here $\Z$ is the evaluation of the path integral of $\SU(2)_\k$ Chern--Simons theory on the three-sphere. One can check that $\mathbb{R}\ni d_j\geq1$ for all the spins $j\in\{0,\dots,\k/2\}$, with in particular $d_0=1=d_{\k/2}$. The fusion rules between representations of spin $j\in\{0,\dots,\k/2\}$ are given in the main text in \eqref{Admissibility} and \eqref{FusionRule}.

To introduce the algebraic expression for the $F$-symbols let us first define, for any integer $n\geq1$, the factorial $[n]!\coloneqq[n][n-1]\dots[2][1]$, and set $[0]!=[0]=1$. With this, we can then define for any admissible triple $(i,j,k)$ the quantity
\be
\Delta(i,j,k)=\delta_{ijk}\sqrt{\f{[i+j-k]![i-j+k]![-i+j+k]!}{[i+j+k+1]!}},
\ee
which is therefore defined to be zero when the triple is non-admissible. The Racah--Wigner quantum $6j$ symbol is then given by the formula
\be
\left\{
\begin{array}{ccc}
i&j&m\\
k&l&n
\end{array}\right\}
=&\:\Delta(i,j,m)\Delta(i,l,n)\Delta(k,j,n)\Delta(k,l,m)\sum_z(-1)^z[z+1]!\nonumber\\
&\times\f{\Big([i+j+k+l-z]![i+k+m+n-z]![j+l+m+n-z]!\Big)^{-1}}{[z-i-j-m]![z-i-l-n]![z-k-j-n]![z-k-l-m]!},\q
\ee
where the sum runs over
\be
\max(i+j+m,i+l+n,k+j+n,k+l+m)\leq z\leq\min(i+j+k+l,i+k+m+n,j+l+m+n).
\ee
Finally, the $F$-symbols are defined as
\be\label{FDefinition}
F^{ijm}_{kln}
=
(-1)^{i+j+k+l}\sqrt{[2m+1][2n+1]}
\left\{
\begin{array}{ccc}
i&j&m\\
k&l&n
\end{array}\right\}.
\ee
The explicit expression for the $R$-matrix of $\SU(2)_\k$ is
\be\label{R matrix expression}
R^{ij}_k=(-1)^{k-i-j}\left(q^{k(k+1)-i(i+1)-j(j+1)}\right)^{1/2},
\ee
from which one can see that $R^{ij}_k=R^{ji}_k$ and $R^{i0}_i=1$. Now, the $S$-matrix has entries defined via the $R$-matrix as
\be\label{SM1}
\D S_{ij}=s_{ij}=\pic{Sij}\hs,
\ee
and using expression \eqref{resolution of crossing} we get the explicit expression
\be\label{s matrix in terms of R}
s_{ij}=\sum_kv_k^2R^{ij}_kR^{ji}_k=(-1)^{2(i+j)}[(2i+1)(2j+1)].
\ee
With this notation, we see that $s_{ij}$ corresponds to the evaluation of the Hopf link in the three-sphere, and we understand the normalization coefficient $S_{00}=\Z=\D^{-1}$. The $S$-matrix satisfies the properties
\be\label{s matrix identities}
S_{ij}=S_{ji},
\q\q
\sum_kS_{ik}S_{kj}=\delta_{ij},
\q\q
S_{ik}S_{kj}=S_{0k}\sum_l\delta_{ijl}S_{lk},
\ee
and can be used to compute the fusion coefficients via the famous Verlinde formula\footnote{The $S$-matrix is real and symmetric in the case of $\SU(2)_\k$, but we indicate the Hermitian conjugate for the sake of generality.}
\be
\delta_{ijk}=\sum_l\f{S_{il}S_{jl}S^*_{kl}}{S_{0l}}.
\ee
When written in the form
\be\label{diagN}
(N_i)_{jk}=\delta_{ijk}=\sum_lS_{jl}\left(\f{S_{il}}{S_{0l}}\right)S^*_{lk},
\ee
the Verlinde formula shows that the $S$-matrices diagonalize the fusion matrices.

\section{Basis states for a two-surface of genus $\boldsymbol{g\geq1}$}
\label{appendix:higher-genus}

\noindent In this appendix, we briefly discuss the choice of basis for the graph Hilbert space on a punctured surface of genus $g\geq1$.

\begin{figure}[h]
$\pic{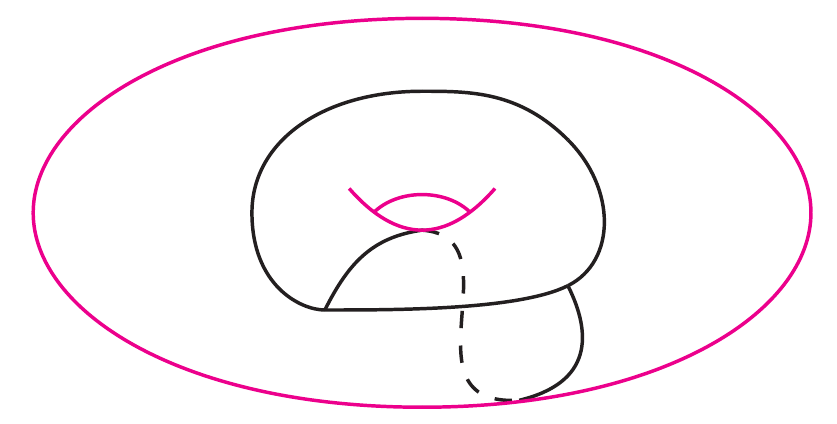}$
\caption{Example of a minimal graph on a torus without punctures.}
\label{torus}
\end{figure}

Let us start by considering the case of a torus with no punctures. A minimal graph for the torus has two three-valent nodes and three strands, and covers the two independent non-contractible cycles. A possible choice of minimal graph is represented on figure \ref{torus}.
By analogy with the construction of the basis for the punctured sphere, one could a priori assume that a basis for the torus is given by all the admissible spin colorings of this minimal graph. However, this set of states is over-complete\footnote{For a fixed level $\k$, the dimension of the graph Hilbert space on the torus without punctures (known as the ground state degeneracy) can be shown to be $\dim\H_{\mathbb{T}}=(\k+1)^2$ \cite{GSD}. For $\k=1$, one finds that there are indeed $(1+1)^2=4$ admissible colorings of the minimal graph of figure \ref{torus}. However, for $\k=2$, this formula gives $\dim\H_{\mathbb{T}}=9$ while one can check that there are 10 admissible colorings of the minimal graph of figure \ref{torus}. As we mentioned, this is due to the fact that not all these colorings are independent. The colorings of the minimal graph forms an over-complete basis.} since the graph features a face which covers the entire surface of the torus (this face corresponds to the dual vertex which results from seeing the torus as the gluing of a parallelogram). One therefore has to impose on the states a flatness constraint for this face. This results in the fact that some linear combination of states is equivalent to other linear combinations.

\begin{figure}[h]
$\pic{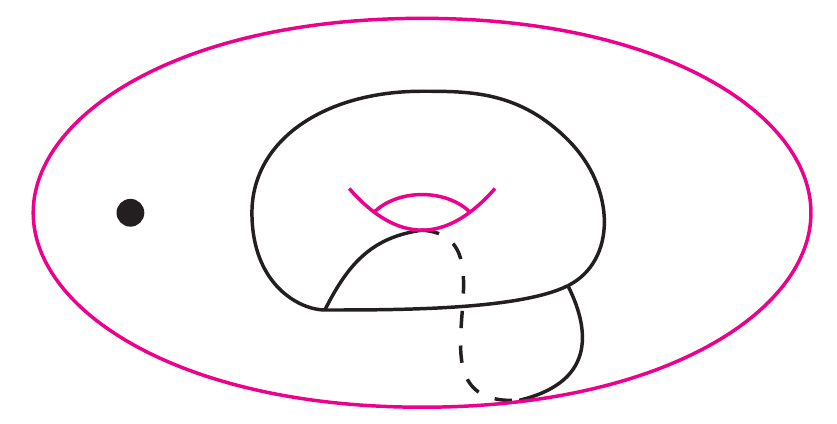}$
\caption{Minimal graph on a torus with one puncture and without violations of the Gauss constraint.}
\label{torus1p}
\end{figure}

There exists a procedure which enables us to get an independent set of states. This is borrowed from Dirac's procedure to quantize constrained systems, and it consists in considering a one-punctured torus without violations of the Gauss constraint. Then, the set of states based on colorings of the graph of figure \ref{torus1p} does indeed form an independent basis. Therefore, we can also define an inner product by declaring these basis states to be orthonormal.
The resulting Hilbert space would be equivalent to a so-called kinematical Hilbert space in Dirac's constrained quantization procedure. Note that this kinematical Hilbert space is finite-dimensional. One has now to impose the flatness constraint for the single face of the graph of figure \ref{torus1p}. The solution space (i.e. the null space of the constraints) defines the physical Hilbert space which we are looking for. The inner product on this physical Hilbert space is just the one induced from the kinematical Hilbert space, which is well-defined since we are in a finite-dimensional context.

This procedure generalizes to all surfaces of higher genus. Any such surface can be triangulated with at most one vertex, which results in one face for the dual graph. The basis for the kinematical Hilbert space (in the above-mentioned sense of Dirac quantization) is based on the graph dual to this triangulation with one vertex, and one has to impose once again a flatness constraint in order to get to the physical Hilbert space.

Higher genus surfaces with punctures are comparably less complicated. One starts with the basis based on the graph dual to the minimal triangulation with one vertex. For states on a one-punctured surface, which also include Gauss constraint violation, we add one strand going from the puncture to any strand of the graph. For any additional puncture, we then add the same $\Q$-shaped piece of graph as in the case of the punctured sphere. States based on these graphs (i.e. states obtained by allowing all admissible colorings) define a basis of the graph Hilbert space on the punctured surface.

This defines a basis which is analogous to the $\Q$ basis for the punctured sphere. One can then also define basis states which have similar properties as the Ocneanu (or fusion) basis. For this discussion, we refer the reader to \cite{KKR}.

\section{Proof of graphical calculus relations}
\label{appendix2}

\begin{proof}[Proof of \eqref{orthonormality of p-punctured Q}]
\be\label{Proof of orthonormality of p-punctured Q}
\Big\la\textbf{Q}^{\vec{\imath}\,\vec{\jmath}}_{n\,\vec{r}\,\vec{s}}\Big|\textbf{Q}^{\vec{\imath}\,'\,\vec{\jmath}\,'}_{n'\,\vec{r}\,'\,\vec{s}\,'}\Big\ra_\tr
&=\Big\la\pic{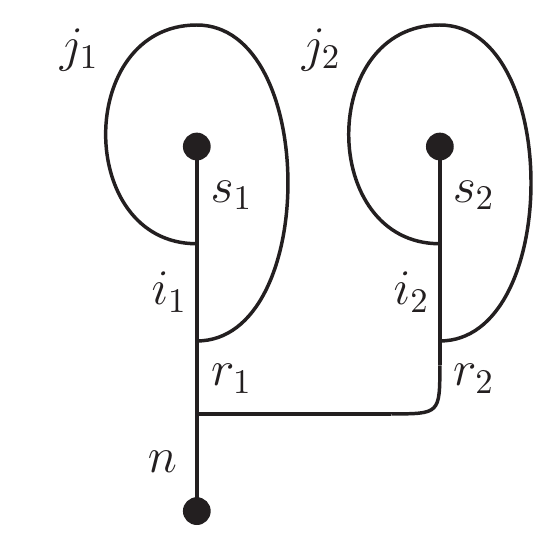}\Big|\pic{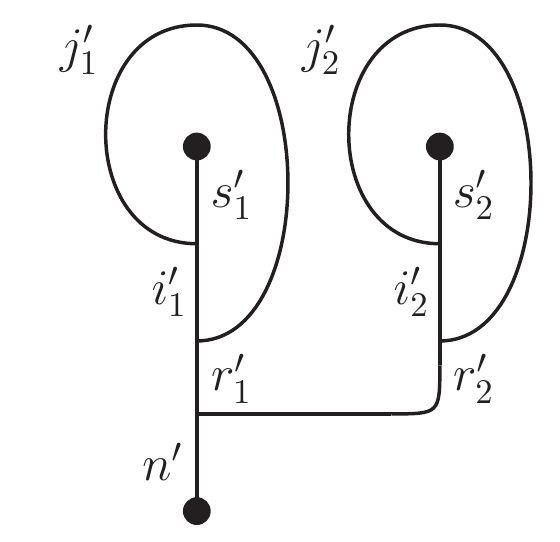}\Big\ra_\tr\nn\\
&=\f{1}{\D}\f{1}{v_n}\delta_{nn'}\pic{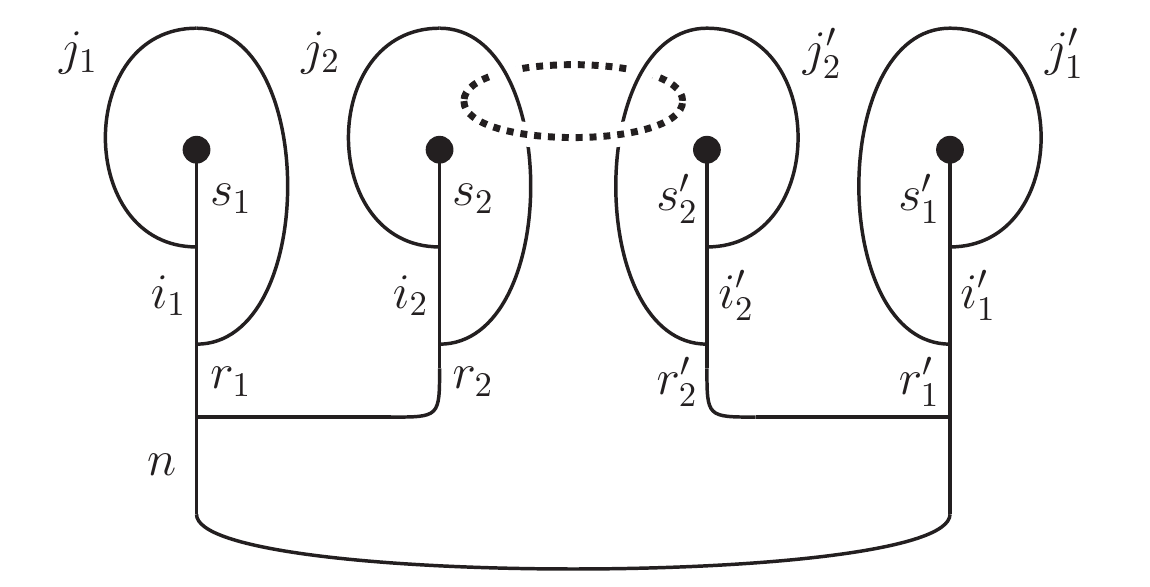}\nn\\
&\stackrel{\tr}{=}\f{1}{\D}\f{1}{v_{j_2}^2v_nv_{s_2}}\delta_{j_2j_2'}\delta_{nn'}\delta_{s_2s_2'}\pic{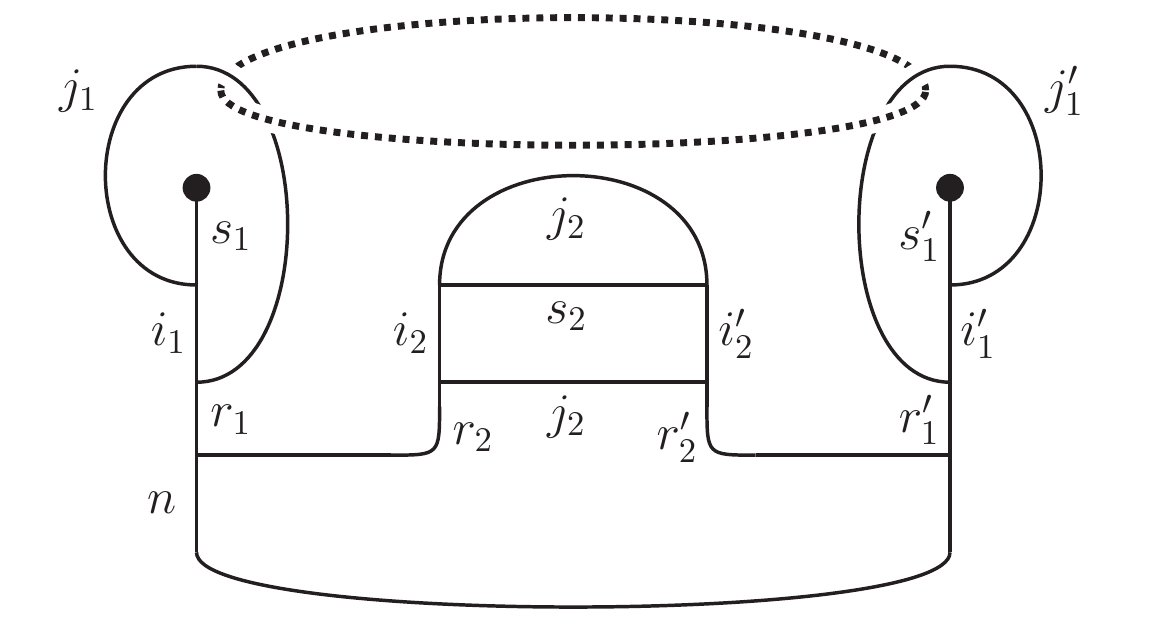}\nn\\
&\stackrel{\tr}{=}\f{1}{v_{j_1}^2v_{j_2}^2v_nv_{s_1}v_{s_2}}\delta_{\vec{\jmath}\,\vec{\jmath}\,'}\delta_{nn'}\delta_{\vec{s}\,\vec{s}\,'}\pic{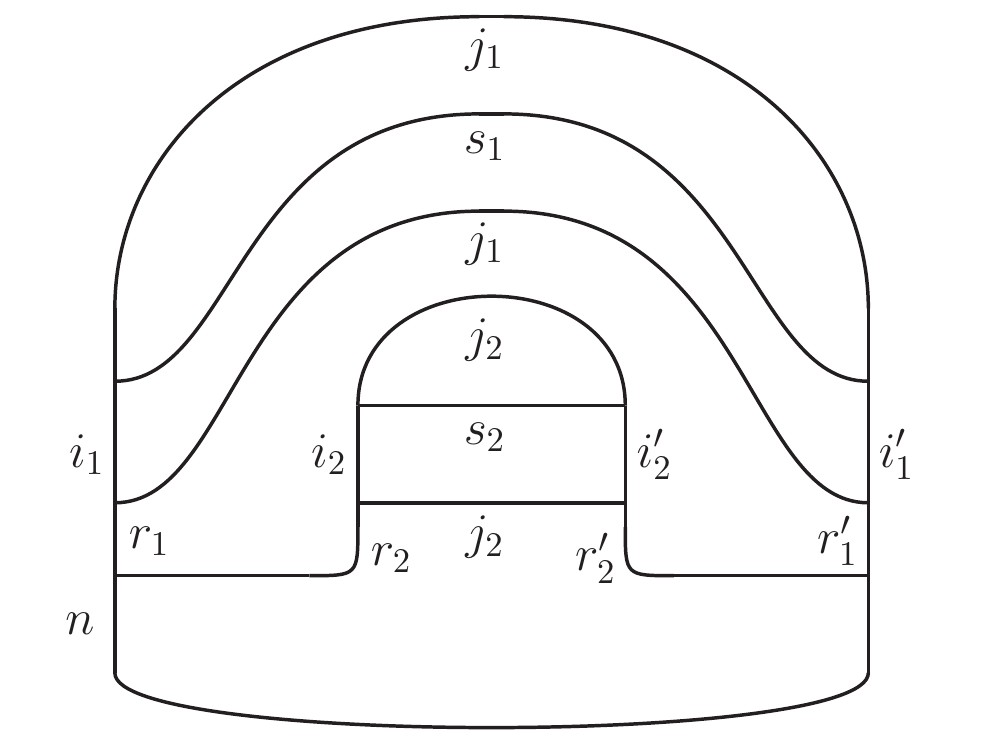}\nn\\
&=\delta_{\vec{\imath}\,\vec{\imath}\,'}\delta_{\vec{\jmath}\,\vec{\jmath}\,'}\delta_{nn'}\delta_{\vec{r}\,\vec{r}\,'}\delta_{\vec{s}\,\vec{s}\,'}.
\ee
Here we have used the notation $\stackrel{\tr}{=}$ to denote equalities which do not follow from the basic rules of graphical calculus, but rather from the evaluation of the trace inner product. For the second and third equalities we have used \eqref{vacuum F-move} and connected the strands $s_{1,2}$ and $s'_{1,2}$, and for the last equality we have used repeatedly the bubble-move \eqref{bubble move}.
\end{proof}

\begin{proof}[Proof of the sliding property \eqref{strand sliding}]
On the one hand, we have that
\be
\pic{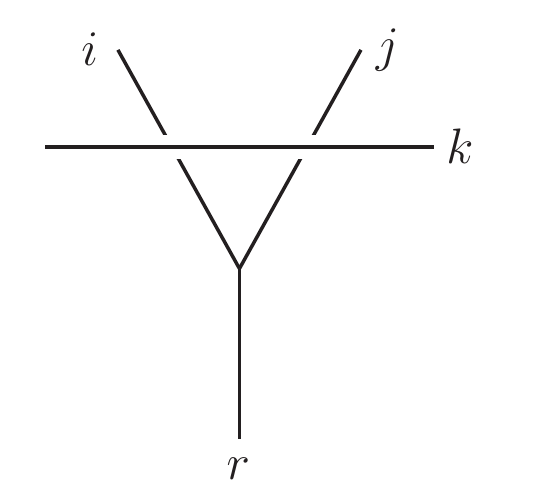}
&=\sum_{mn}\f{v_m}{v_jv_k}\f{v_n}{v_iv_k}R^{kj}_mR^{ki}_n\pic{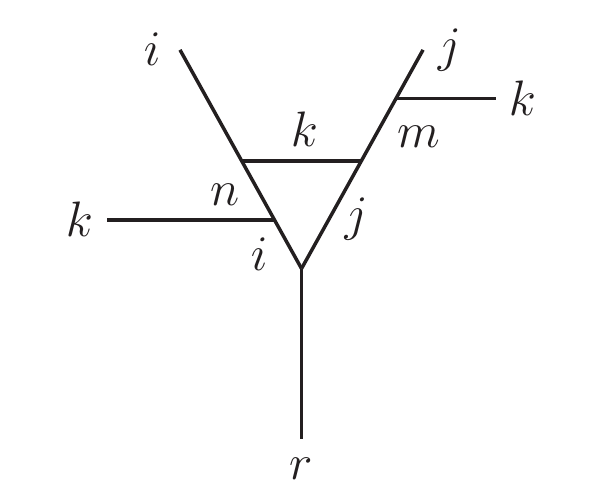}\nn\\
&=\sum_{mnp}\f{v_m}{v_jv_k}\f{v_n}{v_iv_k}R^{kj}_mR^{ki}_nF^{nki}_{rjp}\pic{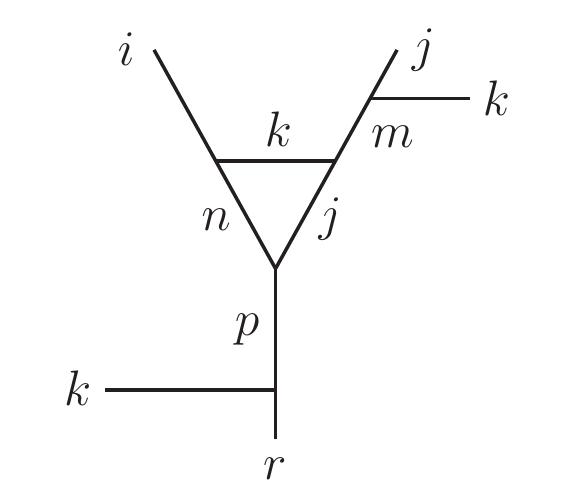}\nn\\
&=\sum_{mnp}\f{v_m}{v_jv_k}\f{v_n}{v_iv_k}\f{v_nv_j}{v_p}R^{kj}_mR^{ki}_nF^{nki}_{rjp}F^{imp}_{jnk}\pic{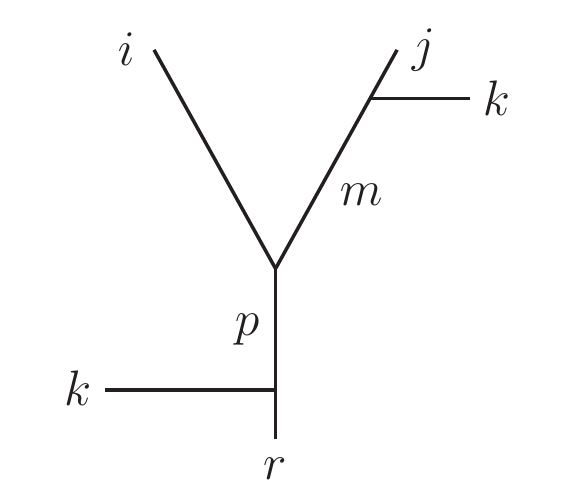}\nn\\
&=\sum_{mnpq}\f{v_n}{v_iv_k}R^{kj}_mR^{ki}_nF^{nki}_{rjp}F^{ipm}_{jkn}F^{ipm}_{kjq}\pic{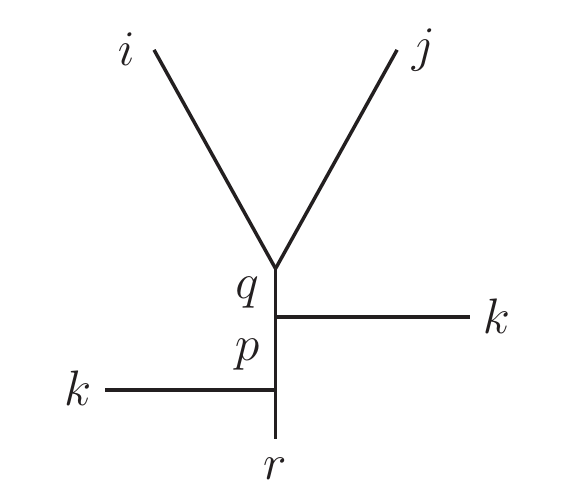}.
\ee
On the other hand, we have that
\be
\pic{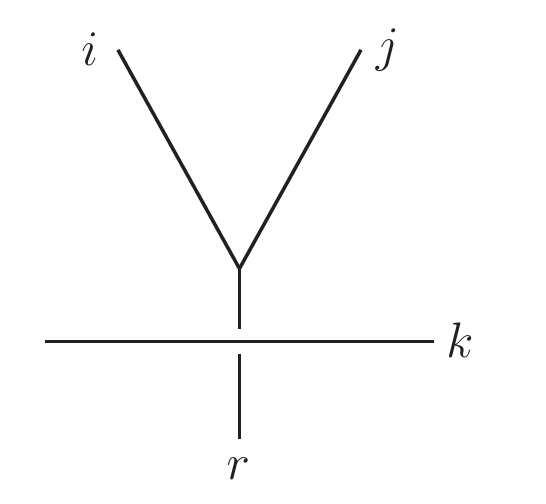}=\sum_p\f{v_p}{v_kv_r}R^{kr}_p\pic{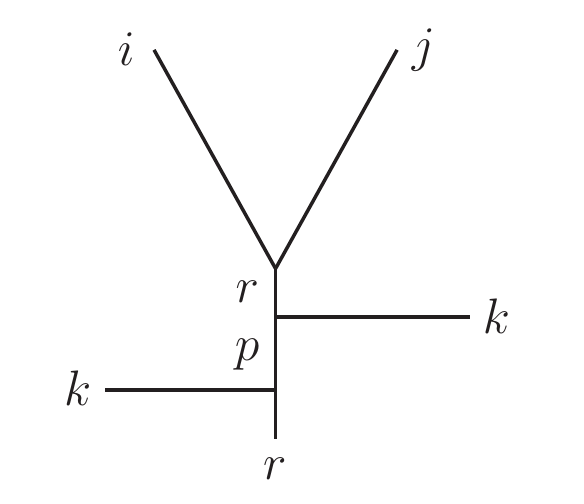}.
\ee
Therefore, in order for the sliding property \eqref{strand sliding} to hold, a necessary and sufficient condition is that
\be
\sum_{mn}\f{v_nv_r}{v_iv_p}R^{kj}_mR^{ki}_nF^{nki}_{rjp}F^{ipm}_{jkn}F^{ipm}_{kjq}\stackrel{?}{=}\delta_{qr}R^{kr}_p.
\ee
Using first the hexagon identity \eqref{hexagon identity} and then the orthogonality relation \eqref{F-orthogonality}, one can show that this relation does indeed hold, thereby ensuring that the sliding property is true.
\end{proof}

\begin{proof}[Proof of \eqref{Bp on face} in the case $s=6$]
Let us denote the flatness projection operator acting on a face by
\be
B_\text{face}=\f{1}{\D^2}\sum_kv_k^2B^k_\text{face}.
\ee
Then we have that
\be\label{Proof of Bp on face}
B^k_\text{face}\triangleright\pic{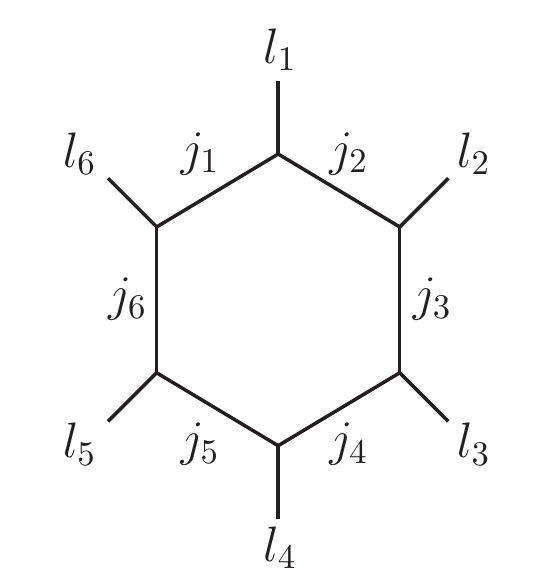}&\coloneqq\pic{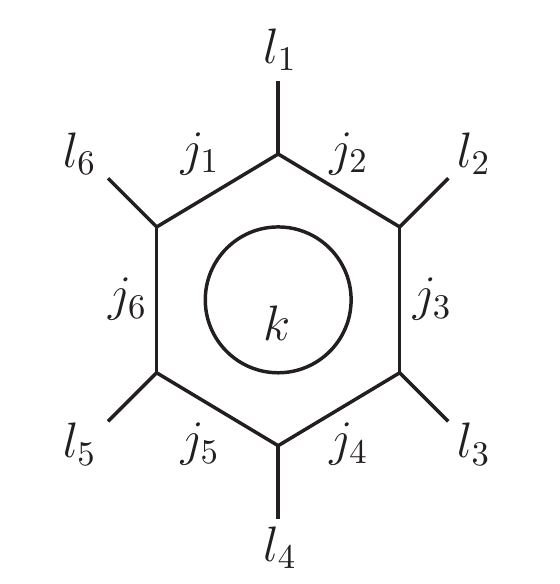}\nn\\
&=\sum_{n_1\dots n_6}\f{v_{n_1}}{v_kv_{j_1}}\f{v_{n_2}}{v_kv_{j_2}}\f{v_{n_3}}{v_kv_{j_3}}\f{v_{n_4}}{v_kv_{j_4}}\f{v_{n_5}}{v_kv_{j_5}}\f{v_{n_6}}{v_kv_{j_6}}\pic{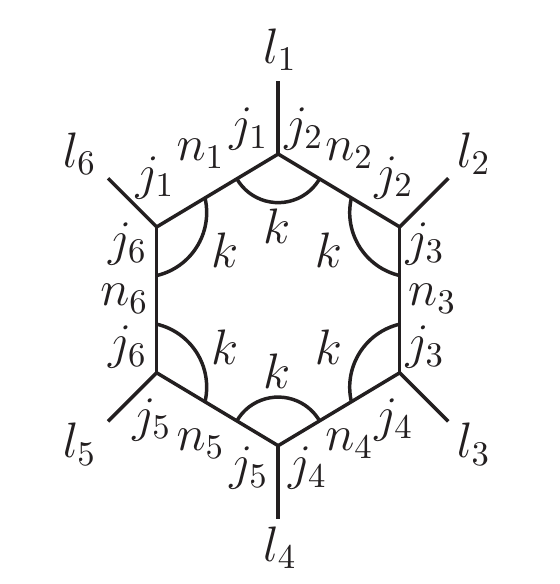}\nn\\
&=\sum_{n_1\dots n_6}F^{n_2l_1n_1}_{j_1kj_2}F^{n_3l_2n_2}_{j_2kj_3}F^{n_4l_3n_3}_{j_3kj_4}F^{n_5l_4n_4}_{j_4kj_5}F^{n_6l_5n_5}_{j_5kj_6}F^{n_1l_6n_6}_{j_6kj_1}\pic{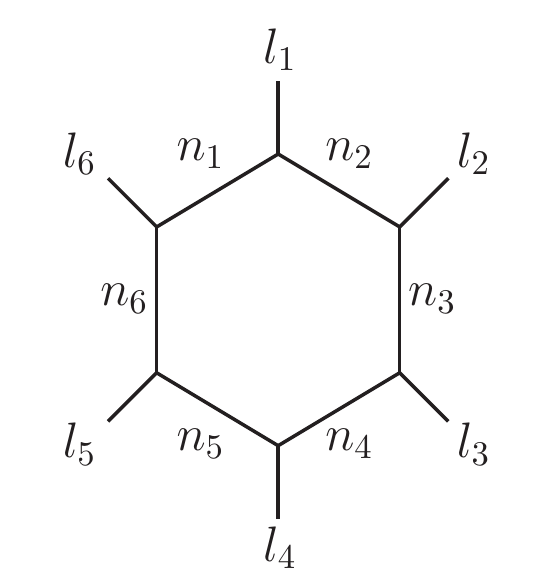}.
\ee
\end{proof}

\begin{proof}[Proof of \eqref{OjtoQ}]
\be\label{Proof of OjtoQ}
\O^{jj}_{00}
&=\f{1}{\D}\sum_kv_k^2\pic{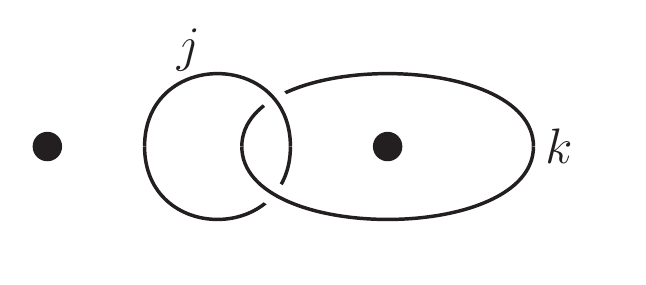}\nn\\
&=\f{1}{\D}\sum_{kmn}v_k^2\f{v_m}{v_jv_k}\f{v_n}{v_kv_j}R^{jk}_mR^{kj}_n\pic{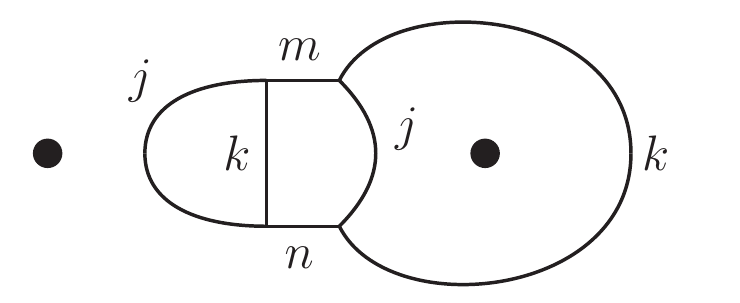}\nn\\
&=\f{1}{\D}\sum_{km}v_m^2\big(R^{jk}_m\big)^2\pic{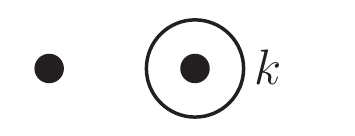}\nn\\
&=\f{1}{\D}\sum_{km}v_m^2\big(R^{jk}_m\big)^2\Q^{kk}_{00}.
\ee
\end{proof}

\begin{proof}[Proof of \eqref{pulling over O}]
\be\label{proof of pulling over O}
\pic{PullingThroughOii001}
&=\pic{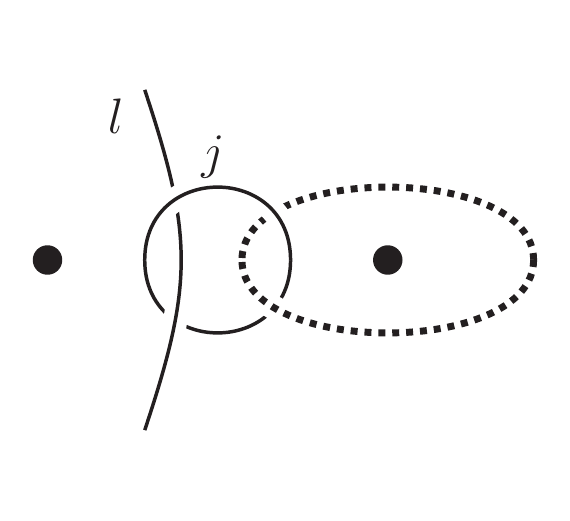}\nn\\
&=\sum_{mn}\f{v_mv_n}{v_j^2v_l^2}R^{jl}_mR^{lj}_n\pic{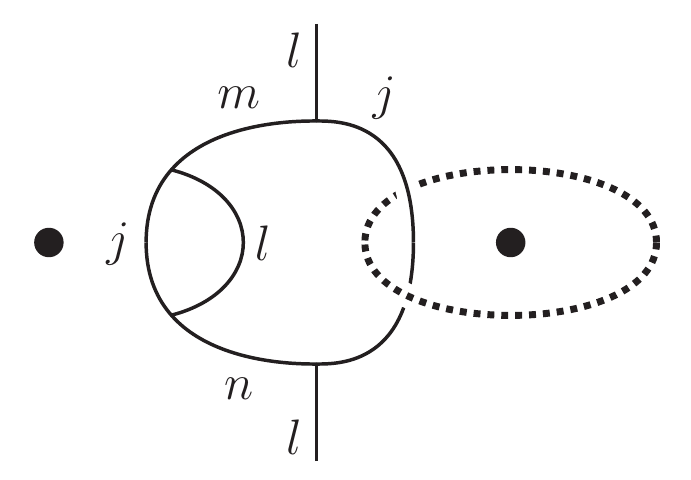}\nn\\
&=\sum_m\f{v_m}{v_jv_l}\big(R^{jl}_m\big)^2\pic{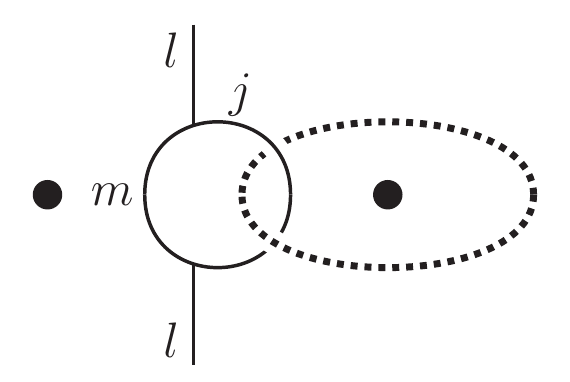}\nn\\
&=\sum_{mp}\f{v_m}{v_jv_l}\big(R^{jl}_m\big)^2F^{jlm}_{ljp}\pic{PullingThroughOii005}\nn\\
&=\sum_p\Omega^{0p}_{ll,jj}\pic{PullingThroughOii005}.
\ee
\end{proof}

\begin{proof}[Proof of \eqref{QAlgebra}]
\be\label{proof of QAlgebra}
\Q^{ij}_{rs}\times\Q^{kl}_{tu}
&=\delta_{st}\pic{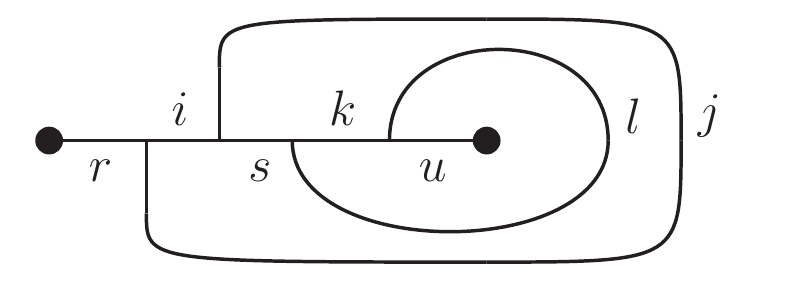}\nn\\
&=\delta_{st}\sum_mF^{jis}_{lkm}\pic{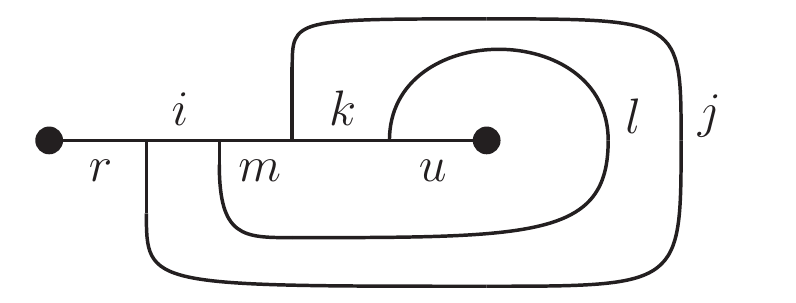}\nn\\
&=\delta_{st}\sum_{mn}\f{v_n}{v_jv_l}F^{jis}_{lkm}\pic{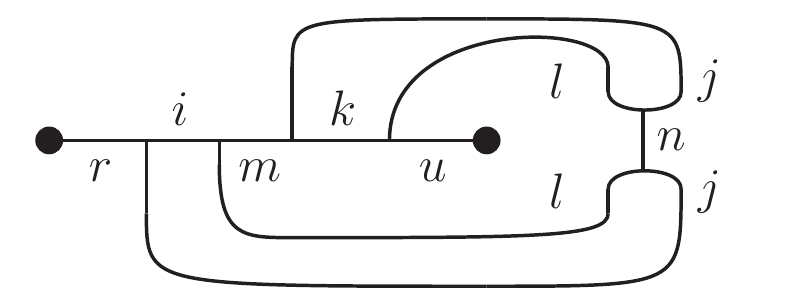}\nn\\
&=\delta_{st}\sum_{mn}\f{v_jv_l}{v_n}F^{jis}_{lkm}F^{rmn}_{lji}F^{umn}_{jlk}\pic{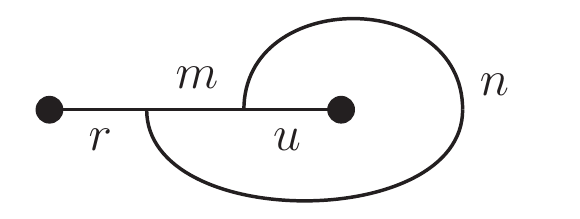}\nn\\
&=\delta_{st}\sum_{mn}\f{v_jv_l}{v_n}F^{jis}_{lkm}F^{rmn}_{lji}F^{umn}_{jlk}\Q^{mn}_{ru}\nn\\
&=\delta_{st}\sum_{mn}\f{v_kv_l}{v_u}F^{jis}_{lkm}F^{rmn}_{lji}F^{mnu}_{lkj}\Q^{mn}_{ru}\nn\\
&=\delta_{st}\sum_{mn}v_s\f{v_mv_n}{v_rv_u}F^{jsi}_{lmk}F^{mnr}_{jil}F^{mnu}_{lkj}\Q^{mn}_{ru}.
\ee
\end{proof}

\begin{proof}[Proof of \eqref{OtoQ}]
\be\label{proof of OtoQ}
\O^{ij}_{rs}
&=\pic{Oijrs}\nn\\
&=\f{1}{\D}\sum_lv_l^2\pic{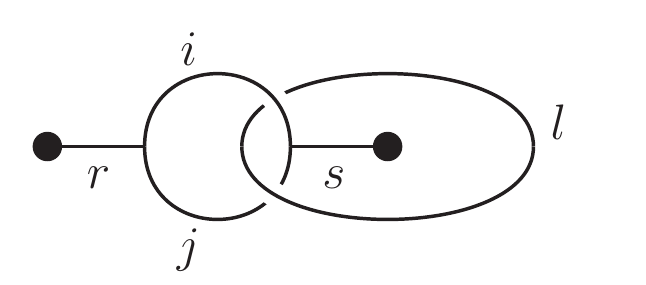}\nn\\
&=\f{1}{\D}\sum_{lmn}\f{v_mv_n}{v_iv_j}R^{il}_mR^{lj}_n\pic{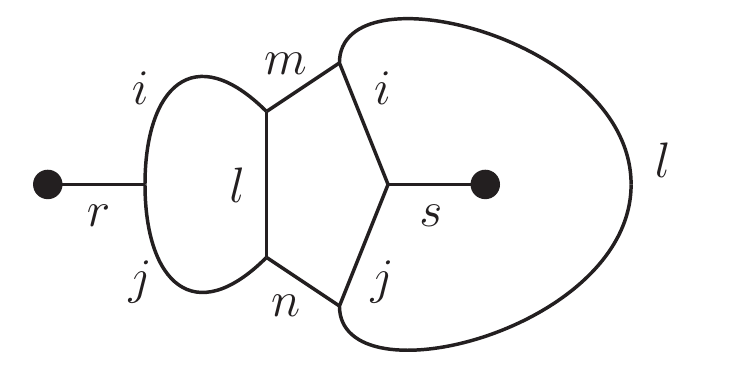}\nn\\
&=\f{1}{\D}\sum_{klmn}\f{v_mv_n}{v_iv_j}R^{il}_mR^{lj}_nF^{mil}_{jnk}\pic{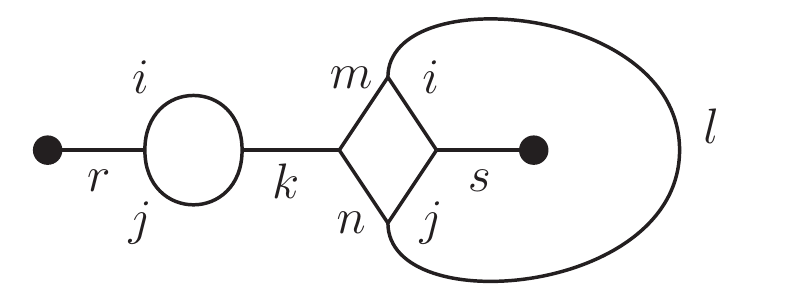}\nn\\
&=\f{1}{\D}\sum_{lmn}\f{v_mv_n}{v_r}R^{il}_mR^{lj}_nF^{mil}_{jnr}\pic{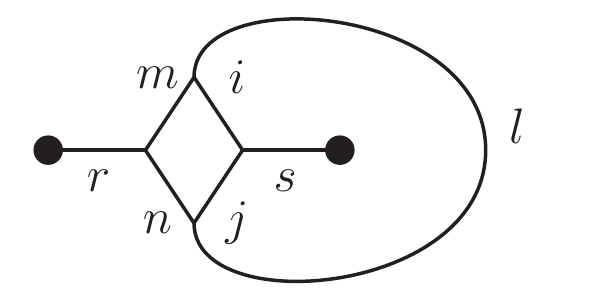}\nn\\
&=\f{1}{\D}\sum_{klmn}\f{v_mv_n}{v_r}R^{il}_mR^{lj}_nF^{mil}_{jnr}F^{lmi}_{jsk}\pic{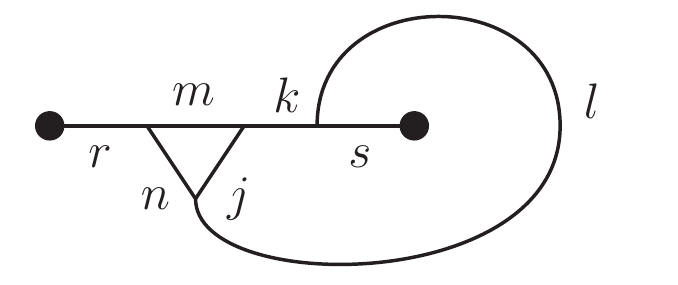}\nn\\
&=\f{1}{\D}\sum_{klmn}\f{v_mv_n}{v_r}\f{v_nv_j}{v_l}R^{il}_mR^{lj}_nF^{mil}_{jnr}F^{lmi}_{jsk}F^{rkl}_{jnm}\pic{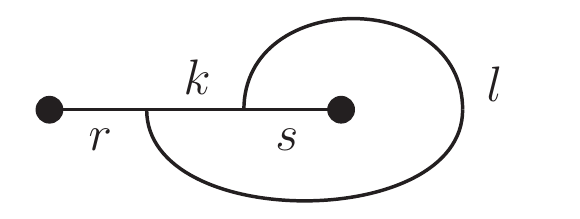}\nn\\
&=\f{1}{\D}\sum_{klmn}\f{v_mv_n}{v_r}\f{v_nv_j}{v_l}R^{il}_mR^{lj}_nF^{mil}_{jnr}F^{lmi}_{jsk}F^{rkl}_{jnm}\Q^{kl}_{rs}\nn\\
&=\f{1}{\D}\sum_{klmn}v_mv_n\f{v_iv_j}{v_rv_s}R^{il}_mR^{lj}_nF^{nmr}_{ijl}F^{kls}_{ijm}F^{rlk}_{jmn}\Q^{kl}_{rs}\nn\\
&=\f{1}{\D}\sum_{kl}v_iv_j\f{v_l^2}{v_s}\Omega^{rs}_{kl,ij}\Q^{kl}_{rs}.
\ee
\end{proof}

\begin{proof}[Proof of \eqref{O times O}]
\be\label{proof of O times O}
\O^{ij}_{rs}\times\O^{i'j'}_{s'u}
&=\delta_{ss'}\pic{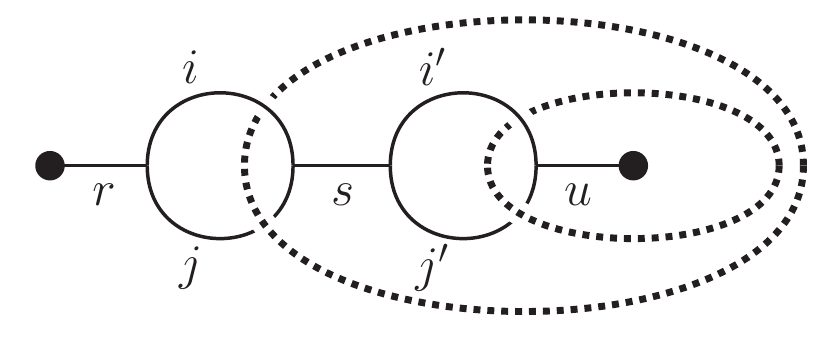}\nn\\
&=\delta_{ss'}\pic{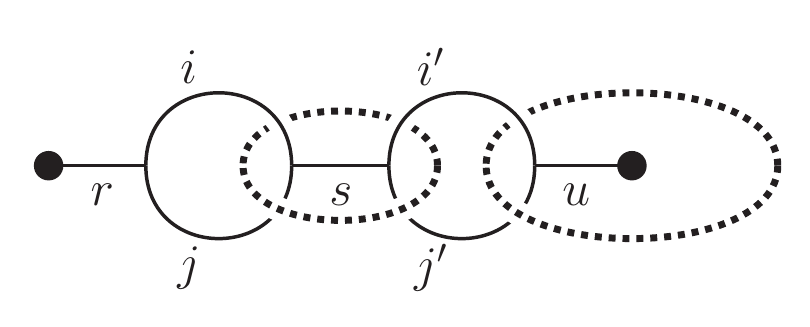}\nn\\
&=\D\delta_{ii'}\delta_{jj'}\delta_{ss'}\f{v_s}{v_iv_j}\pic{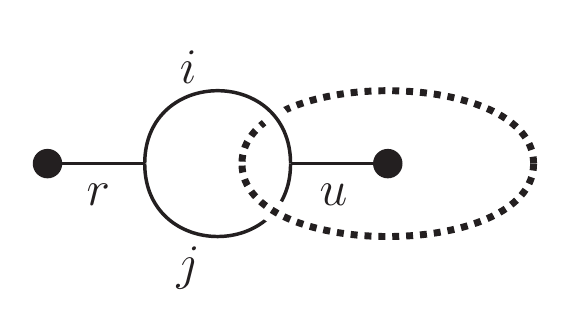}\nn\\
&=\D\delta_{ii'}\delta_{jj'}\delta_{ss'}\f{v_s}{v_iv_j}\O^{ij}_{ru}.
\ee
\end{proof}

\begin{proof}[Proof of \eqref{orthonormality of Os}]
\be\label{proof of orthonormality of Os}
\Big\la\pic{Oijrs}\Big|\pic{OijrsPrime}\Big\ra_\tr
&=\f{1}{\D}\f{1}{v_r}\delta_{rr'}\pic{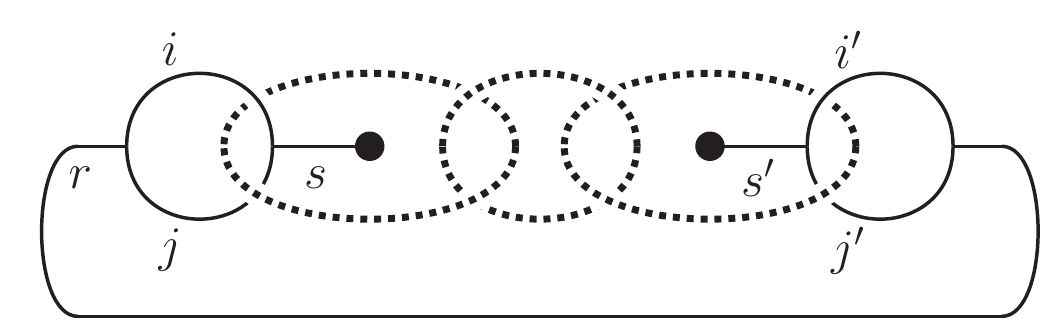}\nn\\
&\stackrel{\tr}{=}\f{1}{\D}\f{1}{v_rv_s}\delta_{rr'}\delta_{ss'}\pic{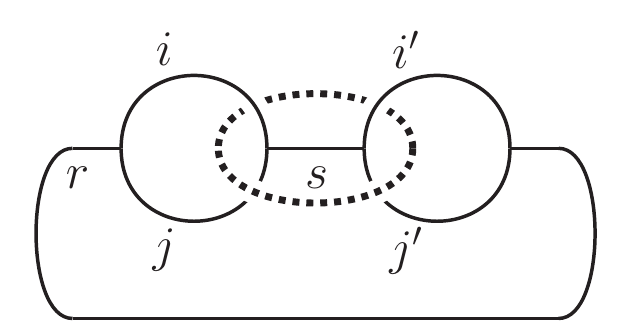}\nn\\
&=\f{1}{v_rv_iv_j}\delta_{ii'}\delta_{jj'}\delta_{rr'}\delta_{ss'}\pic{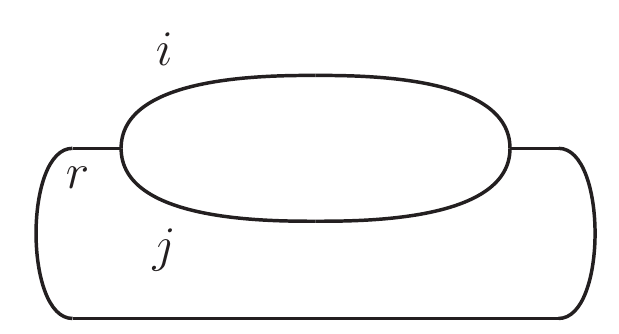}\nn\\
&=\delta_{ii'}\delta_{jj'}\delta_{rr'}\delta_{ss'}.
\ee
\end{proof}

\begin{proof}[Proof of \eqref{pulling through Oijrs}]
\be\label{proof of pulling through Oijrs}
\pic{PullingThroughOijrs1}
&=\pic{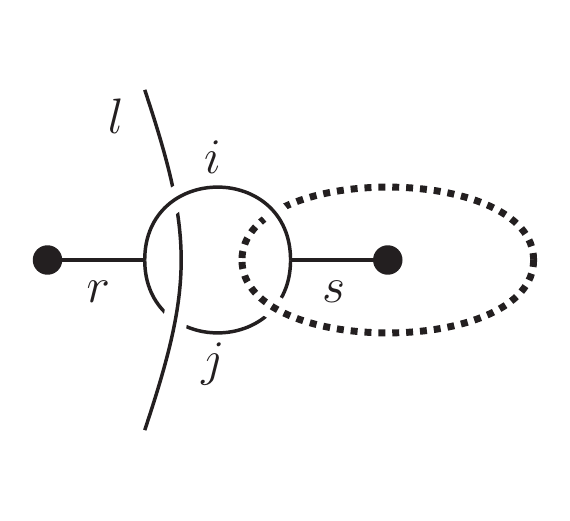}\nn\\
&=\sum_{mn}\f{v_m}{v_iv_l}\f{v_n}{v_jv_l}R^{il}_mR^{lj}_n\pic{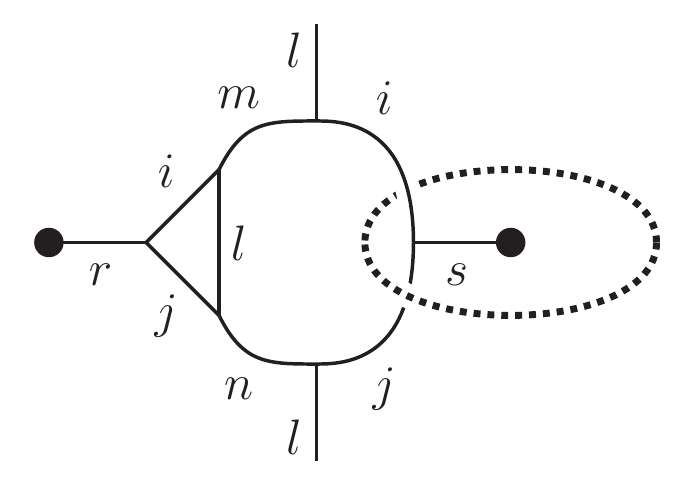}\nn\\
&=\sum_{mn}\f{v_mv_n}{v_rv_l^2}R^{il}_mR^{lj}_nF^{mnr}_{jil}\pic{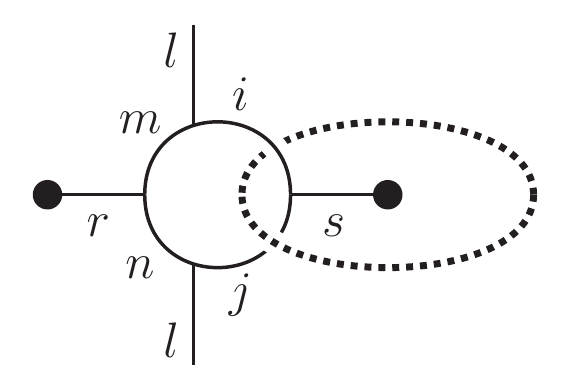}\nn\\
&=\sum_{mnq}\f{v_mv_n}{v_rv_l^2}R^{il}_mR^{lj}_nF^{mnr}_{jil}F^{mrn}_{ljq}\pic{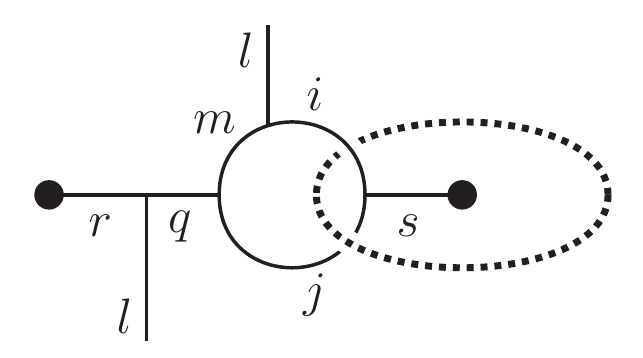}\nn\\
&=\sum_{mnpq}\f{v_mv_n}{v_rv_l^2}R^{il}_mR^{lj}_nF^{mnr}_{jil}F^{mrn}_{ljq}F^{qjm}_{ilp}\pic{PullingThroughOijrs6}\nn\\
&=\sum_{pq}\Omega^{rp}_{ql,ij}\pic{PullingThroughOijrs6}.
\ee
\end{proof}

\begin{proof}[Proof of \eqref{pulling through Qijrs}]
\be\label{proof of pulling through Qijrs}
\pic{PullingThroughQijrs1}
&=\pic{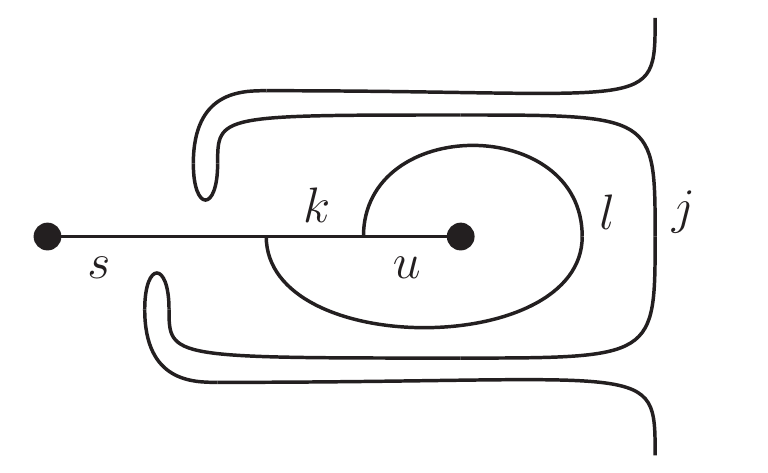}\nn\\
&=\sum_{im}\f{v_iv_m}{v_j^2v_s^2}\pic{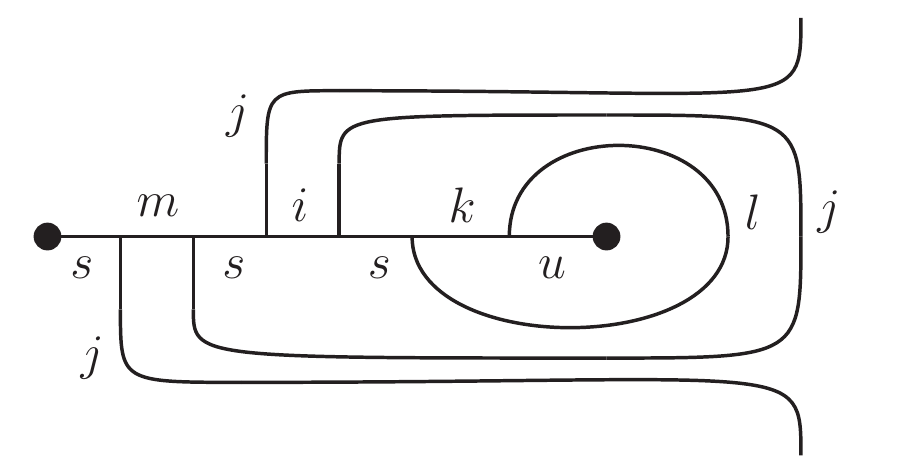}\nn\\
&=\sum_{imr}\f{v_r}{v_j^2v_s}F^{jsm}_{jri}\pic{PullingThroughQijrs4}.
\ee
\end{proof}

\begin{proof}[Proof of orthonormality of the $\bO$ states on $\mathbb{S}_3$]
\be\label{proof of orthonormality of p-punctured O}
\Big\la\bO^{a\,b\,\vec{\imath}\,\vec{\jmath}}_{r\,\vec{s}}\Big|\bO^{a'\,b'\,\vec{\imath}\,'\vec{\jmath}\,'}_{r'\,\vec{s}\,'}\Big\ra_\tr
&=\f{1}{\D^2}\f{1}{v_r}\delta_{rr'}\pic{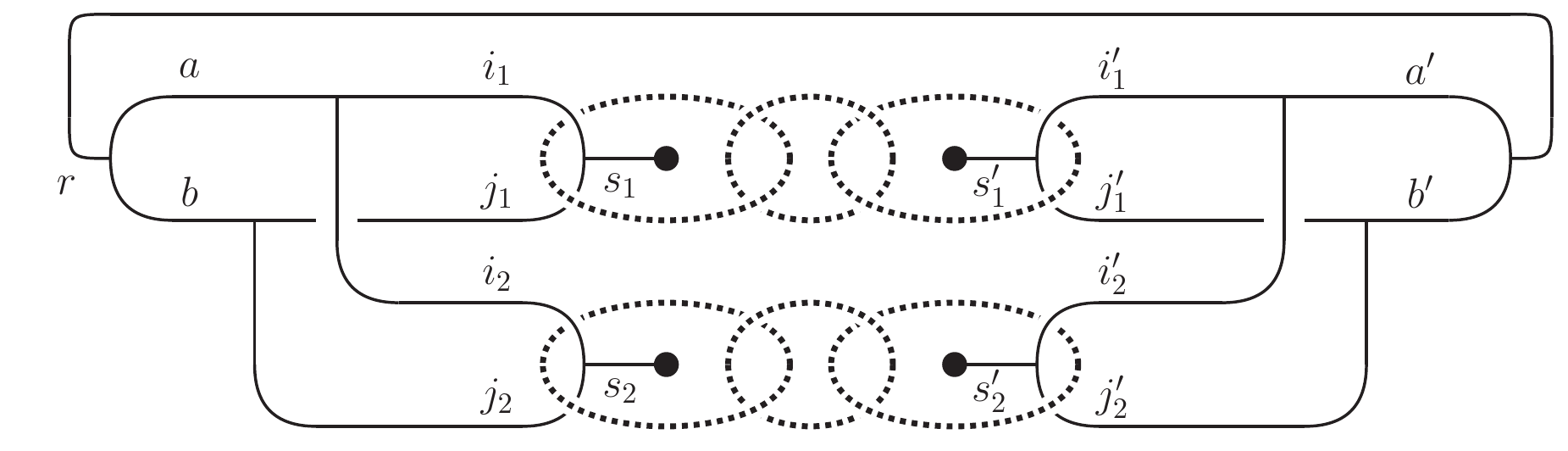}\nn\\
&\stackrel{\tr}{=}\f{1}{\D^2}\f{1}{v_rv_{s_1}v_{s_2}}\delta_{rr'}\delta_{\vec{s}\,\vec{s}\,'}\pic{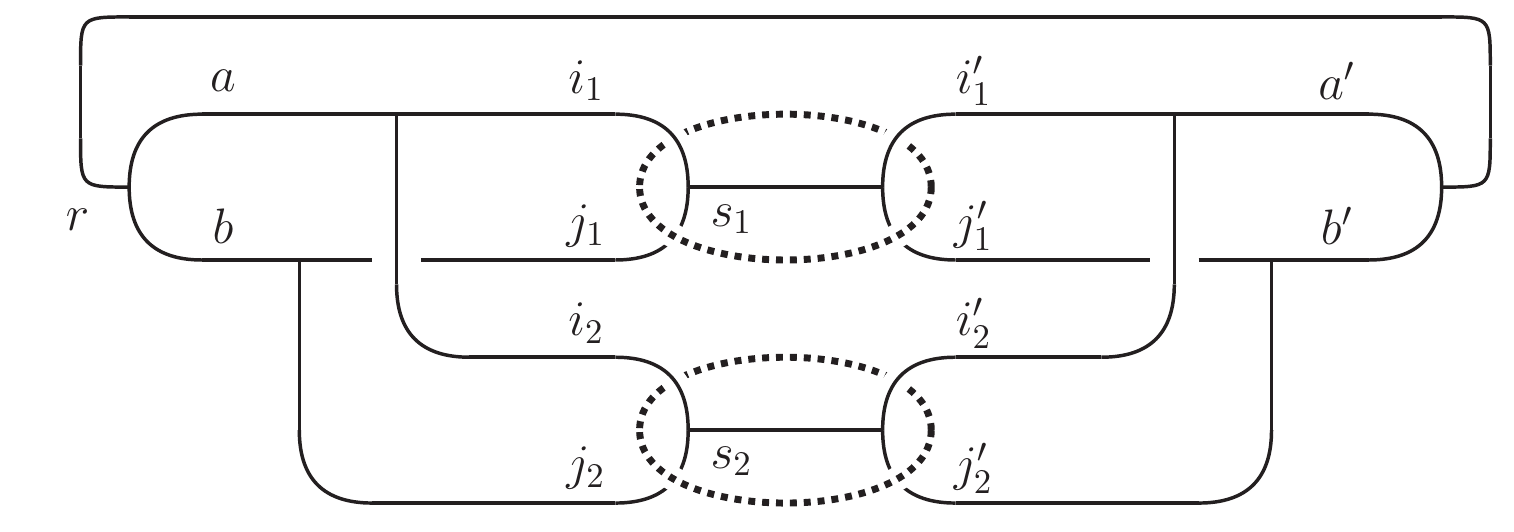}\nn\\
&=\f{1}{v_rv_{i_1}v_{i_2}v_{j_1}v_{j_2}}\delta_{\vec{\imath}\,\vec{\imath}\,'}\delta_{\vec{\jmath}\,\vec{\jmath}\,'}\delta_{rr'}\delta_{\vec{s}\,\vec{s}\,'}\pic{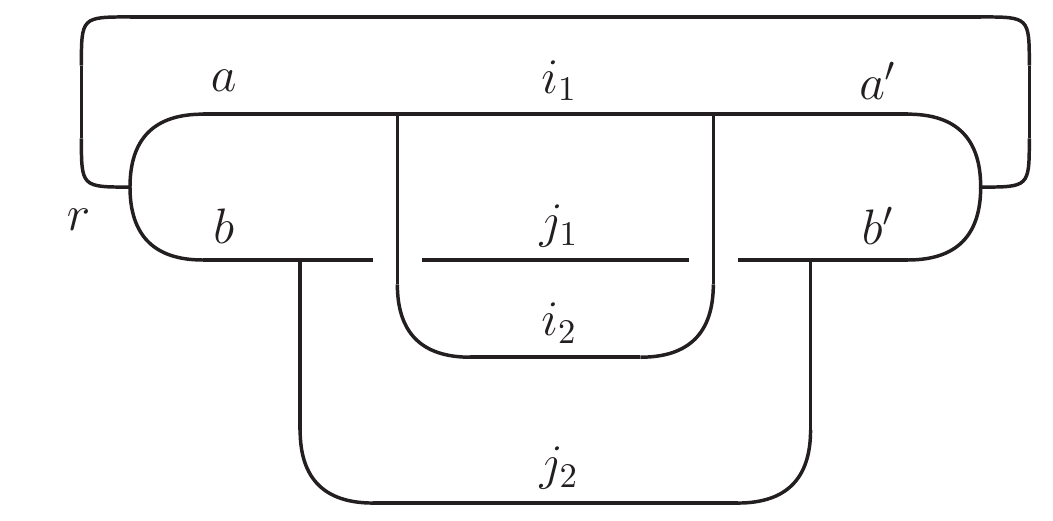}\nn\\
&=\f{1}{v_rv_{i_1}v_{i_2}v_{j_1}v_{j_2}}\delta_{\vec{\imath}\,\vec{\imath}\,'}\delta_{\vec{\jmath}\,\vec{\jmath}\,'}\delta_{rr'}\delta_{\vec{s}\,\vec{s}\,'}\pic{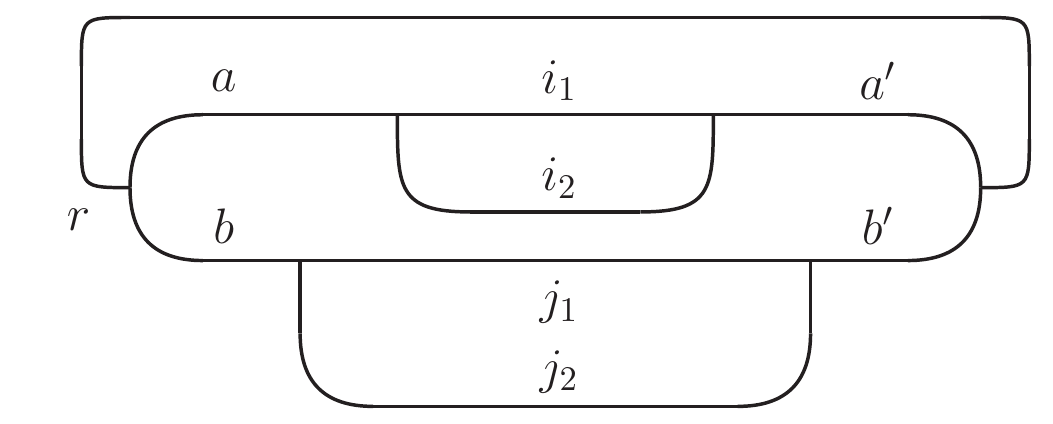}\nn\\
&=\delta_{aa'}\delta_{bb'}\delta_{\vec{\imath}\,\vec{\imath}\,'}\delta_{\vec{\jmath}\,\vec{\jmath}\,'}\delta_{rr'}\delta_{\vec{s}\,\vec{s}\,'}.
\ee
\end{proof}

\begin{proof}[Proof of \eqref{eigenvalue of closed ribbon}]
\be\label{proof of eigenvalue of closed ribbon}
\sum_p\f{v_p}{v_iv_j}\f{1}{v_av_bv_t}\pic{ClosedRibbonEigenvalue1}
&=\f{1}{v_av_bv_t}\pic{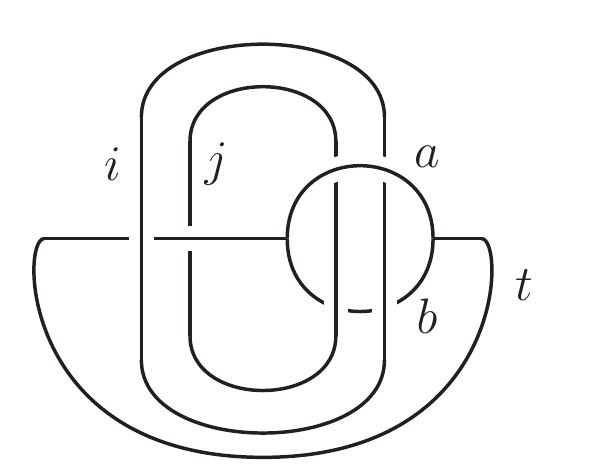}\nn\\
&=\f{1}{v_av_bv_t}\pic{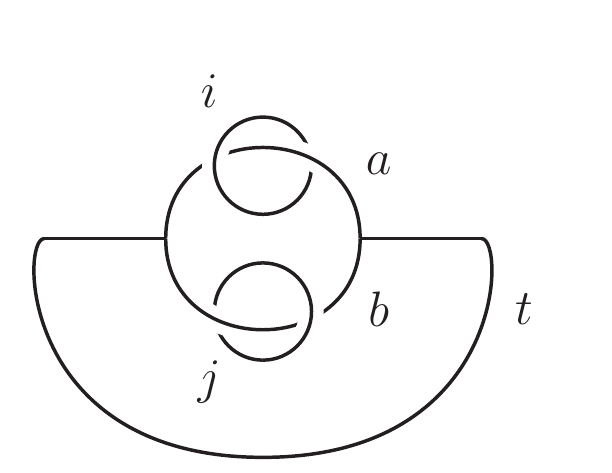}\nn\\
&=\frac{s_{ia}}{s_{0a}} \frac{s_{jb}}{s_{0b}}\delta_{abt}.
\ee
\end{proof}

\begin{proof}[Proof of \eqref{double S matrix}]
\be\label{proof of double S matrix}
\pic{DoubleS1}
&=\sum_{pr}\pic{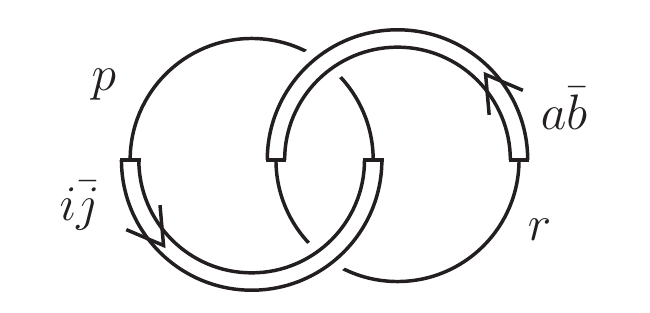}\nn\\
&=\sum_{pr}\f{v_p}{v_iv_j}\f{v_r}{v_av_b}\pic{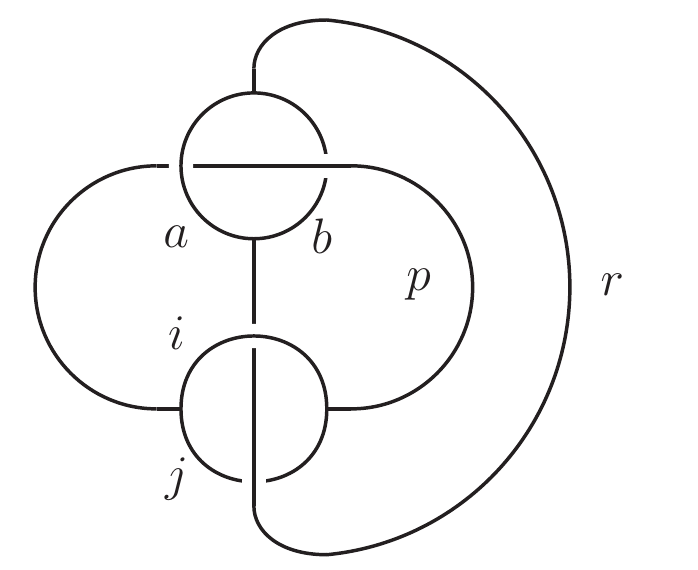}\nn\\
&=\pic{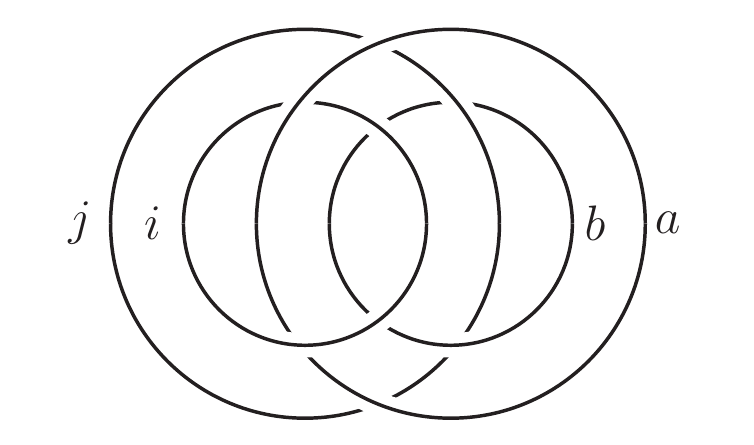}\nn\\
&=\pic{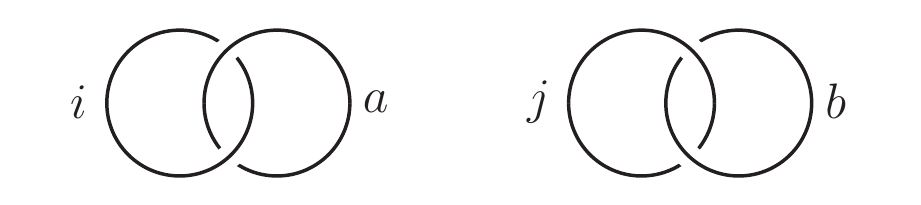}\nn\\
&=s_{ia}s_{jb}
\ee
\end{proof}

\begin{proof}[Proof of \eqref{cribbonEV}]
By using the half-braiding \eqref{ribbon braiding strand} to resolve the crossing between a ribbon and a strand, we get that
\be
\pic{RibbonBraidingOabtu1}=\sum_{pqr}\Omega^{rp}_{qt,ij}\pic{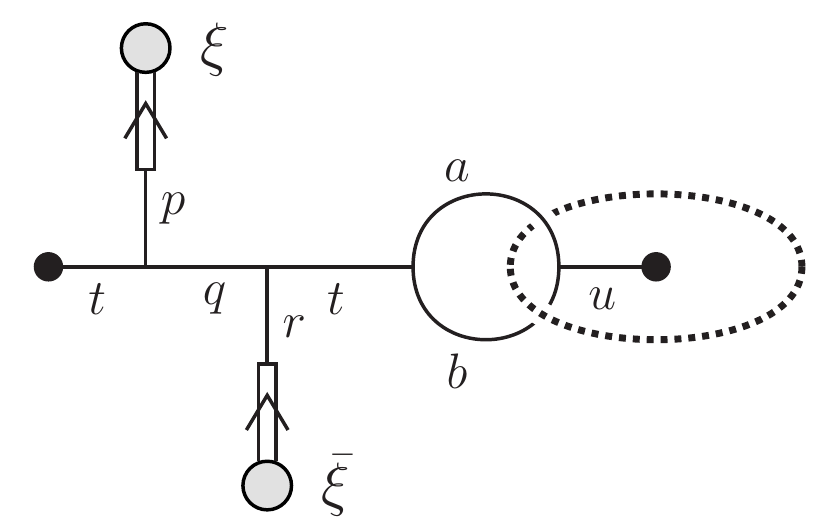}.
\ee
Now, the ribbon quasi-particles $\xi$ and $\bar{\xi}$ can be fused on the right of the puncture, leading to
\be\label{closing ribbon on Oabtu with half-braiding}
\pic{ClosingRibbonOnOabtu1}=\sum_{epq}F^{ptq}_{pte}\Omega^{pp}_{qt,ij}\pic{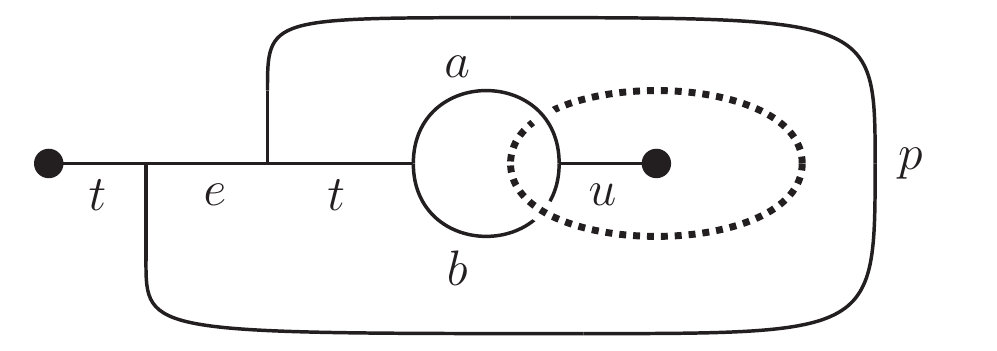}.
\ee
The graph on the right-hand side of this equality is noting but the stacking $\Q^{ep}_{tt}\times\O^{ab}_{tu}$ of two cylinder basis states. By using the inverse of the basis transformation \eqref{OtoQ}, which is given by
\be\label{QtoO}
\Q^{ij}_{rs}=\sum_{kl}\big(\widetilde{\Omega}^{rs}_{ij,kl}\big)^*\O^{kl}_{rs},
\ee
we get that \eqref{closing ribbon on Oabtu with half-braiding} is equal to
\be
\sum_{eklpq}F^{ptq}_{pte}\big(\widetilde{\Omega}^{tt}_{ep,kl}\big)^*\Omega^{pp}_{qt,ij}\O^{kl}_{tt}\times\O^{ab}_{tu}=\sum_{epq}v_p^2F^{ptq}_{pte}\big(\Omega^{tt}_{ep,ab}\big)^*\Omega^{pp}_{qt,ij}\O^{ab}_{tu}=\sum_{pq}v_p^2\Omega^{tt}_{qp,ab}\Omega^{pp}_{qt,ij}\O^{ab}_{tu},
\ee
where in the first equality we have used \eqref{Omega definition} together with the projection property \eqref{O times O} for the $\O$ states. To obtain the last equality, we have used the fact that
\be
\sum_eF^{ptq}_{pte}\big(\Omega^{tt}_{ep,ab}\big)^*
&=\sum_{emn}F^{ptq}_{pte}\f{v_mv_n}{v_tv_p^2}\big(R^{ap}_m\big)^*\big(R^{pb}_n\big)^*F^{nmt}_{abp}F^{ept}_{abm}F^{tpe}_{bmn}\nn\\
&=\sum_{mn}\f{v_mv_n}{v_tv_p^2}\big(R^{ap}_m\big)^*\big(R^{pb}_n\big)^*F^{nmt}_{abp}F^{ptq}_{nam}F^{abt}_{pqn}\nn\\
&=\sum_{kmn}\f{v_kv_n}{v_tv_p^2}R^{ap}_k\big(R^{pb}_n\big)^*F^{apk}_{apm}F^{nmt}_{abp}F^{ptq}_{nam}F^{abt}_{pqn}\nn\\
&=\sum_{kn}\f{v_kv_n}{v_tv_p^2}R^{ap}_k\big(R^{pb}_n\big)^*F^{apk}_{qbt}F^{bnp}_{akq}F^{abt}_{pqn}\nn\\
&=\sum_{kln}\f{v_kv_l}{v_tv_p^2}R^{ap}_kR^{pb}_lF^{pbl}_{pbn}F^{apk}_{qbt}F^{bnp}_{akq}F^{abt}_{pqn}\nn\\
&=\sum_{kl}\f{v_kv_l}{v_tv_p^2}R^{ap}_kR^{pb}_lF^{bpl}_{tkq}F^{kap}_{blt}F^{apk}_{qbt}\nn\\
&=\Omega^{tt}_{qp,ab},
\ee
which is here derived using respectively the pentagon relation on $e$, the relation \eqref{from R* to R}, the pentagon relation on $m$, the replacement \eqref{from R* to R} again, and finally the pentagon relation on $n$.

We have therefore shown that the eigenvalues $\lambda_{ij,ab}$ of the closed ribbon operators acting on $\O$ basis states can be written as
\be
\lambda_{ij,ab}=\frac{s_{ia}}{s_{0a}}\frac{s_{jb}}{s_{0b}}=\f{1}{v_a^2v_b^2}s_{ia}s_{jb}=\sum_{pq}v_p^2\Omega^{tt}_{qp,ab}\Omega^{pp}_{qt,ij},
\ee
as announced in \eqref{cribbonEV}.
\end{proof}

\section{Commutativity of $\boldsymbol{B_n}$ and $\boldsymbol{B_p}$}
\label{appendix:BnBp}

\noindent In this appendix, we show by an explicit calculation that $B_n$ and $B_p$ commute when acting on the cylinder basis states $\Q$. First, one has that
\be
\left(\prod_nB_n\right)\triangleright\Q^{ij}_{rs}=\delta_{ijr}\delta_{ijs}\delta_{r00}\delta_{s00}\pic{Qijrs}\hs=\delta_{ij}\delta_{r0}\delta_{s0}\pic{Qjj00}\hs,
\ee
which implies that
\be\label{BpBvOnQ}
\left(B_p\prod_nB_n\right)\triangleright\Q^{ij}_{rs}=\f{1}{\D}\delta_{ij}\delta_{r0}\delta_{s0}\pic{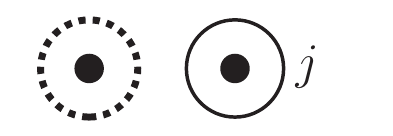}\hs=\f{1}{\D}v_j^2\delta_{ij}\delta_{r0}\delta_{s0}\pic{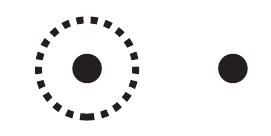},
\ee
where in the last step we have used the fact that the graph is defined on the punctured sphere. Note that, since we are on the sphere, in the $p$-punctured case it is sufficient to act with $B_p$ on $(p-1)$ punctures. If we choose to act on the left puncture of $\Q$, we get that
\be\label{BpOnQ}
B_p\triangleright\Q^{ij}_{rs}
&=\f{1}{\D}\pic{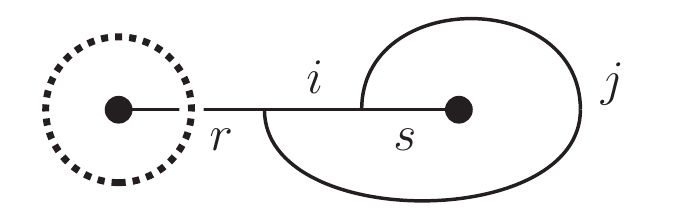}\nn\\
&=\f{1}{\D^2}\sum_kv_k^2\pic{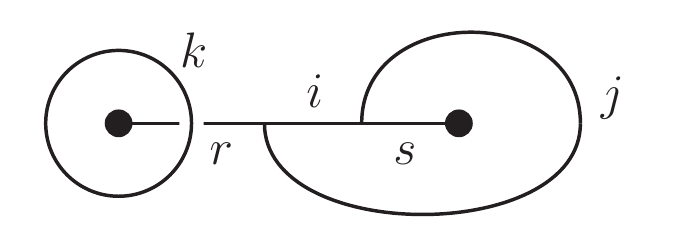}\nn\\
&=\f{1}{\D^2}\sum_{kl}\f{v_kv_l}{v_r}R^{kr}_l\pic{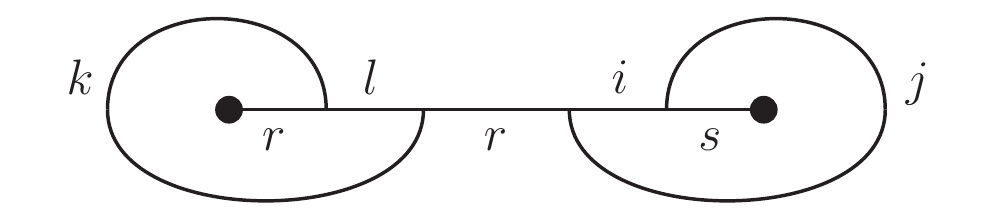}\hs,
\ee
which in turn implies that
\be
\left(\prod_nB_nB_p\right)\triangleright\Q^{ij}_{rs}
&=\f{1}{\D^2}\sum_{kl}\f{v_kv_l}{v_r}R^{kr}_l\delta_{ijr}\delta_{ijs}\delta_{klr}\delta_{r00}\delta_{s00}\pic{QijrsVacuumLoop3}\nn\\
&=\f{1}{\D^2}\sum_kv_k^2\delta_{ij}\delta_{r0}\delta_{s0}\pic{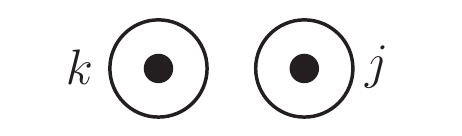}\nn\\
&=\f{1}{\D}v_j^2\delta_{ij}\delta_{r0}\delta_{s0}\pic{Q0000VacuumLoopLeft},
\ee
in agreement with \eqref{BpBvOnQ}. This calculation shows that the unique vacuum state on the cylinder is given by
\be\label{cylinder vacuum state}
\D B_p\triangleright\Q^{00}_{00}=\pic{Q0000VacuumLoopLeft}=\pic{Q0000VacuumLoopRight}.
\ee
It is crucial to recognize that the third equality of \eqref{BpOnQ} is well-defined because of the sliding property
\be\label{strand sliding}
\pic{StrandSliding1}=\pic{StrandSliding2},
\ee
which ensures that we can evaluate the crossing of the loop $k$ in \eqref{BpOnQ} either by going over the strand $r$ or over the strands $i$ and $j$. This sliding property follows simply from the definition and the properties of the $F$-symbols and the $R$-matrix. An explicit proof is given in appendix \ref{appendix2}.

\section{Turaev--Viro state sum}
\label{appendix:TV}

\noindent In this appendix, we briefly recall the definition of the TV state sum in the case of modular data coming from $\SU(2)_\k$. Let $M$ be a compact three-dimensional manifold with boundary $\partial M$ (with possibly disjoint components), $\Delta$ a triangulation of $\partial M$, and $\boldsymbol\Delta$ a triangulation of $M$ which agrees with $\Delta$ on the boundary. The triangulation $\boldsymbol\Delta$ consists of tetrahedra $\boldsymbol\Delta_3$, triangles $\boldsymbol\Delta_2$, edges $\boldsymbol\Delta_1$, and vertices $\boldsymbol\Delta_0$. Let $\psi$ be an admissible coloring with spins of the edges $\Delta_1$ of $\Delta$, and $\phi$ an admissible coloring of the edges $\boldsymbol\Delta_1$ that agrees with $\psi$. Then the TV state sum is defined as
\be\label{TV invariant}
\Z_\text{TV}(M,\Delta,\psi)=\sum_\phi\prod_{\Delta_0}\D^{-1}\prod_{\boldsymbol\Delta_0\backslash\Delta_0}\D^{-2}\prod_{\Delta_1}v\prod_{\boldsymbol\Delta_1\backslash\Delta_1}v^2\prod_{\boldsymbol\Delta_3}G,
\ee
where $G^{ijm}_{kln}\coloneqq F^{ijm}_{kln}/(v_mv_n)$ is the so-called totally tetrahedral-symmetric $6j$ symbol.

As suggested by the notation, $\Z_\text{TV}$ depends on a choice of boundary triangulation $\Delta$ and spin coloring $\psi$. Given such a choice, the state sum \eqref{TV invariant} returns a number which, by its topological nature, does not depend on the choice of bulk triangulation. We can think of this input data $\Delta$ and $\psi$ for the TV state sum as living in state spaces $\K_{\partial M,\Delta}$. Since there exist infinitely-many triangulations of a same two-dimensional manifold, there are infinitely-many spaces $\K_{\partial M,\Delta}$. However, for a given $\Delta$ each of these spaces is finite-dimensional and its elements correspond to a choice of spin coloring of the edges of $\Delta$. If the boundary has two components, say $\Sigma$ and $\Sigma'$ (or $M$ is a cobordism between $\Sigma$ and $\Sigma'$), the TV state sum can be seen either as a bi-linear functional on $\K_{\Sigma,\Delta}\otimes\K^*_{\Sigma',\Delta'}$ or as an operator $\K_{\Sigma,\Delta}\rightarrow\K_{\Sigma',\Delta'}$. In particular, if the three-manifold is of the form $M=\Sigma\times[0,1]$, then the state sum is a projector $\K_{\Sigma,\Delta}\rightarrow\K_{\Sigma,\Delta}$.

\section{Alternative choices of fusion basis}
\label{appendix:other fusion basis}

\be
\pic{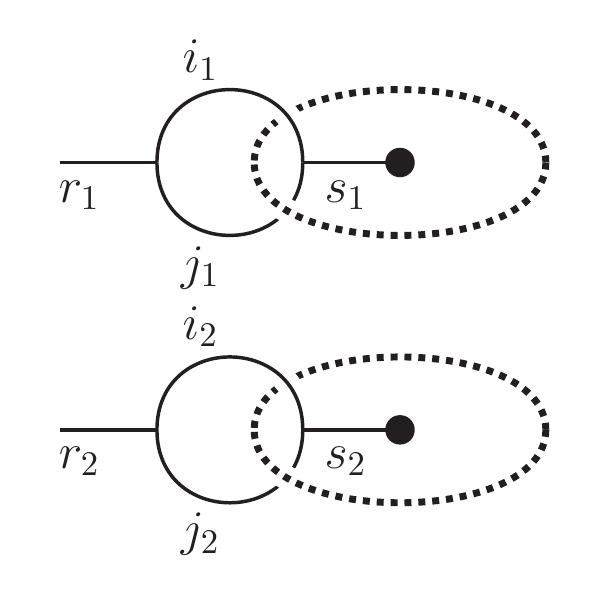}=\sum_{abmt}\f{v_a}{v_{i_1}v_{i_2}}R^{i_2j_1}_mF^{ar_1m}_{j_1i_2i_1}F^{j_1mi_2}_{r_2j_2b}F^{abt}_{r_2r_1m}\pic{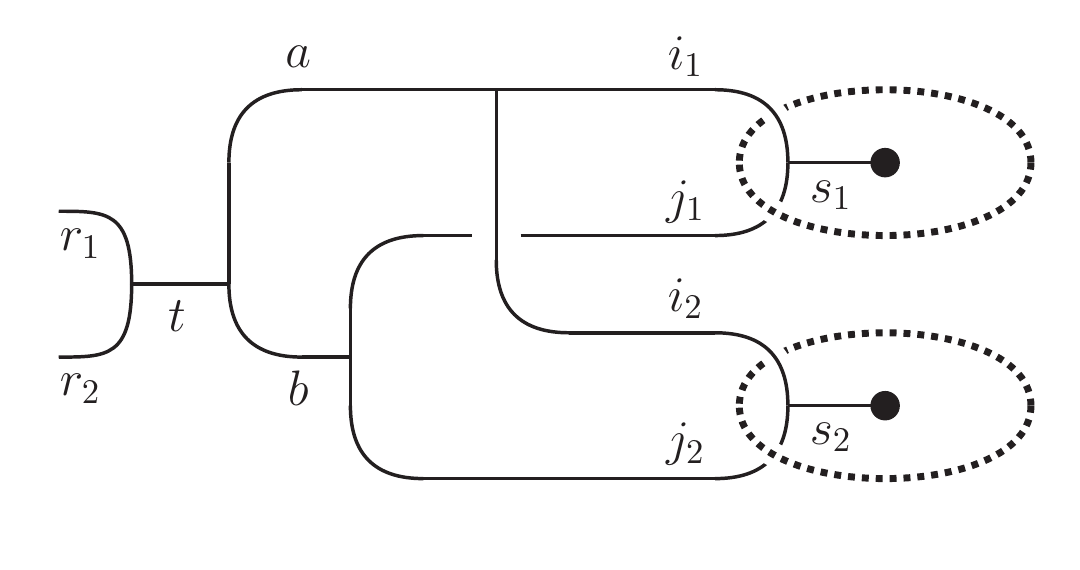}.
\ee
\be
\pic{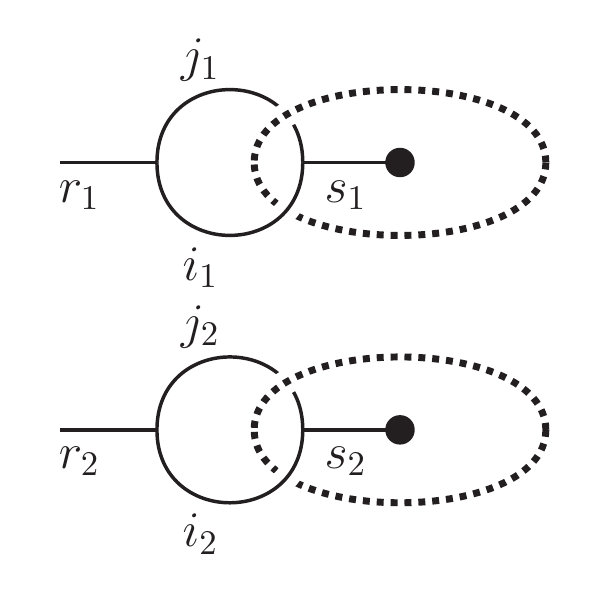}=\sum_{abmt}\f{v_a}{v_{i_1}v_{i_2}}\big(R^{i_1j_2}_m\big)^*F^{ar_2m}_{j_2i_1i_2}F^{j_2mi_1}_{r_1j_1b}F^{abt}_{r_1r_2m}\pic{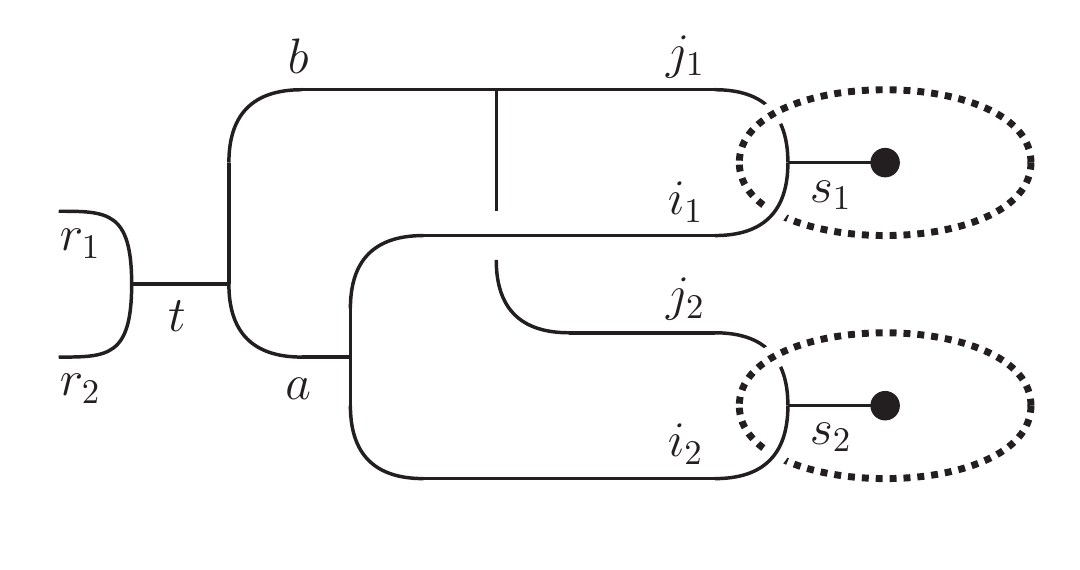}.
\ee
\be
\pic{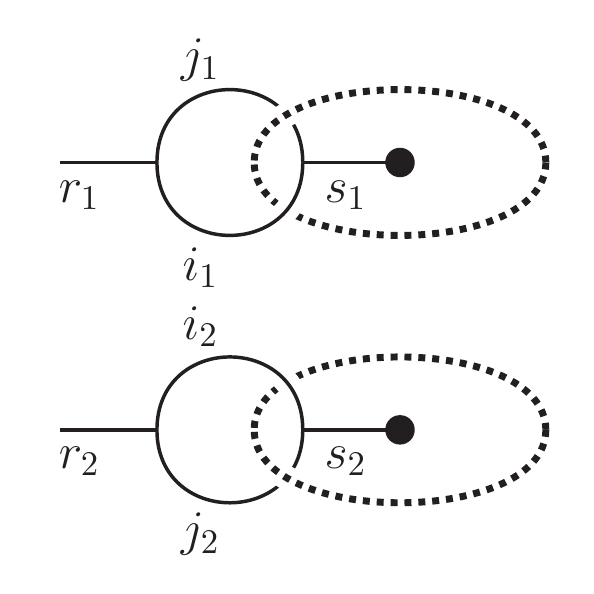}=\sum_{abmt}\f{v_a}{v_{i_1}v_{i_2}}\big(R^{i_1j_1}_{r_1}\big)^*R^{i_2j_1}_mF^{ar_1m}_{j_1i_2i_1}F^{j_1mi_2}_{r_2j_2b}F^{abt}_{r_2r_1m}\pic{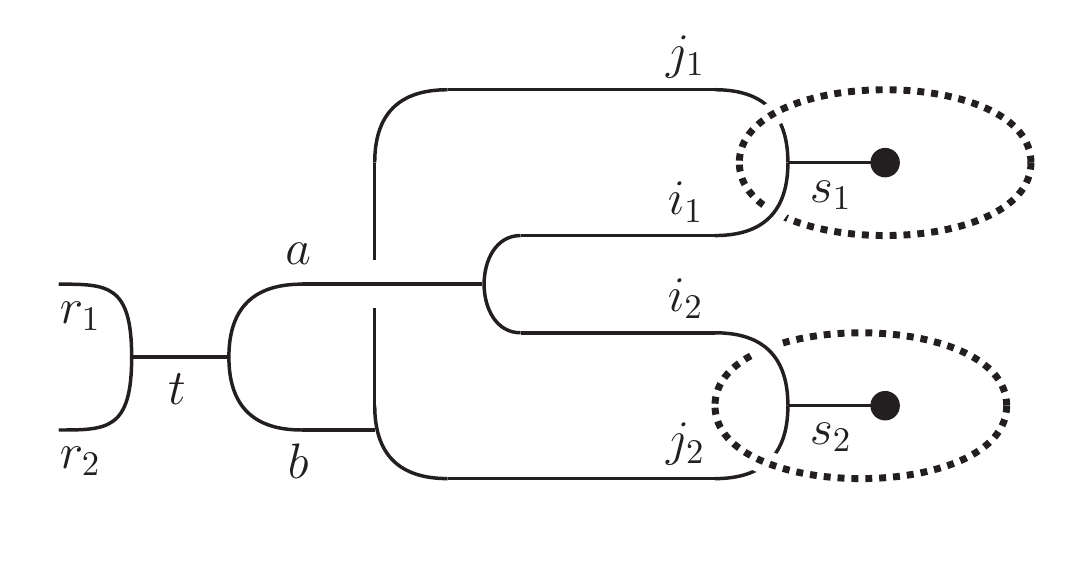}.
\ee

\section{Marked points, action of diffeomorphisms, and extended Hilbert spaces}\label{App_Last}

\noindent In this appendix, we briefly discuss subtleties related to the presence of marked points and the action of diffeomorphisms. The main structure which we have used in order to introduce degrees of freedom is that of the punctures, which are obtained by removing an infinitesimally small disk from the surface $\Sigma$ and by marking a point on the resulting $\mathbb{S}^1$ boundary (equivalently, one can also think of the punctures as points in$\Sigma$ with a tangent vector attached \cite{Kir}).

The marked points indicate the direction from which the open strands have to approach the punctures. Gauge-invariant states (which do not have open strands) therefore do not require the information provided by the marked points. The following discussion is only relevant if we do also allow torsion excitation at the punctures, i.e. open strands ending at the punctures. When allowing for the presence of open strands, the marked points provide important information allowing us to properly define the winding of an open strand reaching the puncture. The marked points are also needed in order to properly define the gluing of states along puncture boundaries.

Note also that the punctures are embedded, which includes also an embedding of the marked points. This has the important consequence two Hilbert spaces $\H_{\Sigma_p}$ and $\H_{\Sigma_{p'}}$ which agree in the number and positioning of the punctures, but disagree in the marked points, do define different Hilbert spaces. Moreover, these Hilbert spaces are not refinements of each other and, with our definitions so far, there also does not exist a common refinement. Further below we will sketch a framework introducing extended Hilbert spaces $\H^\text{ext}_{\Sigma_p}$, for which an arbitrary number of open strands can end at a given puncture. 

A diffeomorphism will in general also change the embedding of the punctures, and here we would like to briefly discuss the resulting action on a given Hilbert space $\H_{\Sigma_p}$. To this end, recall that diffeomorphisms can be thought of either as so-called active or passive transformations. These represent respectively a dragging of the geometrical points of the manifold, or a change of coordinates. In order to make the discussion clearer, let us focus on active diffeomorphisms, and fix once and for all a coordinate atlas for the punctured manifold (which we choose for simplicity to be a two-sphere). Two cases of the action of these active diffeomorphisms can already be discussed without ambiguity: the case of diffeomorphisms acting as the identity on the boundary of the punctures, and the case of diffeomorphisms rotating a puncture by $\pm2\pi$.

In the first case, when a (small) diffeomorphism acts as the identity on the boundary of the punctures, one is effectively considering a smooth deformation of the strands as in \eqref{StrandDeformation}. Therefore, these diffeomorphisms do not change the states since we have defined these latter to be equivalence classes under such smooth deformations.

Next, consider a rotation of $\pm2\pi$ of a puncture (also known as a Dehn twist). In this case, we still remain in the same Hilbert space since the location of the marked points before and after the diffeomorphism does agree. However, the rotations have a non-trivial action on the states. For example, when rotating the right puncture of a $\Q$ basis state by $\pm2\pi$ one obtains respectively
\be
\pic{Qijrs}\q\stackrel{+2\pi\vphantom{\f{1}{2}}}{\longrightarrow}\q\pic{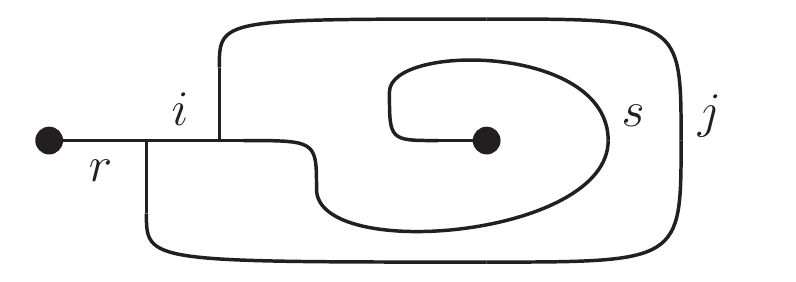}\hs=\Q^{ij}_{rs}\times\Q^{0s}_{ss}
\ee
and
\be
\pic{Qijrs}\q\stackrel{-2\pi\vphantom{\f{1}{2}}}{\longrightarrow}\q\pic{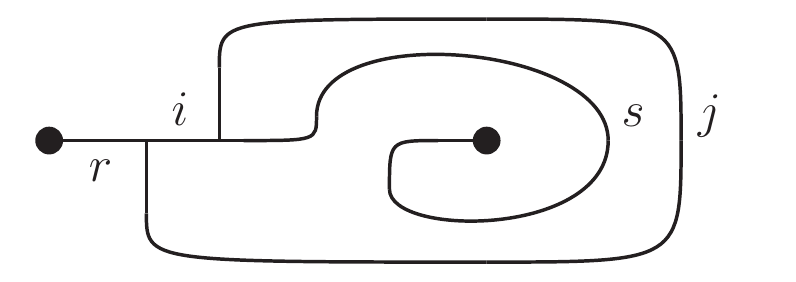}\hs=\sum_m\f{v_m}{v_s^2}\Q^{ij}_{rs}\times\Q^{ms}_{ss},
\ee
where the product of $\Q$ states on the right-hand sides can be computed using \eqref{QAlgebra}. In particular, because of the result of this product, one can see that the $\Q$ basis states are not eigenstates of the action of Dehn twists. As pointed out in \cite{KKR}, the states which diagonalize the action of rotations by $\pm2\pi$ are the $\O$ states. Indeed, one can check that, when rotating the right puncture, one obtains
\be
\pic{Oijrs}\q\stackrel{+2\pi\vphantom{\f{1}{2}}}{\longrightarrow}\q\O^{ij}_{rs}\times\Q^{0s}_{ss}=\left[\big(R^{is}_j\big)^*R^{sj}_i\right]\O^{ij}_{rs}=\big[\theta_i\theta_j^*\big]\O^{ij}_{rs}
\ee
and
\be
\pic{Oijrs}\q\stackrel{-2\pi\vphantom{\f{1}{2}}}{\longrightarrow}\q\sum_m\f{v_m}{v_s^2}\O^{ij}_{rs}\times\Q^{ms}_{ss}=\left[R^{is}_j\big(R^{sj}_i\big)^*\right]\O^{ij}_{rs}=\big[\theta_i^*\theta_j\big]\O^{ij}_{rs},
\ee
where we have first used \eqref{QtoO} to express the $\Q$ states in terms of the $\O$ states, and then the stacking property \eqref{O times O}. In these two calculations, the last equality has been obtained by using the explicit expression \eqref{R matrix expression} for the elements of the $R$-matrix. By doing so, we see that the eigenvalues are actually independent of the label $s$, and can be written in terms of the so-called topological phases $\theta_i\coloneqq\big(R^{ii}_0\big)^*$, in agreement with the results of \cite{KKR}. Therefore, one can see that the action of a Dehn twist on an $\O$ state produces a phase factor, and that the two above results are consistent with each other since a rotation by $\pm2\pi$ followed by a rotation by $\mp2\pi$ amounts to the identity.

Finally, we should discuss the more subtle case of diffeomorphisms rotating a puncture (and therefore its marked point) by an angle $\varphi\in(0,2\pi)$. One way to treat this case is to consider an extension $\H_{\Sigma_{p,s}}$ of the original Hilbert spaces $\H_{\Sigma_p}$, which consists in allowing for an arbitrary number $s$ of marked points (and therefore incoming strands) at a puncture. By using refinement maps that add a marked point with a strand of spin label $j=0$, we can then consider an inductive limit   on the number (and position) of marked points, resulting in an inductive limit Hilbert space $\H^\text{ext}_{\Sigma_p}$. The states obtained by rotating a given marked point by $\varphi$ then belong to the same (much larger) inductive limit Hilbert space.

Importantly, note that this extension of the Hilbert space does not change the classification of the excitations. At the mathematical level, this is due to a so-called Morita equivalence class of operator algebras. This means that the representations of these operator algebras lead to the same physical interpretation, namely that of a puncture as a quasi-particle carrying mass and spin. These different operator algebras are simply the ones obtained by considering the gluing of cylinder states with higher number of marked points. As explained in \cite{KK,Kong,Lan}, the tube algebra \cite{Ocneanu1,Ocneanu2,Kir} which we have considered in \ref{section:2pg} is the ``smallest'' representative in this Morita equivalence class.

In order to see that it is meaningful to consider a gluing algebra with higher number of marked points, let us consider for example punctures with two marked points and two incoming strands. In this case, we can show, along the lines of \eqref{O times O}, that the generalized $\O$ states also provide modules, i.e. satisfy a projector property under stacking. For this, let us consider the states
\be
\O^{ij}_{(r_1r_2r_3)(s_1s_2s_3)}=\pic{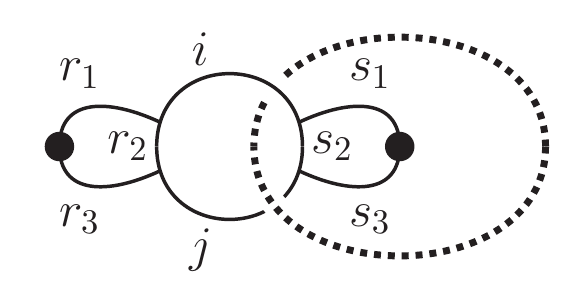}.
\ee
These states can be glued by identifying two by two the marked points of the gluing puncture, and we obtain the generalized stacking property
\be
\O^{ij}_{(r_1r_2r_3)(s_1s_2s_3)}\times\O^{i'j'}_{(s'_1s'_2s'_3)(t_1t_2t_3)}
&=\delta_{s_1s'_1}\delta_{s_3s'_3}\pic{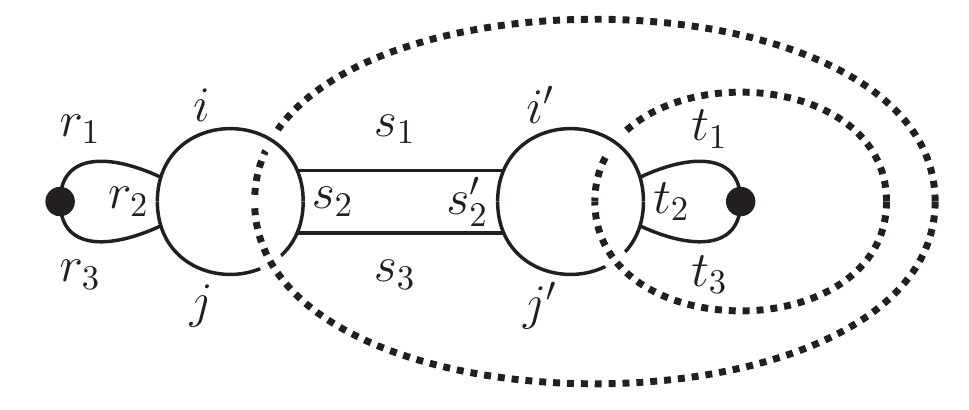}\nn\\
&=\delta_{s_1s'_1}\delta_{s_3s'_3}\pic{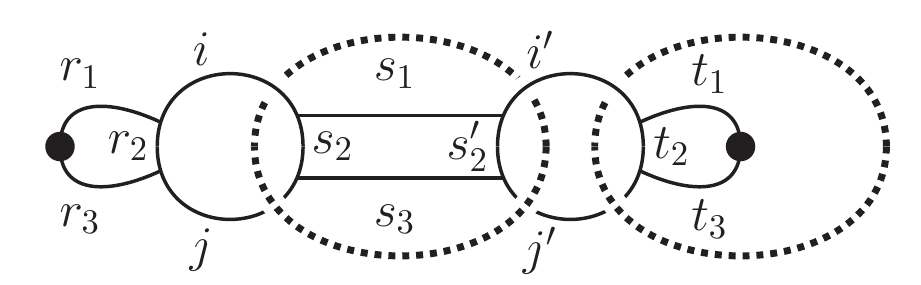}\nn\\
&=\delta_{s_1s'_1}\delta_{s_3s'_3}\sum_{mn}F^{s_1is_2}_{js_3m}F^{i's_1s'_2}_{s_3j'n}\pic{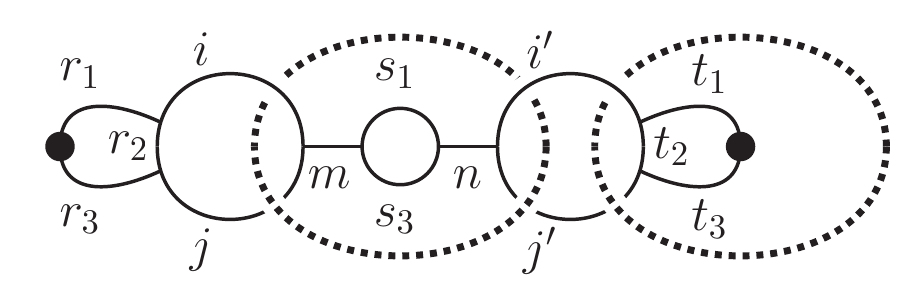}\nn\\
&=\delta_{s_1s'_1}\delta_{s_3s'_3}\sum_m\f{v_{s_1}v_{s_3}}{v_m}F^{s_1is_2}_{js_3m}F^{i's_1s'_2}_{s_3j'm}\pic{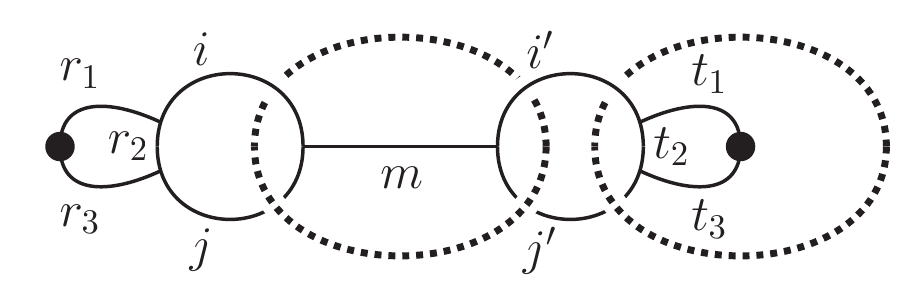}\nn\\
&=\D\delta_{s_1s'_1}\delta_{s_3s'_3}\delta_{ii'}\delta_{jj'}\f{v_{s_1}v_{s_3}}{v_iv_j}\sum_mF^{s_1is_2}_{js_3m}F^{is_1s'_2}_{s_3jm}\pic{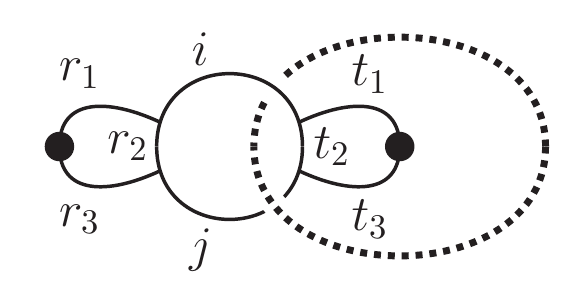}\nn\\
&=\D\delta_{s_1s'_1}\delta_{s_2s'_2}\delta_{s_3s'_3}\delta_{is_1s_2}\delta_{js_2s_3}\delta_{ii'}\delta_{jj'}\f{v_{s_1}v_{s_3}}{v_iv_j}\O^{ij}_{(r_1r_2r_3)(t_1t_2t_3)}.
\ee
These states can then be further generalized to define a fusion basis for the extended Hilbert space $\H^\text{ext}_{\Sigma_p}$. To define a continuum Hilbert space (without the superselection problem mentioned in section \ref{sec:embedding}) one can perform an inductive limit on the number (and positioning) of the punctures.

\end{document}